\documentclass[fleqn,usenatbib]{mnras}

\usepackage{newtxtext,newtxmath}
\usepackage[T1]{fontenc}
\DeclareRobustCommand{\VAN}[3]{#2}
\let\VANthebibliography\thebibliography
\def\thebibliography{\DeclareRobustCommand{\VAN}[3]{##3}\VANthebibliography}

\usepackage[normalem]{ulem}
\usepackage{graphicx, subfigure, algorithm, xcolor, algorithmic, amsmath, amssymb, float, rotating}

\newcommand{\rb}{\mbox{$\rm \text{R}_\text{birth}$}}
\newcommand{\rguide}{\mbox{$\rm \text{R}_\text{guide}$}}
\newcommand{\mgfe}{\mbox{$\rm [Mg/Fe]$}}
\newcommand{\alphafe}{\mbox{$\rm [\alpha/Fe]$}}
\newcommand{\feh}{\mbox{$\rm [Fe/H]$}}
\newcommand{\xfe}{\mbox{$\rm [X/Fe]$}}
\newcommand{\cefe}{\mbox{$\rm [Ce/Fe]$}}
\newcommand{\mnfe}{\mbox{$\rm [Mn/Fe]$}}
\newcommand{\cafe}{\mbox{$\rm [Ca/Fe]$}}
\newcommand{\cfe}{\mbox{$\rm [C/Fe]$}}
\newcommand{\alfe}{\mbox{$\rm [Al/Fe]$}}

\newcommand{\xh}{\mbox{$\rm [X/H]$}}
\newcommand{\fehC}{\mbox{$\rm [Fe/H](R = 0, \tau$)}}
\newcommand{\gradFeh}{\mbox{$\rm \nabla [Fe/H](\tau)$}}
\newcommand{\apogee}{APOGEE}
\newcommand{\galah}{\textsl{\textsc{galah}}}
\newcommand{\aspcap}{ASPCAP}
\newcommand{\sh}{\textsl{\textsc{StarHorse}}}
\newcommand{\logg}{\mbox{$\log g$}}
\newcommand{\teff}{\mbox{$T_{\rm eff}$}}
\newcommand{\rgal}{\mbox{$\rm R_\text{gal}$}}

\newcommand{\numstar}{87,426}
\newcommand{\numrb}{145,447}

\newcommand{\edits}[1]{#1}

\graphicspath{{./}{plots_anders/}}

\title{Unveiling the time evolution of chemical abundances across the Milky Way disk with APOGEE}

\author[Ratcliffe et al.]{Bridget Ratcliffe,$^{1}$\thanks{E-mail: bratcliffe@aip.de}
Ivan Minchev,$^{1}$
Friedrich Anders,$^{2,3,4}$
Sergey Khoperskov,$^{1}$
Guillaume Guiglion,$^{5}$\newauthor
Tobias Buck,$^{5}$
Katia Cunha,$^{6, 7, 8}$
Anna Queiroz,$^{1,9,10}$
Christian Nitschelm,$^{11}$
Szabolcs Meszaros,$^{12, 13}$\newauthor
Matthias Steinmetz,$^{1}$
Roelof S. de Jong,$^{1}$
Samir Nepal,$^{1,10}$
Richard R. Lane,$^{14}$
Jennifer Sobeck$^{15}$
\\
$^{1}$Leibniz-Institut f\"{u}r Astrophysik Potsdam (AIP), An der Sternwarte 16, 14482 Potsdam, Germany\\
$^{2}$Dept. de Física Quàntica i Astrofísica (FQA), Universitat de Barcelona (UB), C Martí i Franqués, 1, 08028 Barcelona, Spain\\
$^{3}$Institut de Ciències del Cosmos (ICCUB), Universitat de Barcelona (UB), C Martí i Franqués, 1, 08028 Barcelona, Spain\\
$^{4}$Institut d’Estudis Espacials de Catalunya (IEEC), C Gran Capità, 2-4, 08034 Barcelona, Spain\\
$^{5}$Max-Planck-Institut f\"{u}r Astronomie, Konigstuhl 17, D-69117 Heidelberg, Germany\\
$^{6}$Steward Observatory, University of Arizona, 933 North Cherry Avenue, Tucson, AZ 85721–0065, USA\\
$^{7}$Observat{\'o}rio Nacional, 77 Rua General Jos{\'e} Cristino, Rio de Janeiro, 20921-400, Brazil\\
$^{8}$Institut d'Astrophysique de Paris, UMR7095 CNRS, Sorbonne Universite, 98bis Bd. Arago, 75014 Paris, France\\
$^{9}$Laboratório Interinstitucional de e-Astronomia - LIneA, Rua Gal. José Cristino 77, Rio de Janeiro, RJ - 20921-400, Brazil\\
$^{10}$Institut f\"{u}r Physik und Astronomie, Universität Potsdam, Haus 28 Karl-Liebknecht-Str. 24/25, D-14476 Golm, Germany\\
$^{11}$Centro de Astronom{\'i}a (CITEVA), Universidad de Antofagasta, Avenida Angamos 601, Antofagasta 1270300, Chile\\
$^{12}$ELTE Gothard Astrophysical Observatory, H-9704 Szombathely, Szent Imre herceg st. 112, Hungary\\
$^{13}$MTA-ELTE Lend{\"u}let "Momentum" Milky Way Research Group, Hungary\\
$^{14}$Centro de Investigación en Astronomía, Universidad Bernardo O’Higgins, Avenida Viel 1497, Santiago, Chile\\
$^{15}$Department of Astronomy, University of Washington, Box 351580, Seattle, WA 98195, USA
}

\date{Accepted XXX. Received YYY; in original form ZZZ}

\pubyear{2023}

\begin{document}
\label{firstpage}
\pagerange{\pageref{firstpage}--\pageref{lastpage}}
\maketitle

\begin{abstract}

Chemical abundances are an essential tool in untangling the Milky Way's enrichment history. However, the evolution of the interstellar medium abundance gradient with cosmic time is lost as a result of radial mixing processes. For the first time, we quantify the evolution of many observational abundances across the Galactic disk as a function of lookback time and birth radius, \rb. Using an empirical approach, we derive \rb\ estimates for \numrb\ \apogee\ DR17 red giant disk stars, based solely on their ages and \feh. We explore the detailed evolution of 6 abundances (Mg, Ca ($\alpha$), Mn (iron-peak), Al, C (light), Ce (s-process)) across the Milky Way disk using 87,426 APOGEE DR17 red giant stars. We discover that the interstellar medium had three fluctuations in the metallicity gradient $\sim 9$, $\sim 6$, and $\sim4$ Gyr ago. The first coincides with the end of high-$\alpha$ sequence formation around the time of the Gaia-Sausage-Enceladus disruption, while the others are likely related to passages of the Sagittarius dwarf galaxy. A clear distinction is found between present-day observed radial gradients with age and the evolution with lookback time for both [X/Fe] and [X/H], resulting from the significant flattening and inversion in old populations due to radial migration. We find the \feh--\alphafe\ bimodality is also seen as a separation in the \rb--\xfe\ plane for the light and $\alpha$-elements. Our results recover the chemical enrichment of the Galactic disk over the past 12 Gyr, providing tight constraints on Galactic disk chemical evolution models. 

\end{abstract}


\section{Introduction} \label{sec:intro}

The precise individual element abundance measurements for large numbers of stars measured by large spectroscopic surveys have advanced the field of Galactic archaeology over the past decade. In particular, the seventeenth data release of the Apache Point Observatory Galactic Evolution Experiment \citep[\apogee;][]{apogeeDR17, Majewski2017} provides 20+ elemental abundance species for over 650,000 stars. Due to its deep coverage in the near infrared, \apogee\ observes stars throughout the Galactic disk \citep{Zasowski2017_apogee2}, and thus is able to provide a detailed chemical map.

The relatively unchanging\footnote{In the present study, we assume that most of the photospheric abundances reflect the abundance pattern in a star's birth cloud.} chemical abundances of stellar atmospheres give insight into the star's birth properties \citep{Tinsley1979, 2002freeman-BH, 2018Minchev_rbirth, 2022Ratcliffe} and can reveal important evolutionary events during the Milky Way's lifetime \citep[e.g.][]{Pagel2009, Matteucci2012, Frankel2018, Helmi2018_gse, horta2022_haloStreams, Belokurov2022, Conroy2022, Khoperskov2022c, Lu2022_Rb}. However, despite the increased number of chemical abundances spanning multiple nucleosynthetic families measured for many stars across the Galactic disk, there are still many unresolved questions.
For instance, the origin of the well-known bimodality in the \feh--\alphafe\ plane is still under debate \citep{Grisoni2017, mackereth2018origin, Clarke2019, 2020_buckchemical, 2020Lian_alphaDichotomy, 2021Sharma, Agertz2021_vintergatanI, Khoperskov2021}.  

It has also been found that some elements capture similar information, meaning the chemical abundance space collapses onto a lower dimensional manifold which roughly corresponds to the different nucleosynthetic families \citep{ting2012principal, PJ2018, Ness2019, Weinberg2019, 2020Ratcliffe, 2021Griffith, 2022Ness}. While a lower-dimensional abundance space may capture the broad strokes of chemical enrichment, each element uniquely contributes to the chemical evolution of the Galaxy (e.g., \citealt{Kobayashi2020, Matteucci2021}). In order to best understand the Milky Way's intricate evolution, it is crucial to describe the way abundances vary across the disk with time. Classical chemical evolution modeling and more modern hydrodynamical simulations have had great success characterizing the Galactic abundance evolution \citep[e.g.][]{Steinmetz1994, Chiappini1997, Minchev2013, sanders15, Kubryk2015, 2020_buckchemical, 2020FIRE_sanderson, 2021Johnson, Bellardini2022}, however there is still tension between observed and expected data modeled with current nucleosynthetic yield tables \citep{blancato2019variations, Rybizki2017_chempy, 2021BuckHD_chemEnrich, Palla2022}. Previous work has utilized open clusters to trace chemical evolution by looking at radial gradients conditioned on age \citep[e.g.][]{Janes1979, Friel1995, Chen2003, Magrini2009, Netopil2016, Zhang2021, Myers2022, Magrini2023, 2022spina}. However, differences in cluster distance estimates lead to variations in the radial gradient measurements of up to 0.03 dex/kpc, even when keeping the set of abundance measurements constant \citep{Donor2018}. Individual element abundance gradients have also been measured for field stars using present-day radii \citep[e.g.][]{Mayor1976, Grenon1987, Daflon2004, Genovali2014, Braganca2019, Eilers2022} or guiding-centre radii \citep[e.g.][]{Casagrande2011, Anders2014}. 

It is now well established that stars migrate away from their birth sites (e.g., \citealt{Selwood2002, Roskar2008, Quillen2009, Schonrich2009, Minchev2010, Grand2012, Frankel2018, Carr2022}), causing the measured present-day abundance gradients to be flatter than the environment they were born in (\citealt{Pilkington2012, Minchev2012a, Minchev2013, Kubryk2013}; see also Figure 11 in \citealt{Vincenzo2020}). Thus, radial migration has begun to wash away chemical signatures in the disk. Knowledge of birth radius (\rb) can help answer fundamental questions still not yet understood, such as the now-lost evolution of the interstellar medium abundance gradient with cosmic time. With knowledge of birth radii, we would be able to go beyond estimating radial abundance gradients from conditioning on age using present Galactic radii --- and therefore getting an incomplete story of the Milky Way's evolution --- and have a clearer understanding of the assembly history of the Galaxy. The origin of the high- and low-$\alpha$ sequences could be understood, in addition to the relative strength of different nucleosynthetic events at specific moments and places in the Galactic disk. 

One way to recover birth radii is to leverage the stars' chemical abundances to recover their birth groups, which are believed to be chemically homogeneous \citep{BH2010}. However, it has become clear in recent years that the hopes for the feasibility of strong chemical tagging (the idea of directly relating the individual members of dispersed star clusters to their birth sites based only on spectroscopy, in the terminology of \citealt{2002freeman-BH}) are slim, at least with regards to the Galactic disc. While some studies used the strong chemical tagging technique to find possibly dispersed star clusters (e.g. \citealt{PriceJones2020}), it has been shown that even in the most favourable conditions (homogeneous high-resolution spectral analysis of only red clump stars with no contamination from field stars), more than 70\% of the recovered statistical groups (using the most advanced clustering methods) contain stars belonging to different real clusters \citep{Casamiquela2021}. 
While strong chemical tagging is thus discarded for the time being, the prospects for weak chemical tagging in the disc are extremely favourable thanks to the ever-growing amount of data from high-resolution spectroscopic surveys.

\cite{2018Minchev_rbirth} devised a method for obtaining stellar birth radii in any observational sample with good age and metallicity measurements. This method consisted of simultaneously recovering stellar birth radii and the metallicity evolution in the Galactic disk with radius and time, \feh(R,$\tau$), by varying the metallicity slope at any given lookback time so that the resulting \rb\ distributions remained physically meaningful. It should be emphasized that this technique was almost completely independent of any analytical or numerical chemo-dynamical models; the only constraint used was that stars were required to be born at physically meaningful radii and the \rb\ distributions of older populations peaked at smaller radii, as expected for inside-out disk formation. Once in possession of \rb, \cite{2018Minchev_rbirth} constrained a number of chemo-kinematic relations with only a handful of stars from AMBRE:HARPS \citep{DePascale2014_ambre, Hayden2017_ambre} and HARPS-GTO \citep{Adibekyan2012, DelgadoMena2017}, combined with stellar ages computed with the \sh\ code \citep{Queiroz2018_starhorse, Anders2018}. Recently, \cite{Lu2022_Rb} discovered the linear relation between metallicity scatter for a given age and the metallicity birth gradient evolution with cosmic time, thus making the \rb\ determination fully self-consistent. The results of the above work have opened up a new way of studying the chemo-dynamical evolution of the Milky Way, where instead of relying on forward modeling over cosmic time (cosmological simulation or semi-analytical modeling), one gets there directly from the data with few prior assumptions on the detailed chemical enrichment. 

\begin{figure*}
     \centering
     \includegraphics[width=.329\textwidth]{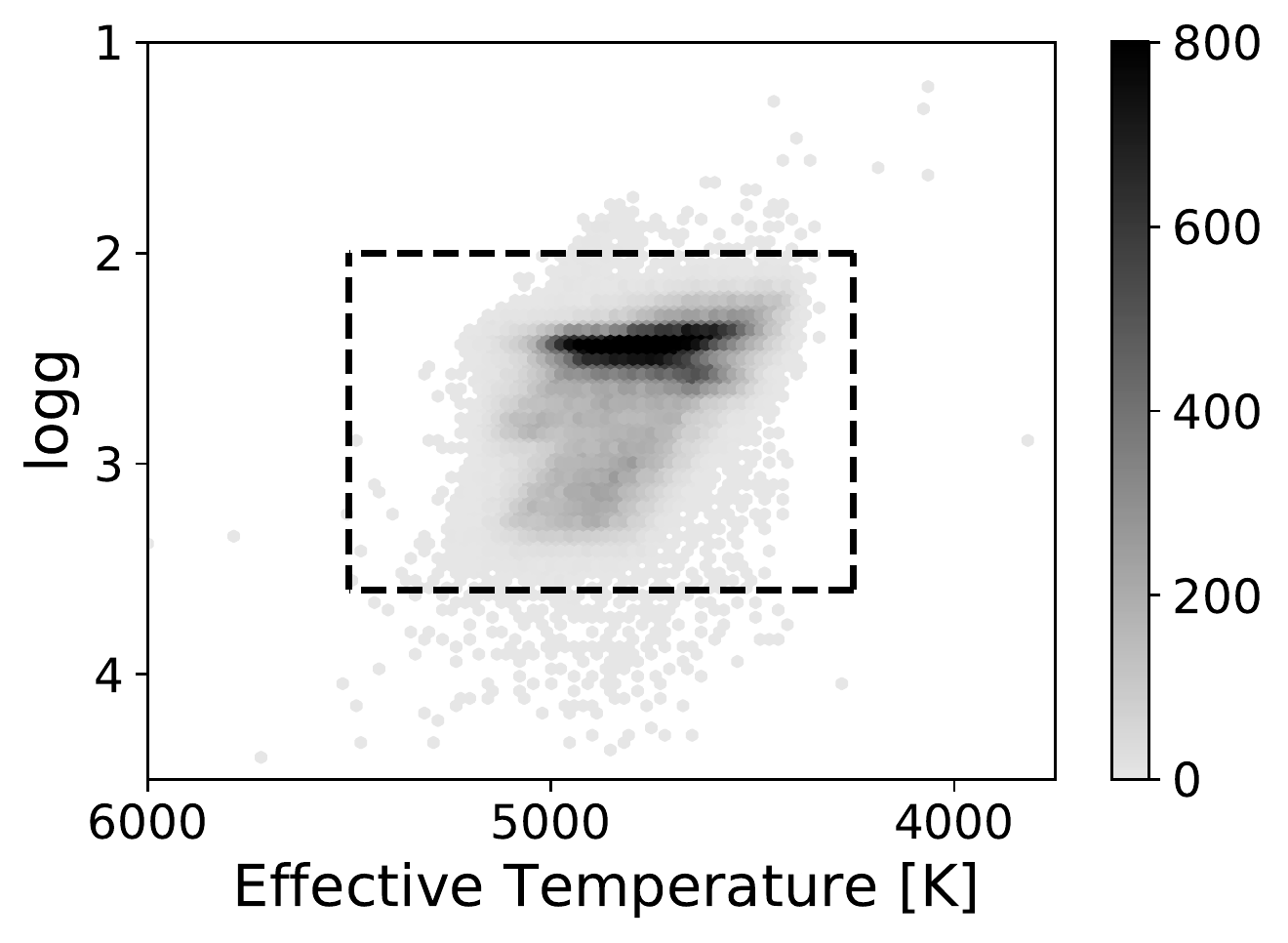}
     \includegraphics[width=.329\textwidth]{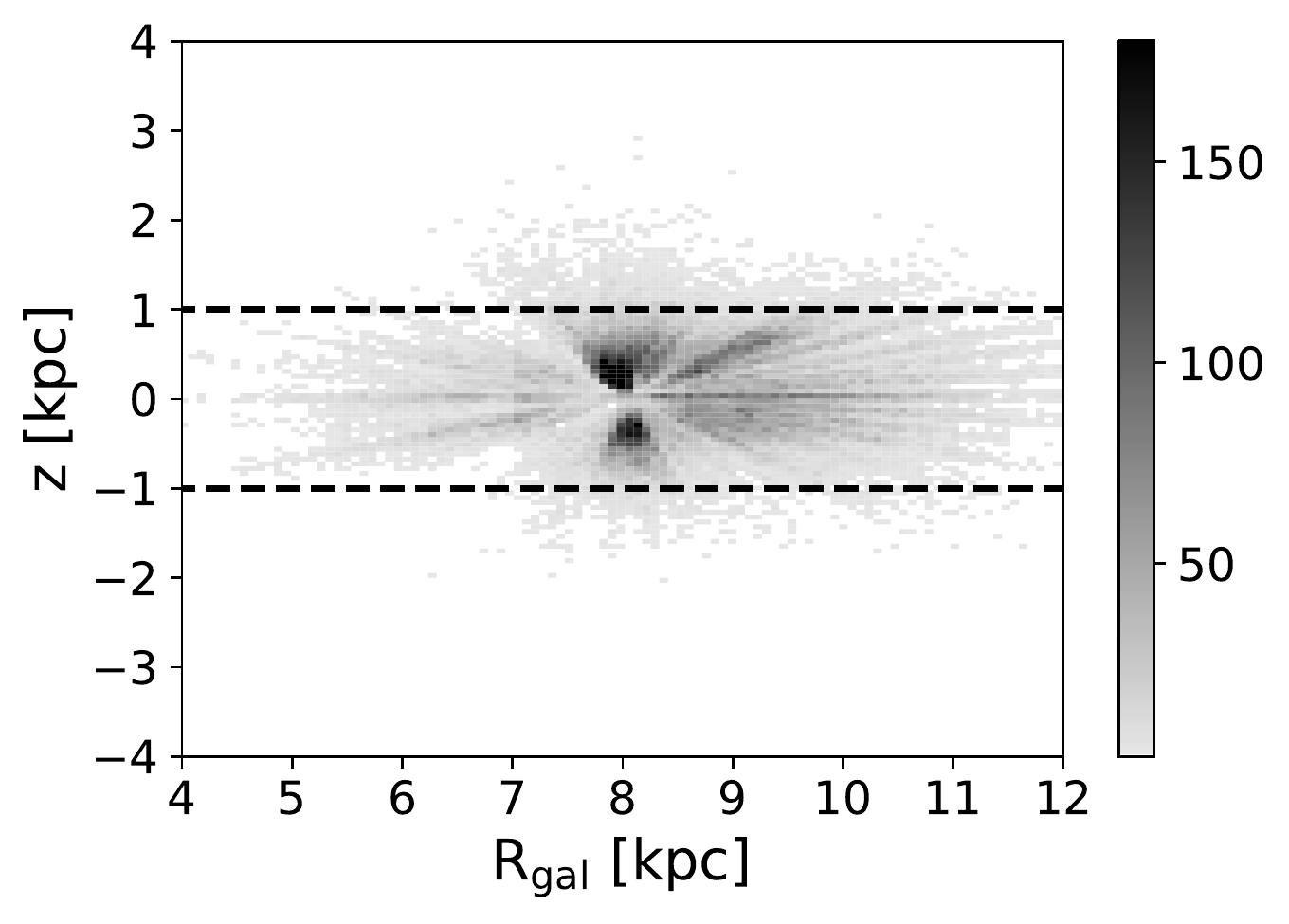}
     \includegraphics[width=.329\textwidth]{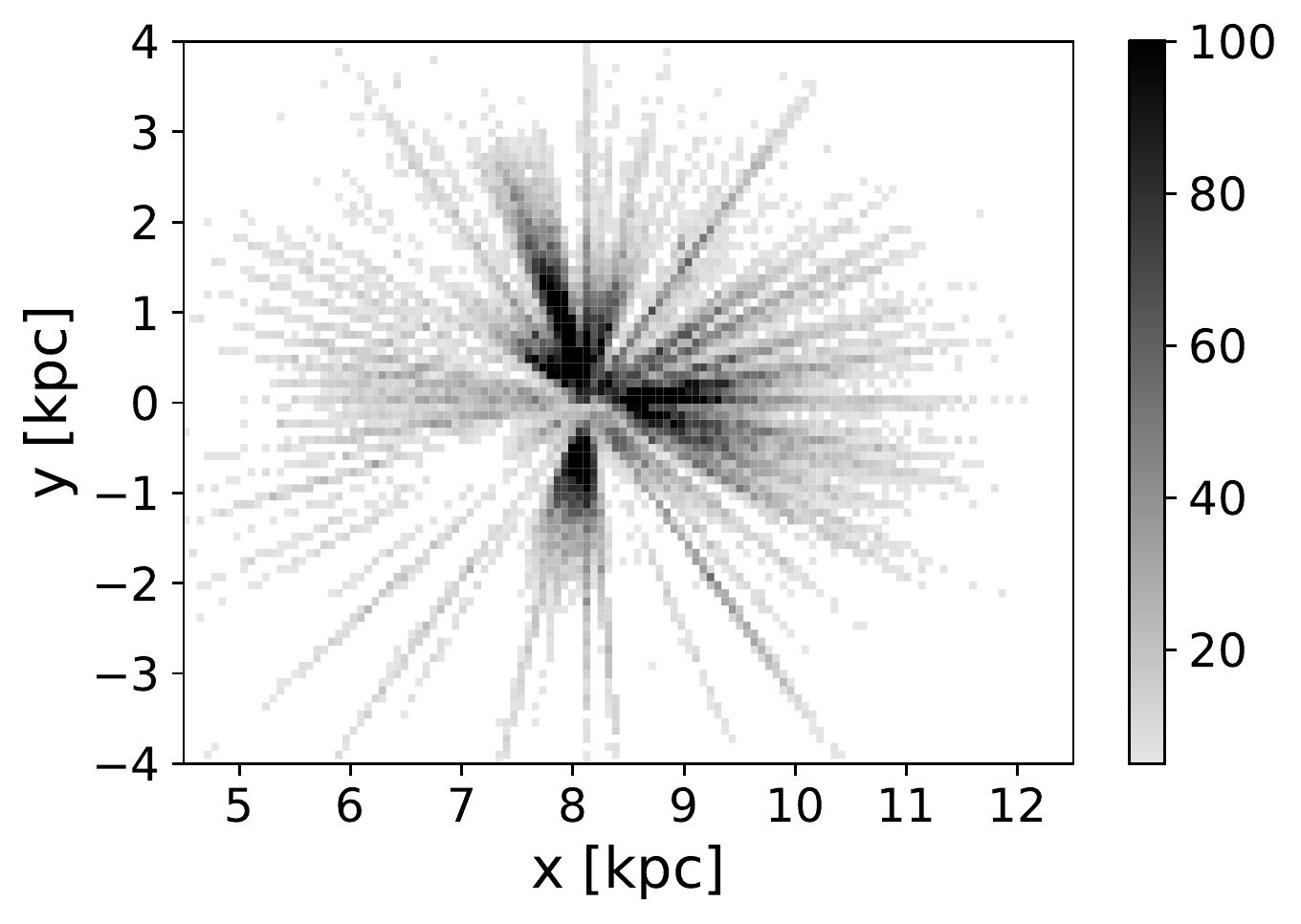}
\caption{\textbf{Left:} 2D histogram of the Kiel diagram of the \apogee\ DR17 catalog with \protect\cite{Anders2023_ages} ages. The black dashed lines indicate the region of the red giant branch that is used in this work, which is chosen to minimize the effects of atomic diffusion and first dredge-up on abundances and ages. \textbf{Middle:}  \rgal--z spatial distribution of \apogee\ DR17 red giant stars with abundance and age quality cuts as described in Section \ref{sec:data}. The black dashed lines show the $z$ cuts we used to choose disk stars. \textbf{Right:} 2D histogram of the Galactic $x-y$ plane of the \numstar\ red giant disk stars used to measure abundance variation across time. The sun is located at (8.125, 0, 0.02) kpc \protect\citep{Gravity2018, Bennett2019}. Only bins with $>10$ stars are shown.}
\label{fig:HR_diagram}
\end{figure*}

If \rb\ can be recovered by using the observed age and \feh\ \citep{2018Minchev_rbirth, Lu2022_Rb}, as it also appears to be the case in cosmological simulations of Milky Way-like galaxy disks \citep{2020Feltzing}, then this should have implications for all other chemical elements. Indeed, \cite{Ness2019}, \cite{Sharma2022}, and \cite{Carrillo2022} (using \apogee, \galah, and cosmological zoom-in simulations respectively) demonstrated that for stars in small bins of \feh\ and age, all other abundances had very small scatter. More recently, \cite{Hayden2022} used the correlation between age, \feh, and elemental abundances to estimate ages for the \galah\ dataset, thus, using properties of stars born at the same time and place without actually recovering \rb\ (see also \citealt{He2022}).

In this paper, we investigate the time evolution of chemical abundances across the Galactic disk using \rb\ derived in this work for \apogee\ DR17 stars with ages derived by \citet{Anders2023_ages} following the method of \cite{Lu2022_Rb}. Specifically, the questions we wish to answer are: (i) what do the $\alpha$-sequences look like across time and \rb, and (ii) how do abundance gradients evolve, and what do they reveal about the different nucleosynthetic processes? \rb\ estimates for a high resolution spectroscopic survey, like \apogee, will provide light on these questions, as we want to probe the Galaxy's evolution in as fine detail as possible. While \rb\ can only be inferred under some modeling assumptions, the relative \rb\ for a given age will be a useful tool in investigating the Milky Way's past.

The paper is organized as follows. The data and method for estimating \rb\ used in this work are discussed in Sections \ref{sec:data} and \ref{sec:Rb_method} respectively. Section \ref{sec:results} details our results, with analyses of \xh\ and \xfe\ abundance gradients across time in Sections \ref{sec:xh} and \ref{sec:xfe}, and the time evolution of the \feh--\mgfe, age--\xfe, and age--\xh\ planes is discussed in Section \ref{sec:enrichment}. Sections \ref{sec:discussion} and \ref{sec:conclusions} provide the discussion and key conclusions of this paper. 

\section{data} \label{sec:data}

We use data provided by the seventeenth data release of \apogee\ \citep{apogeeDR17, Majewski2017} of the fourth phase of the Sloan Digital Sky Survey \citep[SDSS-IV;][]{blanton2017sloan}. The \apogee\ survey collects data from both the northern and southern hemispheres using two high-resolution spectrographs \citep{Wilson2019_apogeeRes} on the 2.5 m Sloan Foundation telescope at Apache Point Observatory \citep{Gunn2006_apogeeTelescope} and the 2.5 m du Pont telescope at Las Campanas Observatory \citep{Bowen1973_apogeeTelescope}. The data reduction pipeline is outlined in \cite{Nidever2015_apogeePipeline}, and stellar parameters and abundances are processed by the \apogee\ Stellar Parameter and Chemical Abundance Pipeline \citep[\aspcap;][]{GP2016, Jonsson2020, Holtzman2015} and the \apogee\ line list \citep{Smith2021}. 

We partner the abundances with kinematics from the astroNN catalog \citep{LeungBovy2019a, Mackereth2018} and 
spectroscopic stellar ages for 185,282 stars from \cite{Anders2023_ages}, \edits{which have a median uncertainty of 1 Gyr and relative uncertainty $<25$\%}. The ages are estimated from stellar parameters and elemental abundance ratios (most importantly $T_{\rm eff}$, [C/Fe], [N/Fe], [Mg/Fe], and $\log g$; not including [Fe/H]) using the supervised machine learning technique {\tt XGBoost} trained on a set of more than 3\,000 red-giant and red-clump stars with asteroseismic ages \citep{Miglio2021}. We calculate the guiding radius (\rguide) from the cylindrical rotational velocity: $$\rguide = \rgal V_\phi / V_0,$$
where $V_\phi$ is the Galactocentric azimuthal velocity and $V_0=229.76$ km/s is the Milky Way rotation curve at solar radius \citep{Bovy2012_velocityCurve, Schonrich2010_lsr}. 

Since our \rb\ determination method is directly dependent on age and \feh, we select a sample with relatively small uncertainties ($\sigma_{\rm age} < 1.5$ Gyr and $\sigma_{\rm \feh}< 0.015$ dex), and remove flagged \feh. In order to avoid systematic abundance trends with evolutionary state \citep{2019Jofre, Liu2019, 2017ApJ...840...99D, Souto2018, Gruyters2014, Shetrone2019, Lian2022}, we focus on red giant ($2 < \logg < 3.6$, and $4,250$ K $ \leq \teff \leq 5,500$ K) disk ($|z| \leq 1$ kpc, eccentricity $ < 0.5$, $|\feh| < 1$) stars. The left panel of Figure \ref{fig:HR_diagram} illustrates the \logg\ and \teff\ cuts employed in this work. This leaves us with \numrb\ stars to use for inferring birth radii. 

For the abundance gradient analysis, we use calibrated \aspcap\ \apogee\ DR17 abundances (\xfe\ for X = Mg, Ca ($\alpha$), Mn (iron-peak), Al (light odd Z), C (light), Ce\footnote{We also examine Ce trends using \cefe\ provided in the BAWLAS catalog. See Section \ref{sec:bawlas} in the Appendix.} (s-process)). To ensure the highest quality sample, we only keep stars with unflagged abundances, and \xfe\ measurement uncertainties $< 0.1$ dex for Ce and $< 0.03$ dex for the other elements. We also remove spurious measurements by using stars with $-1 \leq \xfe \leq 1$, and employ a signal-to-noise ratio cut of $SNREV>$ 65. 

Due to the few stars with ages $>12$ Gyr, we are not able to reliably estimate the metallicity gradient beyond then.\footnote{Additionally, we note that the uncertainties of both the age \citep{Anders2023_ages} and the \rb\ determination methods (Sect. \ref{sec:Rb_method}) grow with age, meaning that for older ages our method becomes less reliable.} Therefore, we focus our analysis on stars with age $\leq 12$ Gyr. This gives us a sample size of \numstar\ stars to measure how abundances vary across the disk with cosmic time. The distribution of these stars in the Galactic $x-y$ plane is shown in the right panel of Figure \ref{fig:HR_diagram}.

\begin{figure*}
     \centering
     \includegraphics[width=.44\textwidth]{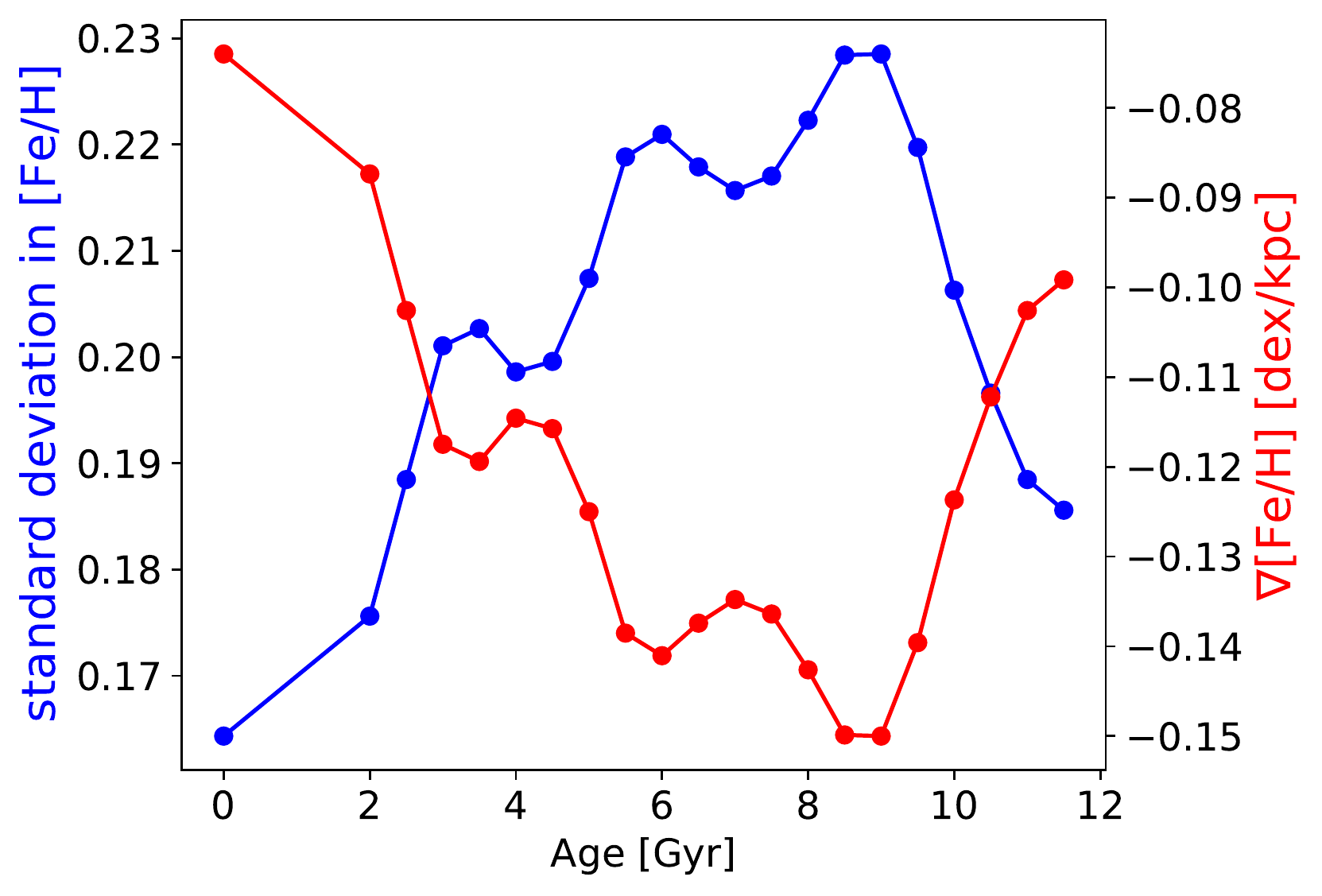}
     \includegraphics[width=.425\textwidth]{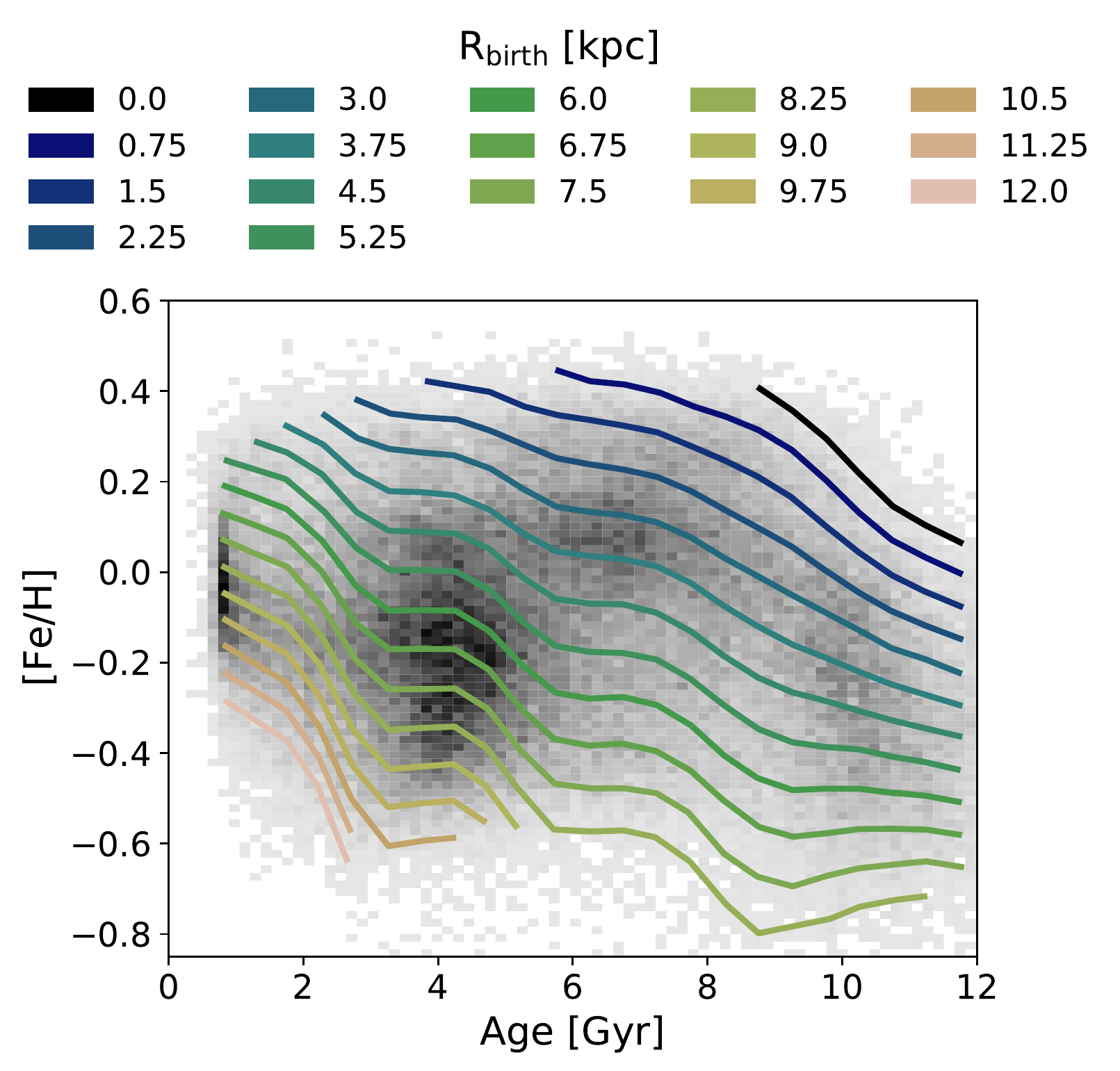}\\
     \caption{\textbf{Left:} The scatter in \feh\ for a given age bin (blue) and the inferred metallicity gradient (red). The standard deviation in metallicity and the gradient are related through Eq. \ref{eqn:Grad_range}, once the \feh\ standard deviation is normalized between 0 and 1. \textbf{Right:} The AMR of \numrb\ \apogee\ DR17 red-giant disk stars with ages from \protect\cite{Anders2023_ages}. The color-coded lines represent the running mean of mono-\rb\ populations with $\Delta$\rb$=0.75$ kpc and mean values given by the color bar. The shape of each \rb\ track is determined from \fehC\ and \gradFeh. The fluctuations in \gradFeh\ at $\sim 6$ and $\sim 4$ Gyr seen in the left panel break the monotonic increase in \feh\ for outer-disk mono-\rb\ populations and even appear to dilute the ISM $4-5$ Gyr ago.}
\label{fig:feh0-range}
\end{figure*}

\section{Method for estimating birth radii} \label{sec:Rb_method}

We follow the method from \cite{Lu2022_Rb} to estimate birth radii using stellar ages and metallicities. \edits{This method assumes that the birth metallicity gradient is always linear in radius, as supported by both observations (\citealt{Deharveng2000, Esteban2017, ArellanoCordova2021}) and simulations (\citealt{Vincenzo2018, Lu2022_sims}). Therefore for any lookback time ($\tau$) we can write \feh\ as a function of the metallicity gradient at that time (\gradFeh), birth radius, and the metallicity at the Galactic center (\fehC): } 
\begin{equation} \label{eqn:feh_Rb}
\feh(\rb, \tau) = \gradFeh\rb + \fehC.
\end{equation}
\edits{Rearranging Equation \ref{eqn:feh_Rb}, we are able to estimate \rb\ as a function of metallicity and age:}
\begin{equation} \label{eqn:rb}
\rb(\text{age}, \feh) = \frac{\feh - \fehC}{\gradFeh},
\end{equation}
where $\tau$ is lookback time and age is observable. \edits{This method also assumes the star-forming gas in the Milky Way is azimuthally chemically homogeneous, in agreement with observations \citep{ArellanoCordova2021, Esteban2022}, and that the full intrinsic metallicity range is well sampled in each age bin.}

The time evolution of the metallicity at the Galactic center, \fehC, is estimated from the upper envelope of the sample's age-metallicity relation (AMR) for $\geq 10$ Gyr. The upper envelope is defined as the maximum \feh\ for a given age bin, after removing outliers using the interquartile method. The smooth enrichment of the Galactic center for $\tau < 10$ Gyr is modeled by a log function, where \feh(R = 0, $\tau = 0)$ is chosen under the assumptions that the youngest stars near the solar neighborhood have a metallicity of 0.06 dex \citep{Nieva2012, Asplund2021} and the present-day gradient is $-0.074$ dex/kpc for the youngest stars \citep{Zhang2021, Myers2022}.

\edits{Using the cosmological hydrodynamic simulations of \cite{Buck2019, Libeskind2020, Khoperskov2022c}, \cite{Lu2022_Rb} showed that the range of [Fe/H] (defined as the $95^\text{th}$ minus the $5^\text{th}$ percentile) in a given age bin over time represents the shape of the metallicity gradient as a function of lookback time:}
\begin{equation} \label{eqn:Grad_range}
    \gradFeh = a\, \text{Range}\widetilde{\feh}(age) + b,
\end{equation}
\edits{where parameters $a$ and $b$ are constants estimated from the boundary conditions and $\text{Range}\widetilde{\feh}(age)$ is the metallicity range normalized to have a minimum of 0 and a maximum of 1 across all age bins. When the scatter in \feh\ is the smallest across all ages ($\text{Range}\widetilde{\feh}(age) =0$), $\gradFeh = b$, which sets $b$ as the current gradient of star-forming gas in the Milky Way, or -0.074 dex/kpc \citep{Zhang2021}. When the scatter in [Fe/H] is the largest across all ages ($\text{Range}\widetilde{\feh}(age) =1$), $\gradFeh = a + b$. This value is the steepest we would expect the metallicity gradient to have ever been, which we set to $-0.15$ dex/kpc as found in other works \citep{2018Minchev_rbirth, Lu2022_Rb}, and thus forces $a = -0.076$ dex/kpc. Different values for the steepest metallicity gradient (and therefore $a$) are explored in Appendix \ref{sec:varyGrad}}. After removing outliers using the interquartile method, we employ the standard deviation in \feh\ as a proxy for the metallicity range for $2< age <12$ Gyr. The present-day standard deviation in \feh\ is then extrapolated using a linear fit from the $2-6$ Gyr values. Finally, to estimate \rb\ we plug Eq. \ref{eqn:Grad_range} into the denominator of Eq. \ref{eqn:rb}. The average uncertainty in \rb\ is 0.7 kpc, estimated from 100 Monte Carlo samples. 

The left panel of Figure \ref{fig:feh0-range} shows the measured scatter in metallicity (blue curve) and the inferred radial gradient (red curve). To best trace the substructure in the data, we estimate \fehC\ and \gradFeh\ using age bins of width 0.5 Gyr and smooth the functions over a 1-Gyr window \edits{(i.e. calculate the running mean with a window size of 3). This ensures a smooth metallicity gradient that removes artificial discontinuities and minimizes artifacts due to noise}. We find a quick steepening in \gradFeh\ between 12 and 10 Gyr ago, which \cite{Lu2022_Rb} linked to the effect of the Milky Way last massive merger -- the Gaia-Sausage-Enceladus (GSE) event \citep{Belokurov2018, Helmi2018_gse}. Instead of the monotonic flattening of the metallicity slope for $age<9$ Gyr found by \cite{Lu2022_Rb}, however, with our dataset we recover two additional fluctuations in \gradFeh\ $\sim6$ and $\sim4$ Gyr ago.
 
The right panel of Figure \ref{fig:feh0-range} presents the AMR stellar density distribution of our \apogee\ red giant sample. The color-coded curves indicate the running mean of mono-\rb\ populations. The fast increase in metallicity between 12 and 10 Gyr ago corresponds to the quick steepening in the gradient, as seen in the left panel. This initial burst in star formation has long been linked to the formation of the high-$\alpha$ population in the \feh--\alphafe\ plane (e.g., \citealt{Chiappini1997, Matteucci2021}). The strong ripples in \gradFeh\ at lookback times of $\sim6$ and $\sim4$ Gyr seen in the left panel interrupt the monotonic increase in \feh\ for mono-\rb\ populations at larger radii, and even appear to dilute the ISM in the range $4\lesssim age\lesssim 5$ Gyr. We discuss the possible causes of these fluctuations in Section \ref{sec:discussion_peaks}.

\begin{figure}
     \centering
     \includegraphics[width=.375\textwidth]{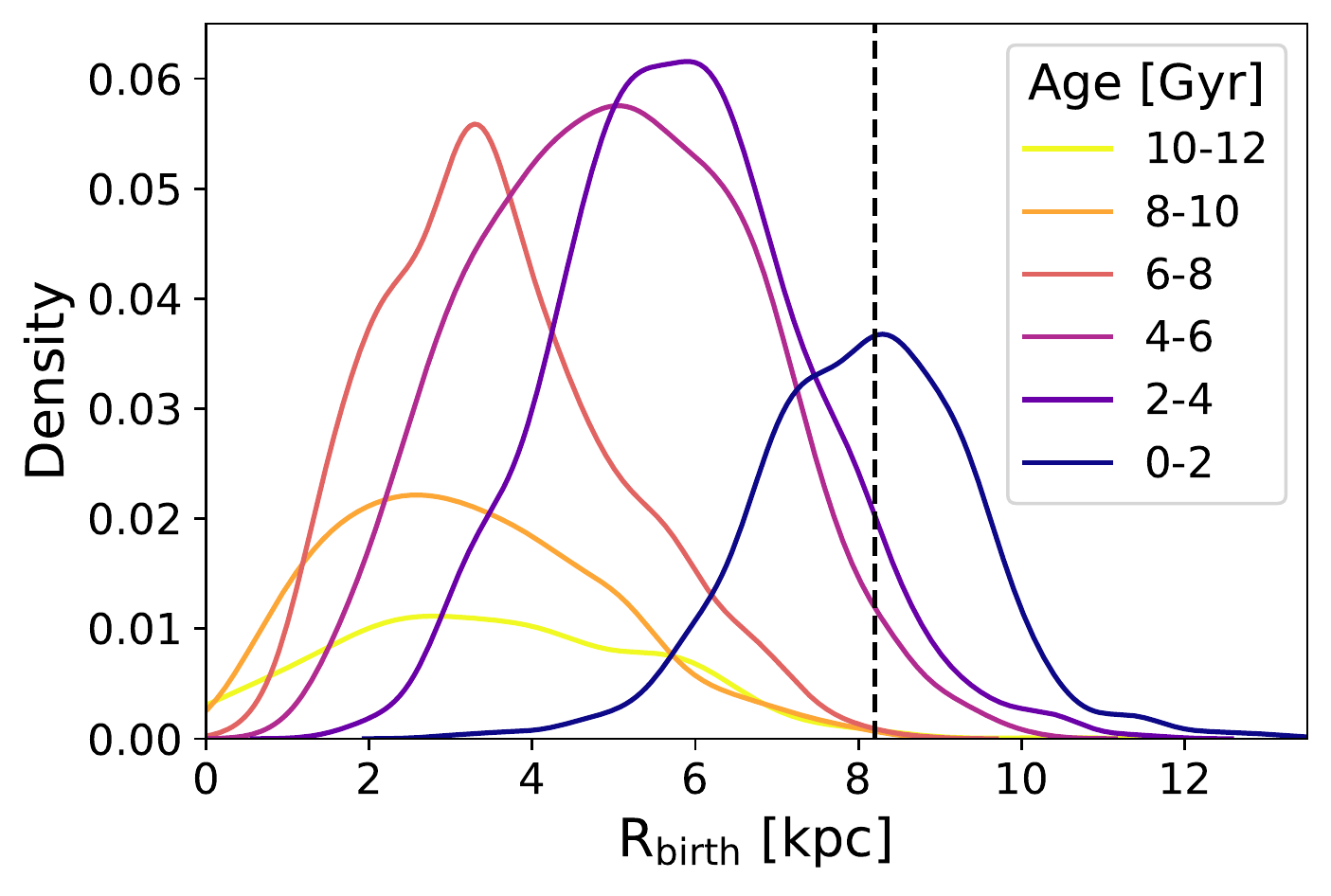}\\
     \includegraphics[width=.375\textwidth]{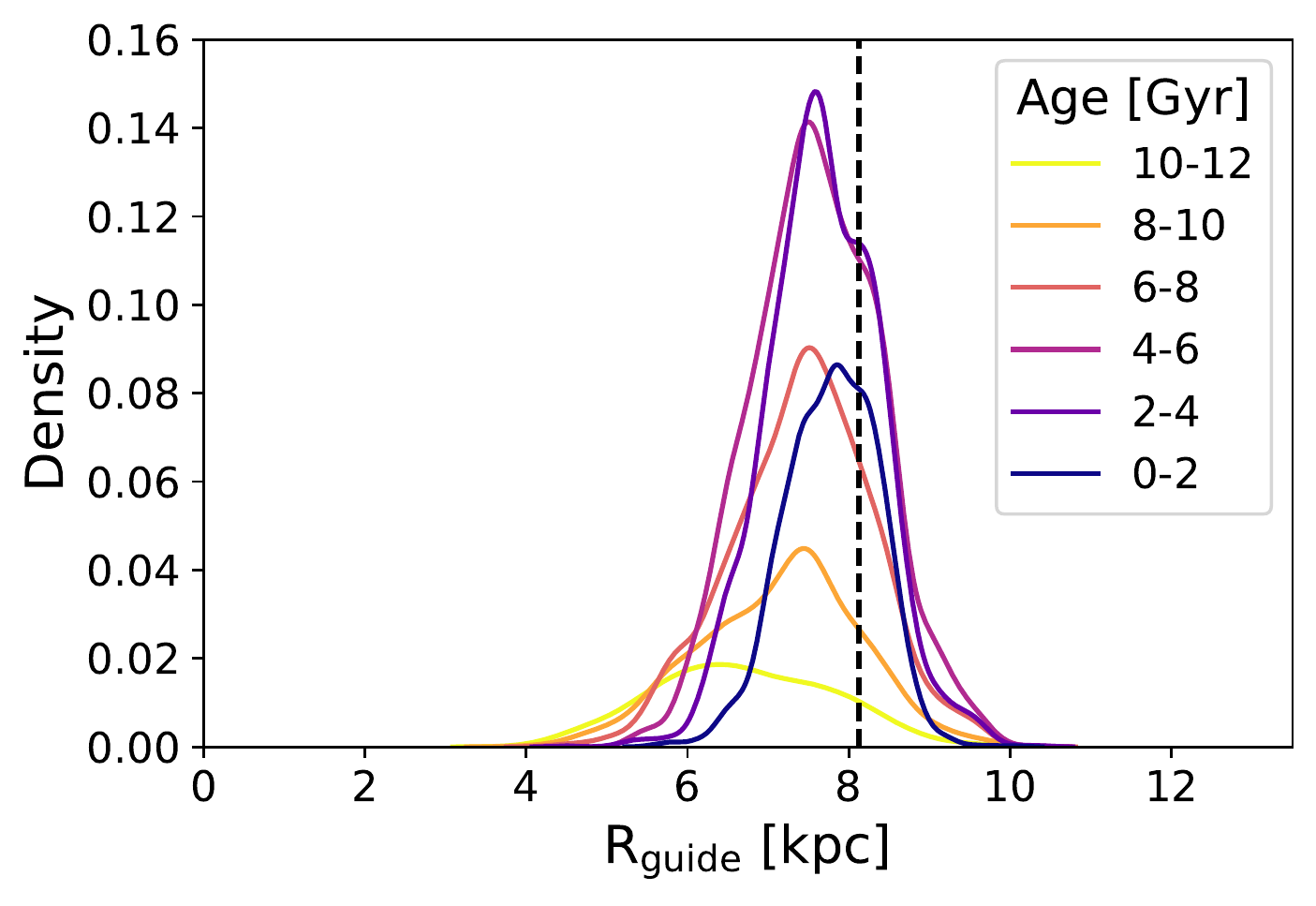}
\caption{\textbf{Top:} \rb\ and \textbf{bottom:} \rguide\ density distributions for mono-age populations in the solar neighborhood ($7.9 < \rgal < 8.35$ kpc, $|z| < .3$ kpc), with each conditional density scaled by the total number of observations such that the total area under all the densities sums to 1. Within the solar neighborhood, we find stars born throughout the disk that have migrated here over time. The youngest stars are found to be born locally, while the oldest stars formed further inwards. This is in agreement with expectations, as young stars have had less time to migrate, while older ones have had longer exposure to radial migration processes.}
\label{fig:SN_Rbdist}
\end{figure}

Figure \ref{fig:SN_Rbdist} contrasts the difference between \rb\ (top) and \rguide\ (bottom) for solar neighborhood stars. \edits{Since \rb\ is not directly observable, we validate our estimates by comparing to our current understanding of how mono-age \rb\ distributions should behave.} In agreement with expectations, we find that the youngest stars peak locally, as they have had less exposure to radial mixing processes (e.g., \citealt{Minchev2013, 2020Feltzing, Netopil2022}), while older population peaks shift to progressively lower radii as expected for inside-out disk formation. Our method also reproduces \rb\ distributions found in state-of-the-art simulations \citep[e.g.][]{Carrillo2022, Agertz2021_vintergatanI}. In contrast to \cite{2018Minchev_rbirth}, who imposed such shifting in the peaks of mono-\rb\ populations in order to estimate the Milky Way's \gradFeh, here we reach a very similar configuration directly from the data, as described above.

\section{Results} \label{sec:results}

With the birth properties of stars, namely their ages and birth radii, we can record the Milky Way disk's evolution with lookback time. Using \rb\ derived by the method in Section \ref{sec:Rb_method}, we start our analysis by investigating the birth radial abundance gradients in \rb--\xh\ (Section \ref{sec:xh}) and \rb--\xfe\ (Sect. \ref{sec:xfe}) planes to understand the overall enrichment of the Galactic disk. In Sect. \ref{sec:enrichment}, we then explore the time evolution of mono-\rb\ populations in the \feh--\mgfe\ and age--\xfe\ abundance planes to understand the relative enrichment of different nucleosynthetic sources. 

\begin{figure*}
     \centering
     \includegraphics[width=\textwidth]{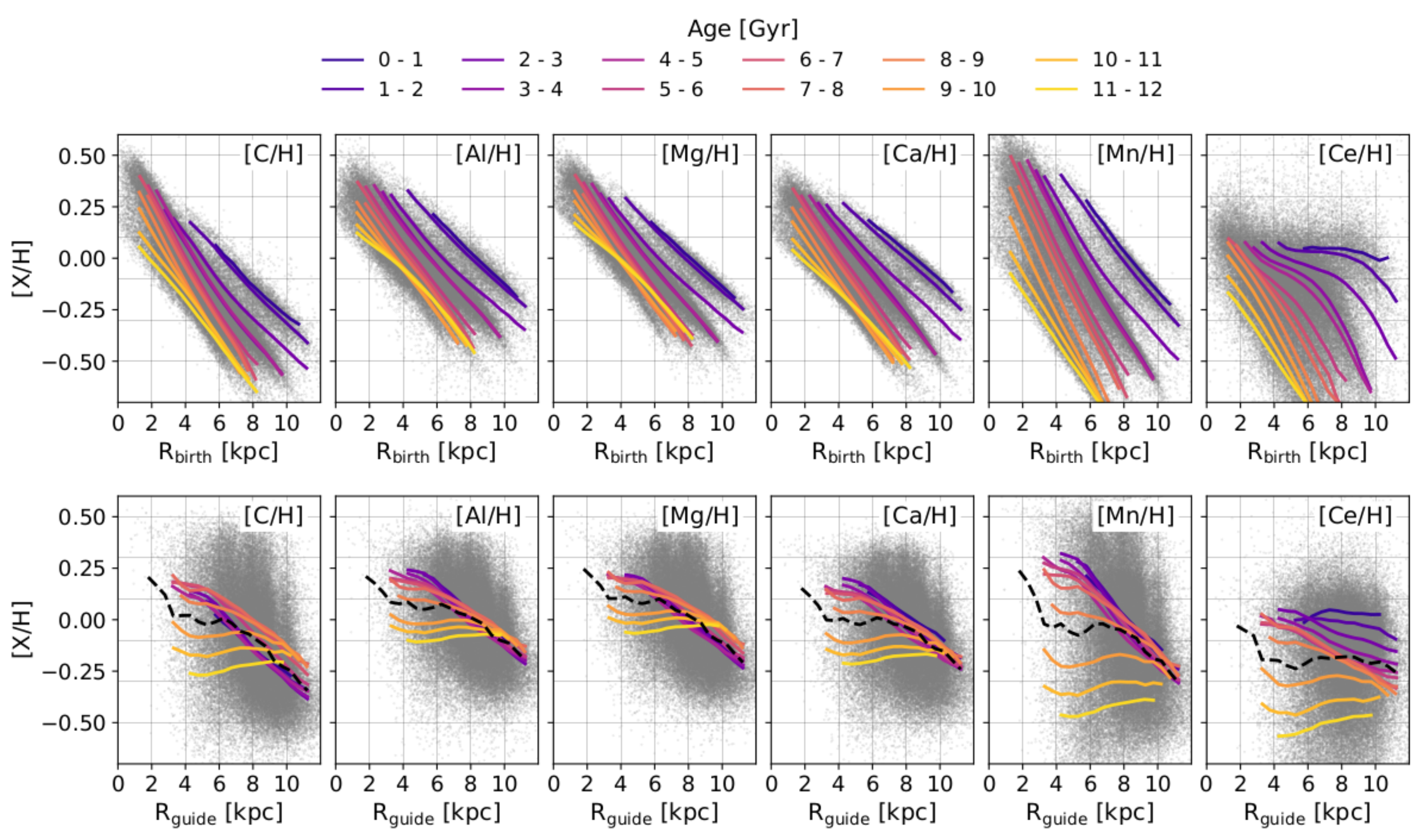}\\
\caption{Time evolution of \xh\ abundance gradients for our sample. \textbf{Top:} Running means of different mono-age populations overlaying the \numstar\ red giant \apogee\ DR17 stars in the \rb--\xh\ plane, determined by calculating the average \xh\ for \rb\ bins of width 0.5 kpc, and then smoothed over 1 kpc. \textbf{Bottom:} Same as the top panels, but looking at the variation over \rguide. The total sample's running mean is provided in each \rguide--\xh\ plane as a black dashed line. The \xh\ gradient with lookback time has been significantly affected by radial migration, especially for the oldest mono-age populations. The younger populations tend to follow the non-conditioned running mean, indicating that there is significant scatter in \xh\ about the mono-age running means. The \rb-- and \rguide--[Ce/H] trends using Ce derived in the BAWLAS catalog are shown in  Figure \ref{fig:bawlas} in the Appendix. }
\label{fig:xh_r}
\end{figure*}

\subsection{[X/H] variation with \texorpdfstring{\rb}{Rb}}
\label{sec:xh}

\begin{figure*}
     \centering
     \includegraphics[width=\textwidth]{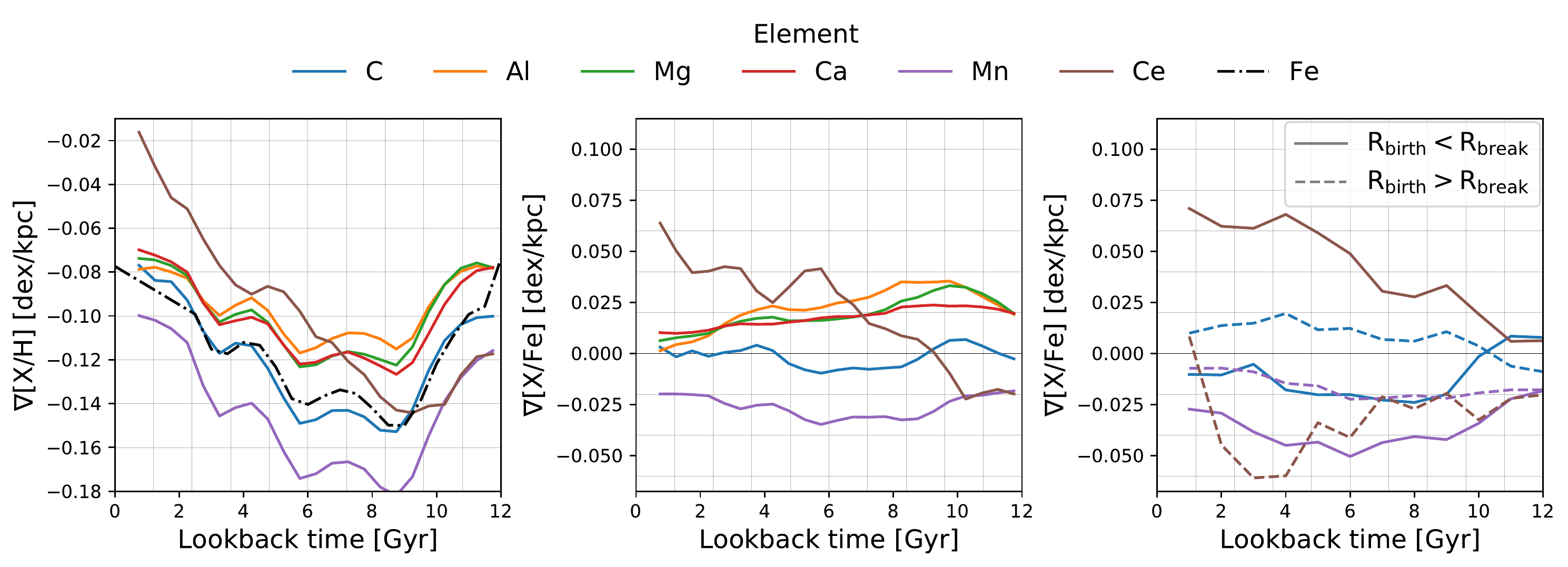}\\
\caption{\textbf{Left:} Measured \xh\ radial gradients across lookback time, for $\Delta$lookback time = 0.5 Gyr. The black dashed line is the derived \gradFeh\ from the left panel of Figure \ref{fig:feh0-range}. Most abundances show similar trends in \rb--\xh\ with time, with slight variations caused by the rate and strength of different enrichment sources. \textbf{Middle:} Measured \xfe\ radial gradients over lookback time assuming a linear fit for each mono-age population. The maximum error on both \xh\ and \xfe\ radial gradients is $<0.005$ dex/kpc, estimated from 100 Monte-Carlo samples where new abundances are drawn from normal distributions with the observed \xfe\ as the mean and their respective uncertainties as the standard deviation. \textbf{Right:} The measured \xfe\ radial gradient over lookback time ($\Delta$lookback time = 1 Gyr) for \mnfe, \cfe, and \cefe\ before (solid line) and after (dashed line) the respective break points ($R_\text{break}$; see Figure \ref{fig:xfe_Rb}). Mono-age populations show non-linear trends between these three abundances and \rb, therefore we a fit a linear function before and after a break point, which is different for different ages and \xfe\ (Figure \ref{fig:xfe_Rb}). The radial \mnfe\ and \cefe\ gradients are steeper before the break points, while \cfe\ sees an inversion in slope before and after the break point.}
\label{fig:slopes}
\end{figure*}

We begin by looking at an array of elements to examine how the different nucleosynthetic families evolved throughout the Galactic disk. The variations of \xh\ with \rb\ for mono-age populations, as color-coded, are plotted in the top panels of Figure \ref{fig:xh_r}. As expected, all \xh\ trends are negative, similar to \feh. The effect of the \gradFeh\ fluctuation $\sim9$ Gyr ago can be readily seen in the oldest three age bins for all elements. The second and third fluctuations ($\sim4$ and $\sim6$~Gyr) are not as apparent, although an overlap can be observed for the bins spanning the range $3-5$ Gyr and $6-8$ Gyr. For all elements and mono-age populations (except for Ce in the younger age bins), the relationships between \xh\ and \rb\ could be fit relatively well with a linear model, and the slope variations are shown in the left panel of Figure \ref{fig:slopes}. Similar to predictions of both semi-analytical \citep[e.g.][]{PrantzosBoissier2000} and chemo-dynamical (e.g., \citealt{Vincenzo2018}) Milky Way-like models, we find that, overall, the slope of \xh\ steepens with increasing lookback time. Most elements show multiple fluctuations in the \xh\ gradient trends over lookback time (left panel of Figure \ref{fig:slopes}), similar to \gradFeh. A notable exception is [Ce/Fe], whose gradient shows a continued steepening with increasing lookback time $\sim6$ Gyr ago, rather than inversion. This can be attributed to the strong overall flattening in its gradient with cosmic time (top rightmost panel of Figure \ref{fig:xh_r}), which counteracts the two fluctuations $\sim9$ and 6 Gyr ago, seen well in the other elements.

We contrast the \xh\ variation with \rb\ to the variation with \rguide\ in the bottom row of Figure \ref{fig:xh_r}. Drastic differences can be seen, especially for the oldest mono-age bins, as also expected from chemo-dynamical simulations. The oldest populations show significantly flatter relations with \rguide, which is expected since older populations have been exposed longer to radial migration processes. The younger populations are most similar to the recovered \xh\ radial birth gradients, though with larger scatter. The dashed black curve in each bottom panel shows the running mean of the total stellar sample, whose slope and shape is strongly dependent on the survey selection function and the quality cuts used for the sample at hand and in general does not provide any useful information. Most of the mono-age populations (age $\lesssim 8$ Gyr) follow the same \rguide--\xh\ trend as the running mean of our sample while the mono-age populations separate in the \rb--\xh\ plane, indicating that there is minimal \rb\ information contained in \rguide.

\begin{figure*}
    \centering
     \includegraphics[width=\textwidth]{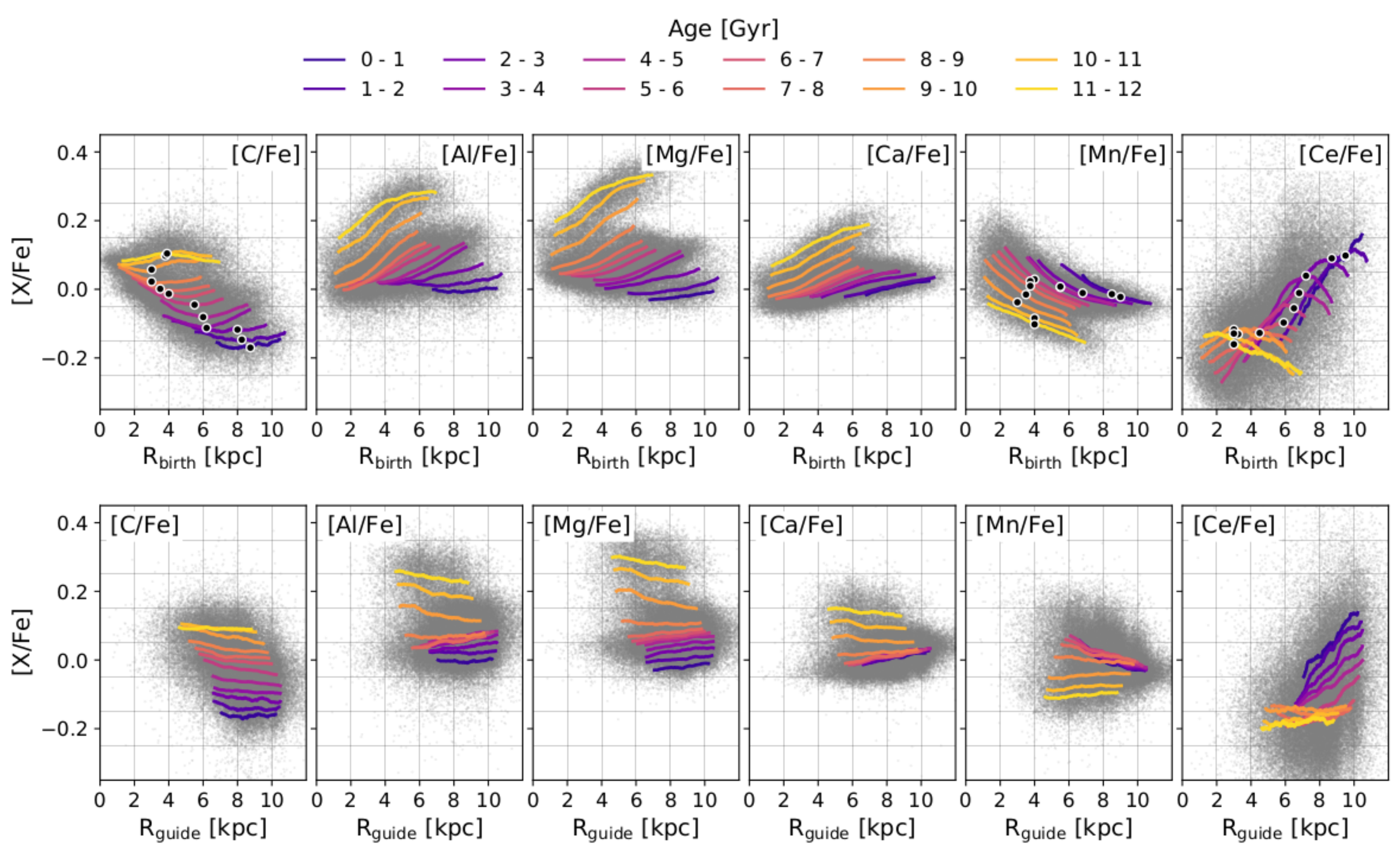}\\
\caption{Same as Figure \ref{fig:xh_r}, but for the \xfe\ abundance gradients. \textbf{Top:} Running means of mono-age populations in the \rb--\xfe\ plane. Remarkably, in all elements except for Ce, a bimodality akin to that found in the \feh--\alphafe\ plane can be seen, with the high-$\alpha$ sequence appearing as a distinct elongated blob composed of the oldest $2-3$ age bins. This time period corresponds exactly to the period during which the initial steepening in \gradFeh\ takes place in Figure \ref{fig:feh0-range}. Most radial mono-age \xfe\ gradients can be adequately fit by a linear model, except for \cfe, \cefe, and \mnfe. In each mono-age population for these elements, we mark the location where the slopes either flatten or invert by a black dot with a white outline. \textbf{Bottom:} Same as top but using the guiding radius. Trends with both radius and time seen in the top panels are almost completely erased due the angular momentum redistribution in the disk that has taken place. The effect is strongest for the oldest populations, resulting in the trends with radius being completely lost and inverted for Mg, Ca, Mg, and Al. See Appendix Figure \ref{fig:bawlas} for the \rb-- and \rguide--\cefe\ relationship using \cefe\ from the BAWLAS catalog. }
\label{fig:xfe_Rb}
\end{figure*}

\subsection{[X/Fe] variation with \texorpdfstring{\rb}{Rb}}
\label{sec:xfe}

We now explore the \xfe\ trends with \rb, resulting from our recovered \feh\ gradient evolution. The top panels of Figure \ref{fig:xfe_Rb} show the evolution of the relative enrichment between Supernovae Ia contribution (as traced by Fe) and other sources over time throughout the Milky Way disk. The middle panel of Figure \ref{fig:slopes} shows the value of each \xfe\ gradient as a function of lookback time. 

Each age bin in Figure \ref{fig:xfe_Rb} has its own unique trend between \xfe\ and \rb\ that does not necessarily capture the overall trend of the whole data. Specifically, starting with \cfe\ (top left panel of Figure \ref{fig:xfe_Rb}), we see that the full sample has a negative correlation between \cfe\ and \rb. However, once conditioned on age, each mono-age population has a primarily weak \cfe\ radial birth gradient. The global near-zero radial gradient in \cfe\ over lookback time is also seen in the middle panel of Figure \ref{fig:slopes}, and may be a consequence of carbon's relationship with mass and age \citep{Salaris2015}, and the relationship between Fe, age, and birth radius. Since \cfe\ has a minimal radial gradient with age, the [C/H] gradient varies most similar to \feh\ over time (left panel of Figure \ref{fig:slopes}). 

The radial \mgfe, \cafe, and \alfe\ birth gradients for (almost) all mono-age populations are positive and flattening with time as seen in Figure \ref{fig:xfe_Rb}. Ca and Mg, being $\alpha$-elements, show very similar trends in both their \xfe\ and [X/H] radial gradient evolution with time, except for the smaller range in \cafe. Mg and Al also have essentially the same radial gradients (Figure \ref{fig:slopes}). 

Unlike the $\alpha$-elements, the \mnfe\ radial gradient is negative, and has stayed at a fairly constant rate throughout time. Despite the fact that Mn, Mg, Ca, and Al have different enrichment sources, the $3-4$ and $4-5$ Gyr age tracks and $6-7$ and $7-8$ Gyr age tracks for these 4 elements live in very similar areas in the \rb--\xfe\ plane, reflecting the \gradFeh\ fluctuations at those times (Figure \ref{fig:xfe_Rb}). 

Another element ratio that stands out is \cefe\, showing a strong positive correlation with \rb\ for age $<8$~Gyr and a weaker relationship in older aged populations, including an inversion for oldest age bin. Unlike the other abundances, the age tracks do not cleanly separate in the \rb--\cefe\ plane. The weak \cefe\ abundance gradient for older aged stars suggests that the production rate of Ce and Fe varied slowly at first, then saw larger differences at later times (consistent with Figure 7 of \citealt{2016Battistini}).



So far, we have focused on the \xfe\ abundance ratios' radial gradients in Figure \ref{fig:xfe_Rb} being linear for each mono-age bin (Figure \ref{fig:slopes}, middle). However, some of the trends with \rb\ cannot accurately be fit with a linear model. Most noticeably, \cfe, \cefe, and the younger aged populations in \mnfe\ have a break point, where the abundance trends begin to flatten, or even invert slope. The locations of the break points are shown in Figure \ref{fig:xfe_Rb} as black dots with a white outline. The right panel of Figure \ref{fig:slopes} shows these three abundances fit linearly before (solid line) and after (dashed line) their respective break points. The younger mono-age populations show an inverse in slope before and after the break point in \cfe\ and \cefe, with a positive and negative radial gradient in the outer region of the disk, respectively. Meanwhile, the \mnfe\ radial gradient is negative, with the gradient after the break point being flatter. These dramatic changes in the \rb\ gradients of \cfe, \cefe, and \mnfe\ for different radii ranges should be considered in modeling the evolution of these elements.

We also show the trends of \xfe\ with guiding radius for different mono-age populations in the bottom panels of Figure \ref{fig:xfe_Rb} to illuminate the additional information learned with knowledge of \rb. Overall, we can see that many birth gradients are lost or altered as a result of radial migration. For instance, most \rguide--\xfe\ relations are flatter or inverted, especially for the older age bins, compared to radial gradients measured with \rb\ (e.g. \alfe, \mgfe, \cafe), and nonlinear gradients can become smoothed by stellar migration (e.g. \cfe). 

Radial gradients with \rb\ also reveal structure in the data that recreates the chemical evolution of the Galaxy which cannot be seen in the present-day disk. The top panels of Fig. \ref{fig:xfe_Rb} show a bimodality in the \rb--\xfe\ plane, particularly for the $\alpha$-elements Mg and Ca, the odd-Z element Al, and the iron-peak element Mn. One of the groups is captured by older ($>$ 9 Gyr) stars, has high \mgfe, and is representative of the high-$\alpha$ sequence, while the other corresponds to the low-$\alpha$ sequence. The clean separation of the high- and low-$\alpha$ sequences is not typically seen in the \rguide--\xfe\ plane (see bottom panels of Figure \ref{fig:xfe_Rb}). With a clear view of the different enrichment histories, we see that the high-$\alpha$ sequence typically has a steeper \xfe\ radial birth gradient, suggesting that it formed in a chemically non-homogeneous environment. Within the low-$\alpha$ sequence, the younger stars appear to separate into their own distinct group in the light element abundances and \mgfe. This is due to a rapid weakening of the \xfe\ gradient $3-4$ Gyr ago.


\subsection{Stellar chemical enrichment with lookback time}\label{sec:enrichment}

Until here, we have explored how the radial gradients of individual chemical abundances and abundance ratios evolve across the Galactic disk with time. In this Section, we now study the time evolution of mono-\rb\ populations in the \feh--\mgfe\ and age--\xfe\ planes to investigate the relative rate of enrichment. As a caveat, we note that our age estimates, while being independent of [Fe/H], do depend to some degree on [Mg/Fe] (see \citealt{Anders2023_ages}). \edits{However, this dependency seems to have a minimal effect on our results as we are able to reproduce the same trends using ages derived using isochrone fitting (see Section \ref{sec:starhorse} in the Appendix).}

\subsubsection{Time evolution of the [Fe/H]--[Mg/Fe] plane}\label{sec:mgfe-feh}

\begin{figure*}
     \centering
     \includegraphics[width=.72\textwidth]{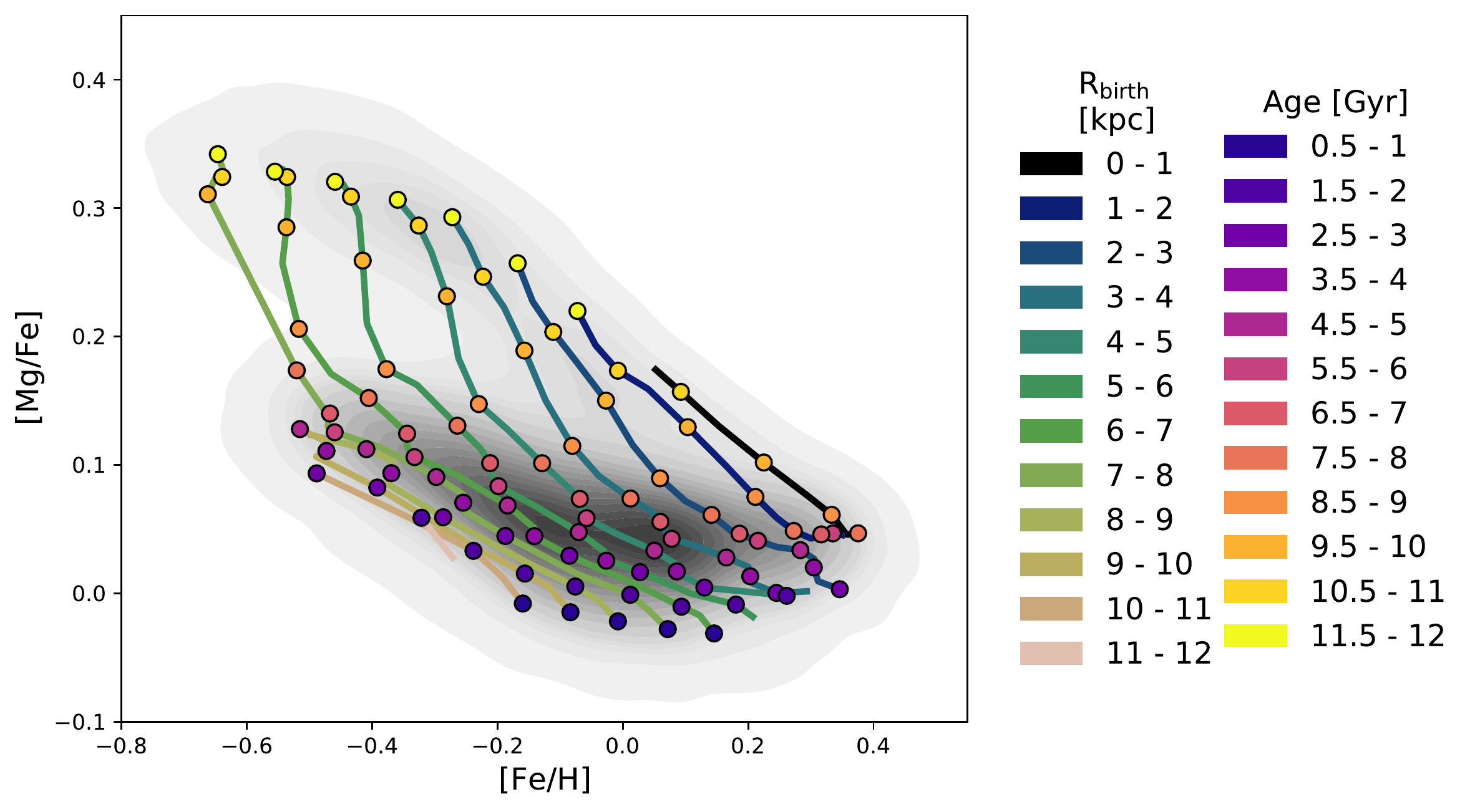}
`\caption{The density distribution of \numstar\ red giant \apogee\ stars in the \feh--\mgfe\ plane with the time evolution of \rb\ tracks overlaid. The tracks are chosen with $\Delta$\rb\ = 1 kpc and $\Delta$age = 0.5 Gyr, with points at every other age bin. The fluctuation in the \feh\ gradient at $\sim 9$ Gyr corresponds to the valley between the high- and low-$\alpha$ sequences, while the fluctuation at $\sim 6$ Gyr corresponds to a period of slower enrichment.}
\label{fig:mgfe_feh}
\end{figure*}

We begin this section by looking at the formation and evolution of the $\alpha$-sequences across \rb\ and age. Figure \ref{fig:mgfe_feh} shows different \rb\ tracks as a function of time in the \feh--\mgfe\ plane, similar to \cite{2020_buckchemical, 2021Sharma, Lu2022_Rb}. For a given \rb, the track begins with a higher value of \mgfe, and over time moves to lower values of \mgfe\ and higher values of \feh\ for younger aged stars. Overall, the evolution through this plane for a given \rb\ is smoothly increasing in \feh. There is, however, a wiggle at a lookback time of $\sim 3.5 - 5$ Gyr where younger stars jump to lower \feh\ at higher \rb, and see minimal changes in \feh\ at smaller \rb. This non-monotonic evolution in the $\alpha$-bimodality is caused by the fluctuation in \gradFeh\ at $\sim 4$ Gyr, seen in Figure \ref{fig:slopes}, and leads to 3 different ages having almost the same \feh\ for a given \rb\ (also seen in the right panel of Figure \ref{fig:feh0-range}). During this time of the \gradFeh\ fluctuation $\sim 6$ Gyr ago, the \feh\ and \mgfe\ gradients stay fairly constant with time
, causing stars born $5.5-8$ Gyr ago to have similar \feh\ and \mgfe\ at a given \rb. Note that the points along each \rb\ track are all separated by 1 Gyr, i.e., a smaller separation corresponds to a slower enrichment.
This is another reason why mono-abundance populations cannot provide a good proxy for mono-age populations, in addition to radial mixing or sample covering larger Galactic radius \citep{Minchev2017}. 

Figure \ref{fig:mgfe_feh} reveals that the high-$\alpha$ sequence (defined in the \feh--\mgfe\ plane such as in Figure \ref{fig:alphaSeq_def}) has formed close to the Galactic center ($\rb \leq 7$ kpc), in agreement with \cite{2020_buckchemical, Lu2022_Rb} and as expected for an inside-out disk formation. \edits{This is also consistent with high-$\alpha$ stars being predominantly found in the inner parts of the Galaxy \citep[e.g.][]{Bensby2011, Anders2014, hayden2015chemical}.} It is uniformly old (yellow dots) up to \feh$\approx0-0.1$~dex and its age decreases to about 6.5~Gyr at the most metal-rich part (where the low- and high-$\alpha$ sequences meet). The majority of the stars composing the high-$\alpha$ sequence were formed at the very inner disk, according to the color-coded curves. As found also by \cite{Lu2022_Rb}, the high-$\alpha$ sequence formed during the early-on steepening of \gradFeh. Going through the steepest point at lookback time $\sim 9$ Gyr caused a large jump in both \feh\ and \mgfe, thus forming the gap in the \feh--\mgfe\ plane.

\begin{figure*}
     \centering
     \includegraphics[width=.78\textwidth]{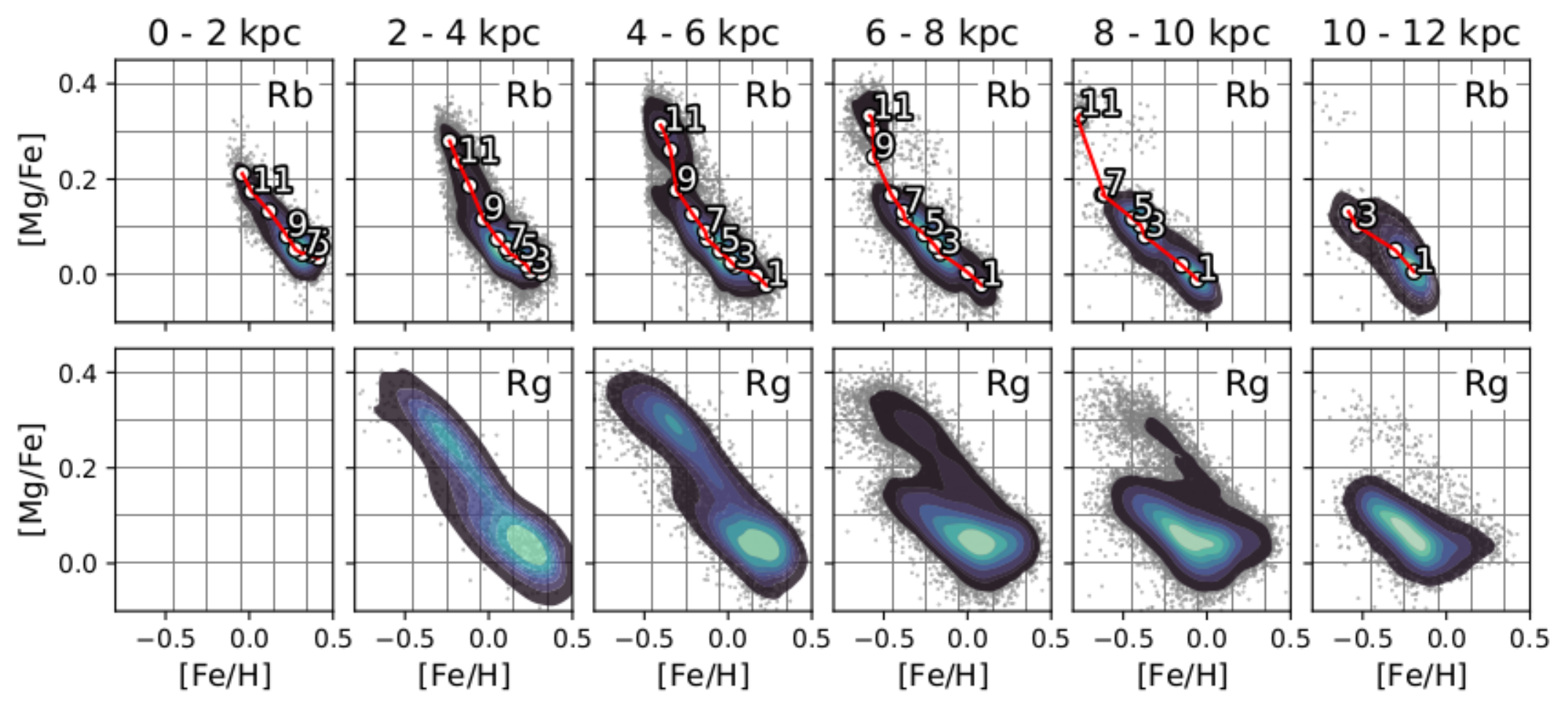}
\caption{Distribution of stars in the \feh--\mgfe\ plane separated as a function of \textbf{Top:} birth radius and \textbf{Bottom:} guiding radius. The grey points are the individual stars within each \rb/\rguide\ bin, with the filled in density contours on top. The spatial separation as a function of \rb\ creates tighter relationships between \mgfe\ and \feh, especially in the low-$\alpha$ sequence. The white points and labels in the top row denote the lookback time in Gyr. 
}
\label{fig:alaphSeqs}
\end{figure*}

Figure \ref{fig:alaphSeqs} shows the distribution of the \feh--\mgfe\ plane as a function of \rb\ (top) and \rguide\ (bottom), allowing us to make direct comparisons between the observable and inherent trends. While the high-$\alpha$ sequence is present beyond the solar neighborhood in \rguide, with a small proportion of stars currently located in the outer disk, it was predominantly formed at lower \rb, with only about $2$\% of its stars born beyond 7 kpc. The formation structure of the high-$\alpha$ sequence, which is typically concluded to be radially uniform in the inner disk and solar neighborhood \citep{2012bensby, nidever2014tracing, hayden2015chemical}, is also masked by radial migration. The most metal-rich high-$\alpha$ stars predominantly formed closer to the Galactic center, while the lower-\feh\ stars were formed further out, suggesting an inside-out formation also for the high-$\alpha$ sequence.

Unlike the high-$\alpha$ sequence, the low-$\alpha$ sequence (see Figure \ref{fig:alphaSeq_def} for our visual definition of the two sequences) was formed throughout the Galactic disk. Figure \ref{fig:alaphSeqs} shows that the overall trend of the low-$\alpha$ sequence evolves with \rb\ similar to \rguide. That is, the low-$\alpha$ stars have lower values of \feh\ for higher \rb\ or \rguide, indicating that the metal-rich low-$\alpha$ stars formed at very low \rb\ (consistent with suggestions by e.g. \citealt{Kordopatis2015}). Unlike with \rguide, separating by \rb\ creates much narrower distributions, revealing a tight relationship between \mgfe\ and \feh\ in the low-$\alpha$ sequence for a given \rb. Specifically, we can see regions of higher density in this plane that show the transition of the location of the low-$\alpha$ sequence with Galactic birth location. This indicates that the higher \mgfe\ tail of the low-$\alpha$ sequence formed first in the inner disk, while the stars composing the lower-\mgfe, higher-\feh\ end formed later on further out in the disk. The unveiling of this spatial structure with \rb\ illustrates that the high-\feh\ region of the low-$\alpha$ sequence migrated outwards, blurring the strong inherent correlations between \mgfe, \feh, and age. 

\subsubsection{Age--abundance plane}

\begin{figure*}
     \centering
     \includegraphics[width=\textwidth]{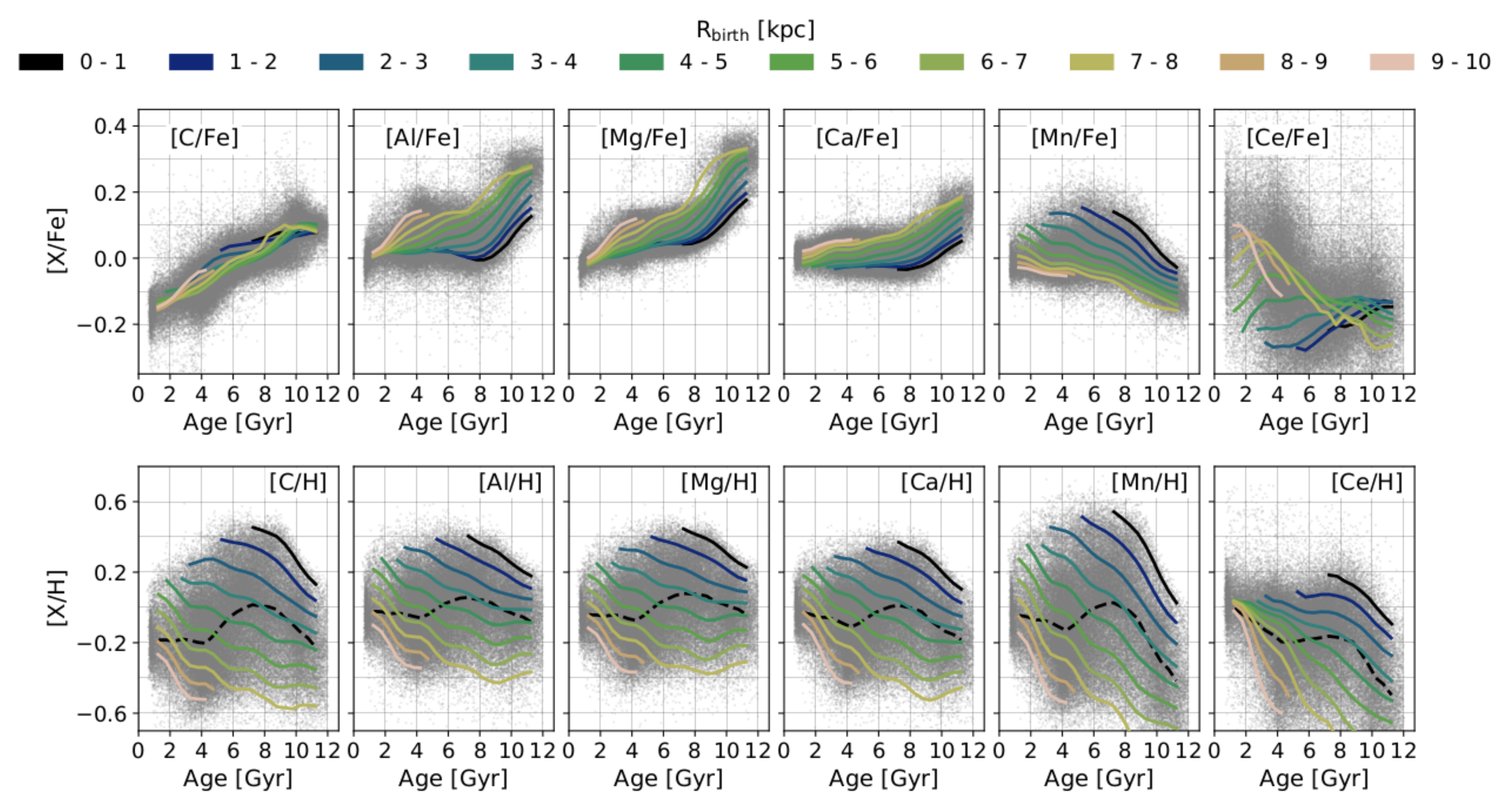}
\caption{Mono-\rb\ populations in the \textbf{top:} age--\xfe\ and \textbf{bottom:} age--\xh\ planes of \numstar\ red giant \apogee\ DR17 stars. The running means are derived using age bins of 0.5 Gyr and averaged over 1 Gyr. The running mean of the total population is given in the bottom panels, calculated the same way as the mono-\rb\ bins.  The mono-\rb\ populations lie on steeper tracks in the age--\xfe\ and --\xh\ planes than the overall population. Similar to \feh, \xh\ has three fluctuations in the gradient, so a given \rb\ is not monotonically increasing in \xh\ with lookback time. Note that when working in terms of mono-\rb\ populations, age becomes lookback time, i.e., corresponds to the actual evolution with cosmic time. See Figure \ref{fig:bawlas} in the Appendix for age--\cefe\ and --[Ce/H] trends using Ce from the BAWLAS catalog. }
\label{fig:xfe_age_r}
\end{figure*}

The age--\xfe\ plane provides insight into the overall evolution of \xfe\ with time. The \xfe\ abundance ratios of Mg, Ca, Mn, and Al have been fairly constant over the past $\sim$9 Gyr, with a distinct steeper slope at $9-12$ Gyr (grey points in top panels of Figure \ref{fig:xfe_age_r}). Conditioning on mono-\rb\ populations changes this perception, as shown by the color-coded curves.
Except for \cfe\ --- whose mono-\rb\ populations show minimal differences due to the already very strong positive age--\cfe\ relation --- the mono-\rb\ populations occupy different \xfe\ for a given age, and show a strong age--\xfe\ correlation even when the overall population does not. Each mono-\rb\ population for $\rb < 9$ kpc increases in \mgfe, \cafe, and \alfe\ until 4 Gyr, when the \xfe\ abundance increases at a slightly faster rate for $\sim$5 Gyr, before proceeding to enrich at a faster rate until 12 Gyr. An exception to this is that the inner most \rb\ see a decrease in \alfe\ until about 8 Gyr ago, then proceed to increase in \alfe. The $\rb > 9$ kpc region is more concentrated towards younger ages, and therefore only sees the steep enrichment at age $<6$~Gyr. \mnfe\ follows a similar pattern, just with a negative correlation with age. Unlike the other abundances, the correlation between \cefe\ and age is strongly dependent on the birth radius. For outer radii, \cefe\ declines with increasing age, while in the inner disk the relationship inverts. The above results illustrate the importance of incorporating \rb\ estimates in studying the Galaxy's evolution, as in their absence it can be concluded that \alphafe\ is not a good tracer for age up to the slope increase at $\sim$9 Gyr, as recently concluded by \cite{Hayden2022}. 

The bottom panels of Figure \ref{fig:xfe_age_r} show the age--\xh\ plane, which is equivalent to the AMR shown in the right panel of Figure \ref{fig:feh0-range}. Similar to the age--\xfe\ plane, separating the age--\xh\ plane by \rb\ provides structure to an otherwise uncorrelated relationship. The black dashed line represents the full sample's running mean in \xh\ across age. \xh\ overall shows lots of scatter and minimal correlation with age, particularly for age $<5$ Gyr. Conditioning the sample on \rb\ creates structure similar to the age--\feh\ plane, given in the right panel of Figure \ref{fig:feh0-range}. Each mono-\rb\ population shows a clear enrichment in \xh\ with time, with the same fluctuations at 4, 6, and 9 Gyr (except perhaps for Ce).

With \rb\ available, it is clear from Figure \ref{fig:xfe_age_r} that both \xh\ and \xfe\ correlate well with age for all birth radii, were it not for radial migration blurring the relations. The statistical phenomenon that causes this flattening and even inversion of relations seen in Figures \ref{fig:xfe_Rb} and \ref{fig:xfe_age_r} is known as Yule-Simpson's paradox \citep{simpson1951}, which was recently shown to be ubiquitous in the field of Galactic Archaeology \citep{2019minchev}.

\section{discussion} \label{sec:discussion}

\subsection{Comparison to radial gradients found in other work}

Due to the angular momentum redistribution of stars within the Milky Way disk throughout its lifetime, the radial abundance gradients measured today for mono-age populations do not reflect those at birth. Observed present-day [X/H] (typically \feh\ or [M/H]) and \xfe\ gradients are found to be flatter for older ages or show only weak evolution with age \citep{Casagrande2011, Anders2017, Myers2022, SalesSilva2022, Ratcliffe2022_conditional, Anders2023_ages}. Using our recovered \rb, we find a distinctly different gradient evolution with age, which in this case can only be thought of as lookback time.

The inherent birth gradients in \xfe\ that we recover are much steeper than the observed gradients of mono-age populations ($>0.02$ dex/kpc for \cefe\ and \alfe, $<-0.02$ dex/kpc for \mnfe, $>0.02$ dex/kpc for lookback time $>9$ Gyr in \mgfe). The \xh\ gradients (left panel of Figure \ref{fig:slopes}) show a clear flattening with time, in agreement with \cite{PrantzosBoissier2000, Vincenzo2018, Agertz2021_vintergatanI, Magrini2023}. This is contradictory to some chemical evolution models and simulations, which assume the abundance gradients are steepening with cosmic time \citep[e.g.][]{Chiappini2001, Vincenzo2020, Bellardini2022}, and illustrates the importance of utilizing the \rb\ information now available.

Figure \ref{fig:xfe_Rb} shows the radial \xfe\ information lost to radial mixing processes, and explains the mildly positive/negative \xfe\ gradients observed in the present-day disk \citep{Carrera2011, Yong2012, Anders2017, Myers2022}. The similarities we find between the evolution of \alfe\ and \mgfe\ with \rb\ agrees with previous findings that Al behaves like an $\alpha$-element in the age--\xfe\ plane \citep{Spina2021} and highlights the relative production of Al in SNII is similar to Mg. 

While we approximated the radial birth \xfe\ gradients as linear, they can best be explained as a combination of two pieces. We fit a linear model both before and after a break point in \cfe, \cefe, and \mnfe, which are the abundances we deem to be the most affected by a break (right panel of Figure \ref{fig:slopes}). For each abundance, we find that the location of the break in the \xfe\ gradient is mostly moving to larger radii over time, starting at $\sim4$ kpc $10+$ Gyr ago and located $>8$ kpc for stars $0-1$ Gyr. For the $\alpha$-elements and Al, their \xfe\ radial gradients steepen slightly after the break. The radial gradient in \mnfe\ weakens after the break, and the break creates an inversion in slope for \cfe\ and \cefe. The location of the break point varies for different \xfe, and should be considered in modeling the chemical evolution of the Milky Way (e.g. a radially varying initial mass function; \citealt{Guszejnov2019, Horta2021}).

\subsection{What causes the fluctuations in \texorpdfstring{\gradFeh}{GradFeh}?}
\label{sec:discussion_peaks}

As discussed in Section \ref{sec:Rb_method}, we find three fluctuations in the metallicity gradient at $\sim$4, $\sim$6 and $\sim$9 Gyr ago. Although more work is needed to understand this, the older one may be attributed to the effect of the last massive merger, GSE \citep{Helmi2018_gse, Belokurov2018}, as also concluded by \cite{Lu2022_Rb} and supported by a very similar \gradFeh\ fluctuation seen in a Milky Way-like cosmological simulation by \cite{Buck2019} during a GSE-like merger (Figure A1 in \citealt{Lu2022_Rb}). In agreement with this picture, \cite{Ciuca2022} found a dip and a stand-alone feature (the ``blob") in the AMR at the same relative time we find the first steepening in \gradFeh, which they also attributed to the GSE. The recent study of \cite{Vincenzo2019} suggested that the GSE merger heated up the gas into the Milky Way halo, thus delaying star formation and forming the gap in the \feh--\alphafe\ plane. The pause in star formation due to the merger would primarily be felt in the outer disk (seen in simulations; \citealt{buck2023}), therefore the metallicity gradient steepens due to the continually enriching inner disk. This can be seen in Figure \ref{fig:xh_r}, where the 3 oldest mono-age populations have similar \xh\ at higher \rb, but show an enrichment over time at the lowest \rb.

\begin{figure}
    \centering
     \includegraphics[width=.375\textwidth]{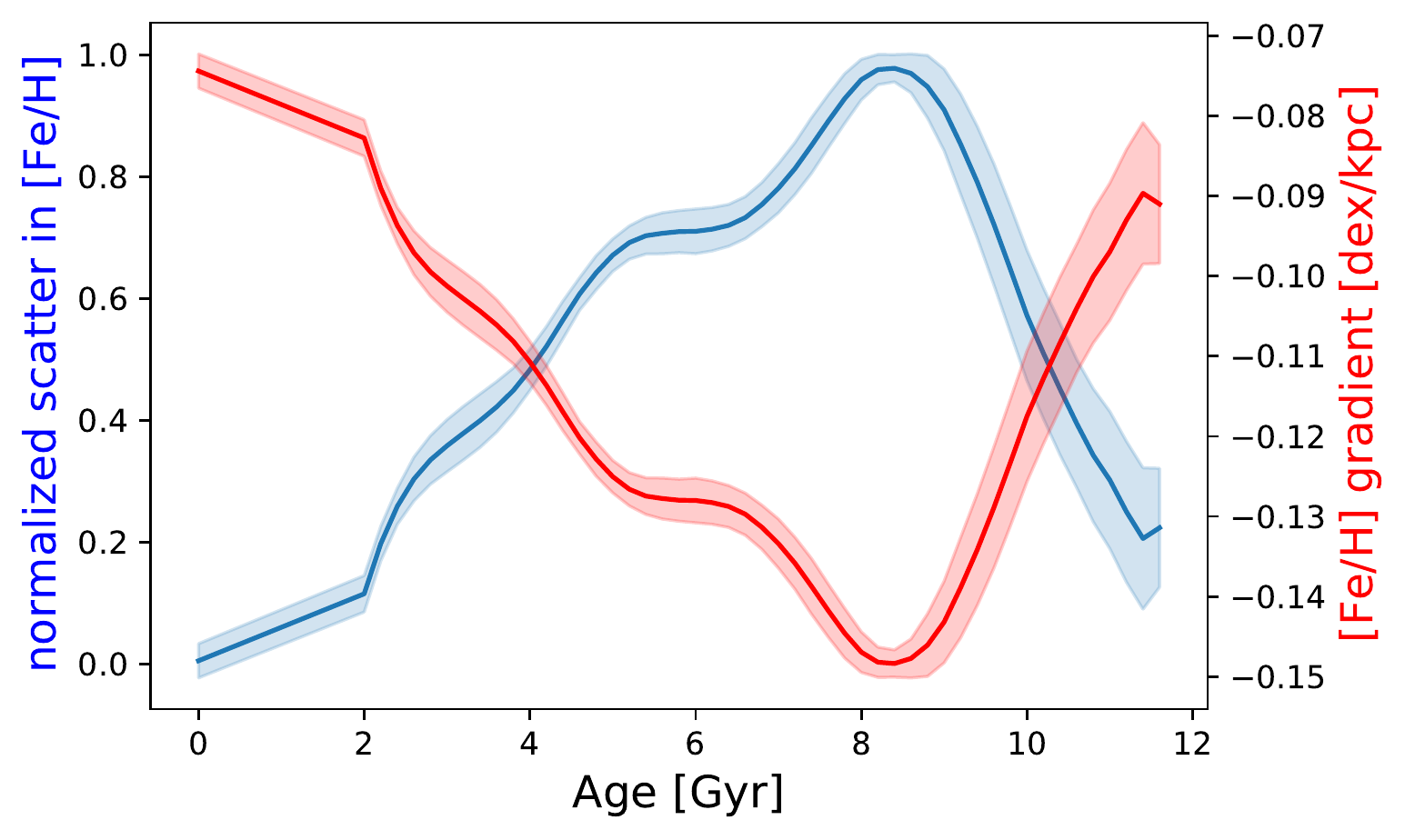}
\caption{\edits{The normalized scatter in \feh\ (blue) and inferred \gradFeh\ (red) averaged over 100 Monte Carlo samples where both the age and \feh\ are drawn from a normal distribution about their measured values and a standard deviation of their measurement uncertainty. To show the robustness of the fluctuations but still ensure a smooth evolution over time, we choose an age bin width of 1 Gyr over a rolling window of 0.2 Gyr (i.e. we calculate the scatter in \feh\ for $1.5-2.5$ Gyr, $1.7 - 2.7$ Gyr, etc.).}}
\label{fig:samplingGrad}
\end{figure}

\edits{We find the fluctuations in \gradFeh\ to be robust to both the range and standard deviation of different age bin widths. When accounting for age and metallicity uncertainty by sampling from a normal distribution with a mean of the observed value and a standard deviation of the measurement error, we find that the fluctuations at $\sim 4,6$ Gyr ago become weaker. However, the non-monotonic flattening of the metallicity gradient is still seen (Figure \ref{fig:samplingGrad}).} \edits{We are also able to recover multiple fluctuations in the AMR using the age catalogs of \cite{Queiroz2023_SH} (APOGEE DR17, GALAH DR3), \cite{Mackereth2019} (APOGEE DR17), \cite{Leung2023} (APOGEE DR17), and \cite{Sharma2018} (GALAH DR3); see also \cite{Lin2020} (GALAH DR2) and Figure 2 in \cite{Sahlholdt2022}}. \edits{While \cite{Lu2022_Rb} did not find multiple fluctuations in their derived \gradFeh\ using LAMOST, we find that performing a stricter cut in [Fe/H] error on their sample recovers a second fluctuation, though the fluctuations are weaker than recovered in the other surveys/datasets we examined. Investigating this further by performing a crossmatch between LAMOST and GALAH, we determine that LAMOST’s lower resolution may be the reason any additional structure beyond the single fluctuation at the time of GSE is concealed.}


It is tempting to attribute the \gradFeh\ fluctuations at $\sim 4$ and 6 Gyr to passages by the Sagittarius dwarf galaxy \citep{Ibata1994, Law2010, Laporte2018}, at which time this minor (yet quite massive) merger is expected to have deposited $\sim 50\%$ of its gas to its host, thus diluting the disk in \feh\ \citep{TepperGarcia2018}. The effect of the gas infall from Sagittarius is expected to be the most prominent at the outskirts, diluting the gas metallicity in the disk, and thus directly affecting the radial metallicity gradient of stellar populations formed soon after close passages~\citep{Annem2022}. Indeed, \edits{for both of these fluctuations}, the metallicity gradient begins to steepen at the same time as Sagittarius’ estimated pericenter passages in literature~\citep{Law2010}. \edits{\cite{Mor2019} found a period of star formation rate enhancement $\sim2-3$ Gyr ago, and suggested that it was possibly due to a recent merger. Indeed, we find a fluctuation just prior to this at $\sim$4 Gyr ago, which suggests that the star formation rate enhancement began after the effects of the merger on the metallicity gradient.} The fluctuation $\sim 6$ Gyr ago also coincides with the oldest of the three episodes of enhanced star formation ($5.7$ Gyr ago) found by \cite{RuizLara2020}, which, in agreement with some recent simulations~\citep{DiCintio2021,Khoperskov2022a}, can be associated with a pericenter passage of Sagittarius.

While the \gradFeh\ wiggle at $\sim 9$ Gyr does not show a strong signature in the AMR of mono-\rb\ populations (Figure \ref{fig:feh0-range}, right), a stagnation or decline with age is found in \feh, as well as in all other elements besides [Ce/H] where the effect is only weak, outside \rb$\sim6$~kpc during the second and third events (Figure \ref{fig:xfe_age_r}, bottom). These stronger signatures in the AMR are due to the slower \feh\ evolution in the innermost disk, after the initial star-formation burst ending around 8 Gyr ago in our data. 
 
Using chemical abundance data from APOGEE and {\it Gaia} RVS \citep{GaiaCollaboration2022RB}, \cite{Spitoni2023} suggest a three-infall chemical evolution model (albeit not taking into account radial migration), with the most recent gas infall starting $\sim2.7$ Gyr ago. While our three fluctuations in \gradFeh\ are synonymous with the gas infalls found by \cite{Spitoni2023}, the most recent one affecting the volume probed by the APOGEE DR17 data was $\sim 4$ instead of $\sim2.7$ Gyr ago.
 
\subsection{The high-\texorpdfstring{$\alpha$}{a} sequence formed throughout the inner disk in a non-mixed environment}

We find that the high-$\alpha$ sequence primarily formed within $\rb \leq 7$ kpc (in agreement with state-of-the-art simulations, e.g. \citealt{2020_buckchemical}), with few high-$\alpha$ stars born at larger radii. \edits{This is resonant with \cite{Bensby2011, Anders2014, hayden2015chemical}}, and \cite{Queiroz2020}, who find few high-$\alpha$ stars located further beyond the solar neighborhood in the present-day Galactic disk, and it suggests that these stars have not had time to migrate further. The few high-$\alpha$ stars we find born at larger \rb\ are younger than the remainder of the high-$\alpha$ sequence, and in fact we find a mean difference in \rb\ of 3 kpc between the high-$\alpha$ sequence as a whole and the young ($<4$ Gyr) $\alpha$-enhanced stars. The existence of these apparently young $\alpha$-enhanced stars \citep{Martig2015, Chiappini2015} has convincingly been suggested to be an artifact of incorrect age assignment due to binary mass transfer \citep{Jofre2016_BSS, Fuhrmann2017, Izzard2018, Jofre2023}. \edits{This suggests that their \rb\ estimates are incorrectly assigned, and explains the young $\alpha$-enhanced stars' inconsistent \rb\ with the remainder of the high-$\alpha$ sequence}. Using the chemical ages of \citet{Anders2023_ages}, we find a \edits{$\sim3$\%} contamination rate of young \edits{($< 6$ Gyr) stars in the high-$\alpha$ sequence}, and therefore conclude they do not affect the results found in this work.

With knowledge of \rb, we find that the high-$\alpha$ sequence is both a sequence of \rb\ and age, where the high-\feh\ end is formed closer to the Galactic center, and the low-\feh\ end is formed further out (Figures \ref{fig:mgfe_feh} and \ref{fig:alaphSeqs}). The high-$\alpha$ sequence began forming along steep \feh\ and \mgfe\ radial gradients, and does not follow one evolutionary track as concluded in some chemical evolution models and works \citep[e.g.][]{2015Haywood, Haywood2019, 2021Sharma}. Given that (\feh, \mgfe) correspond to different \rb\ tracks, this suggests that the gas was not chemically homogeneous during the formation of the high-$\alpha$ sequence (contrasting with what is observed today e.g. \citealt{Eilers2022, Ratcliffe2022_conditional}), and in fact we find that the high-$\alpha$ sequence had larger \xfe\ gradients than the low-$\alpha$ sequence (in agreement with \citealt{buck2023}). While previous works concluded that the high-$\alpha$ sequence evolved along a single chemical track due to efficient gas-phase mixing \citep[e.g.][]{Haywood2013, nidever2014tracing, 2015Haywood, Katz2021}, the differences we see between the \rb\ tracks highlight the information lost to radial mixing processes (see Figure \ref{fig:xfe_Rb}). 

\subsection{The evolution of a given birth radius in the [Fe/H]--[Mg/Fe] plane with time}

Figure \ref{fig:mgfe_feh} reveals that the high-$\alpha$ sequence formed $>10$ Gyr ago\footnote{It should be kept in mind that different age estimates are not necessarily on the same scale (depending on the method and the use of prior).}, which is the period during which the gradient is found to steepen (Figure \ref{fig:feh0-range}, left). The valley between the two sequences formed between 9 and 10 Gyr ago with the gap larger for lower \feh, captured by the steep gradients in [X/H]. The origin of the transition between the high- and low-$\alpha$ sequences is a fundamental question still not well understood. Recent work has proposed a variety of situations in which the $\alpha$-bimodality arises. Using simulations from the NIHAO-UHD project, \cite{2020_buckchemical} found that the high-$\alpha$ sequence formed first, and the low-$\alpha$ sequence was formed after the interstellar medium was diluted with metal-poor gas brought in by a gas-rich merger. A chemical bimodality was also found in the VINTERGATAN simulation and linked to a rapid formation of an outer, metal-poor, low-\alphafe\ disk surrounding the inner, metal-rich, old high-\alphafe\ population  \citep{Agertz2021_vintergatanI, Renaud2021_vintergatanII, Renaud2021_vintergatanIII}. Assuming 2 main starburst episodes $\sim6$ Gyr apart, \cite{2020Lian_alphaDichotomy} found that the $\alpha$-dichotomy can result from the rapid suppression of star formation after the first starburst. Conversely, \cite{Clarke2019} showed that an $\alpha$-bimodality can occur even if the $\alpha$-sequences form non-sequentially, where the high-$\alpha$ sequence forms from rapidly self-enriching star-forming clumps while the low-$\alpha$ sequence is produced by radially distributed star formation. \cite{Khoperskov2021} also suggested that mergers are not necessary to create the bimodality, while \cite{2021Sharma} modeled the valley between the two sequences with time delays between SNII and SNIa. The latter interpretation, however, is at odds with the hybrid chemo-dynamical model by \cite{Minchev2013}, where only a single gas infall was used and strong migration was present; yet, no separation into low- and high-$\alpha$ was found. Our results indicate that the high- and low-$\alpha$ sequences formed consecutively, and are consistent with work proposing that the low-$\alpha$ sequence was created by gas brought in by a gas-rich merger.

The fluctuations in the birth \gradFeh\ found at $\sim4$ and $\sim6$ Gyr (Figure \ref{fig:feh0-range}) create only small wiggles in the mono-\rb\ tracks in the [Fe/H]--[Mg/Fe] plane, and are able to reverse the continuous \feh\ enrichment at each \rb.

\subsection{The weak possibility of chemical tagging in current-day data}

Since the dynamics of stars evolve over time \citep{Selwood2002, Roskar2008, 2009schonrichBinney, Minchev2010}, the goal of chemical tagging is to utilize the chemical homogeneity of stellar birth groups to determine stellar birth populations from the stars' chemical abundances \citep{2002freeman-BH, BH2010}. With estimates of \rb, we directly test this ability under current day observational data. We find that most \xfe\ distributions are normally distributed with different means for mono-age, mono-\rb\ populations, indicating that a given (age, \rb) is reasonably chemically homogeneous (Figure \ref{fig:xfe_dists} in the Appendix). However, particularly for \rb\ close to the Galactic center, there is a fair amount of overlap between the different age distributions. This means that under observational uncertainties, there are a variety of (age, \rb) pairs for a given \xfe. Using bins of 0.05 dex in \mgfe, \cafe, \mnfe, \cfe, \alfe, and 0.1 dex in \cefe\ and \feh, we find there is an average age dispersion of 1 Gyr and an average \rb\ dispersion of 0.74 kpc for a given abundance pattern. Both of these values are on par with their measurement uncertainties, and suggest stronger versions of chemical tagging are infeasible until ages and abundances can be estimated more precisely (see also \citealt{Casamiquela2021}).  

\section{Conclusions} \label{sec:conclusions}

This work explored in detail the enrichment with lookback time of six chemical abundances across the Milky Way disk, using birth radii and stellar age for \numstar\ \apogee\ DR17 red giant stars. Since stars radially migrate away from their birth sites, a powerful tool in reconstructing the evolution of the Milky Way disk is to look at the preserved birth properties of stars --- their abundances. With access to precise ages and \feh, \xfe\ for large samples of stars, we estimated the stellar birth radii without resorting to chemo-dynamical models, using the recent technique of \citet{Lu2022_Rb}. This allowed us to go beyond interpreting a diluted history from the present-day picture of the Galaxy and instead uncover the detailed chemical evolution of the Milky Way disk directly from the data.

In this work we focused on representative elements from the $\alpha$- (Mg, Ca), iron-peak (Mn), light (C), light odd-Z (Al), and s-process (Ce) families. While we performed quality cuts on signal-to-noise, abundance flags, and measurement uncertainties (Section \ref{sec:data}), we remind the reader that the uncertainty in \cefe\ is larger than the uncertainty in the other abundances (0.1 dex compared to 0.03 dex) and is dependent on effective temperature \citep{Cunha2017}. Section \ref{sec:bawlas} in the Appendix briefly looks at the evolution of Ce using \cefe\ from the BAWLAS catalog \citep{Hayes2022_bawlas}, and the evolution of neutron capture elements are discussed in more detail in our work in preparation. Our key findings are summarized below:

\begin{itemize}
    \item We found that the birth metallicity gradient evolution had three steepening phases $\sim4,\sim6,$ and $9$ Gyr ago (Figure \ref{fig:feh0-range}). This is in contrast to previous work using a similar method but finding monotonically flattening \gradFeh\ with time \citep{2018Minchev_rbirth} or only one fluctuation at $\sim8$~Gyr \citep{Lu2022_Rb}. We tentatively associate these fluctuations in \gradFeh\ with dilution in \feh\ from gas brought in by the GSE and Sagittarius mergers.

    \item The \gradFeh\ fluctuations at lookback times $\sim4$ and $\sim6$~Gyr result in flattening (and for some \rb\ and times even inversion) of the AMR for mono-\rb\ populations (Figure \ref{fig:feh0-range}, right). This breaks the expectation that a given \rb\ is monotonically increasing in \feh, even if no scatter is present at the time of birth \citep[e.g.][]{2018Minchev_rbirth}. See also Figure 8 in \cite{2020_buckchemical}.
    
    \item There is structure in the \rb--\xfe\ plane for the $\alpha$-, light odd-Z, and iron-peak elements that is lost in the observable \rguide--\xfe\ plane. This structure corresponds to the evolutionary events that create the high- and low-$\alpha$ sequences (Figure \ref{fig:xfe_Rb}). 

    \item As a whole, the radial \xh\ gradient flattens over time (Figure \ref{fig:slopes}). The shape of the \rb--\xh\ radial gradient is most similar to the \rguide--\xh\ gradient for the youngest stars, while the older stars have significantly different gradients (Figure \ref{fig:xh_r}). This is a consequence of radial mixing, as these younger stars have not had time to be strongly affected by radial migration.

    \item Except for \cefe, the other \xfe\ and \xh\ abundances examined in this work are good tracers of age once conditioned on \rb. Mono-\rb\ populations do not monotonically increase in \xh\ over lookback time due to the fluctuations in the metallicity gradient $< 8$ Gyr ago (Figure \ref{fig:xfe_age_r}).

    \item We find that the gap in the \feh--\alphafe\ plane comes from the quick transition between Type-II and Type-Ia Supernova dominated evolutionary phases that causes a fast drop in \alphafe, in agreement with previous work \citep{Lu2022_Rb}. 

    \item In order to explain the decline in \mgfe\ with increasing \feh\ in the high-$\alpha$ sequence, one needs strong inside-out disk formation very early on (Figure \ref{fig:mgfe_feh}). This causes the transition from Type II to Type Ia Supernova in the inner few kpc sooner than in the outer radii, and can be linked to the strong positive [X/Fe] gradients of $\alpha$-elements for the high-$\alpha$ sequence (top panels of Figure \ref{fig:xfe_Rb}). 
 
\end{itemize}

Our results suggest that the Milky disk should be studied as one evolving entity, rather than splitting it into low- and high-$\alpha$ populations. We find that the $\alpha$-sequences are not discrete disks, but rather they represent different times and birth radii, where the high-$\alpha$ sequence is simply composed of stars born at \rb$<6$~kpc with age $>8$~Gyr and the low-$\alpha$ sequence is a compilation of different ages and \rb\ (see Figure \ref{fig:mgfe_feh}). This conclusion is supported by the fact that except in chemical space, a bimodality in disk stars is not seen in kinematics or morphology (e.g., \citealt{bovy2012milky}). 

The data-driven technique used in this work recovers the trends in chemical relations expected from detailed forward modeling \citep{Matteucci1989, Chiappini1997, Minchev2013, 2020buck_NIHAO-UHD, 2020_buckchemical, Hemler2021, 2021Johnson}. Here, however, we infer these trends directly from the data using a completely different approach with less prior assumptions. 

The detailed evolution of chemical abundances with birth radius and time that we derive in this work can serve as a prior at every time step for chemical evolution models, which are typically constrained by only the present-day metallicity gradient. Such work, however, needs to decipher the contribution from now-dead massive stars, galactic fountains, gas infall from cosmic filaments and satellites, etc., which, although all are naturally accounted for in our results, cannot be quantified by our approach. 

\section*{Acknowledgements}

Funding for the Sloan Digital Sky Survey V has been provided by the Alfred P. Sloan Foundation, the Heising-Simons Foundation, the National Science Foundation, and the Participating Institutions. SDSS acknowledges support and resources from the Center for High-Performance Computing at the University of Utah. The SDSS web site is \url{www.sdss.org}.

SDSS is managed by the Astrophysical Research Consortium for the Participating Institutions of the SDSS Collaboration, including the Carnegie Institution for Science, Chilean National Time Allocation Committee (CNTAC) ratified researchers, the Gotham Participation Group, Harvard University, Heidelberg University, The Johns Hopkins University, L’Ecole polytechnique f{\'e}d{\'e}rale de Lausanne (EPFL), Leibniz-Institut f\"{u}r Astrophysik Potsdam (AIP), Max-Planck-Institut f\"{u}r Astronomie (MPIA Heidelberg), Max-Planck-Institut f\"{u}r Extraterrestrische Physik (MPE), Nanjing University, National Astronomical Observatories of China (NAOC), New Mexico State University, The Ohio State University, Pennsylvania State University, Smithsonian Astrophysical Observatory, Space Telescope Science Institute (STScI), the Stellar Astrophysics Participation Group, Universidad Nacional Aut\'{o}noma de M\'{e}xico, University of Arizona, University of Colorado Boulder, University of Illinois at Urbana-Champaign, University of Toronto, University of Utah, University of Virginia, Yale University, and Yunnan University.

We acknowledge Lucy(Yuxi) Lu for helpful discussions. B.R. and I.M. acknowledge support by the Deutsche Forschungsgemeinschaft under the grant MI 2009/2-1. This work was partially funded by the Spanish MICIN/AEI/10.13039/501100011033 and by the "ERDF A way of making Europe" funds by the European Union through grant RTI2018-095076-B-C21 and PID2021-122842OB-C21, and the Institute of Cosmos Sciences University of Barcelona (ICCUB, Unidad de Excelencia ’Mar\'{\i}a de Maeztu’) through grant CEX2019-000918-M. FA acknowledges financial support from MCIN/AEI/10.13039/501100011033 through grants IJC2019-04862-I and RYC2021-031638-I (the latter co-funded by European Union NextGenerationEU/PRTR). 

\section*{Data Availability}

The datasets used and analysed for this study are
derived from data released by \apogee\ DR17, \cite{Queiroz2023_SH}, and \cite{Anders2023_ages}. The rest of the relevant datasets are available from the corresponding author upon reasonable request.

\appendix
\label{sec:append}

\section{Consistency across age catalogs} 
\label{sec:starhorse}

In this section, we show that the radial abundance gradients throughout time found in the main body of the text are not dependent on age derivation. The age catalog of \cite{Anders2023_ages} estimates age from stellar parameters and elemental abundance ratios using the supervised machine learning technique {\tt XGBoost}. Here, we now use ages derived from isochrone fitting for \apogee\ DR17 subgiant branch stars using the \sh\ catalog \citep{Queiroz2023_SH}. We follow the same methods as in Section \ref{sec:Rb_method} to estimate \rb\ for 13,735 stars with $|\feh| \leq 1$, $|z| \leq 1$ kpc, eccentricity $< 0.5$, $\feh_\text{err} < 0.015$, and $\text{age}_\text{err} < 1.25$ Gyr. 

\begin{figure*}
     \centering
     \includegraphics[width=1\textwidth]{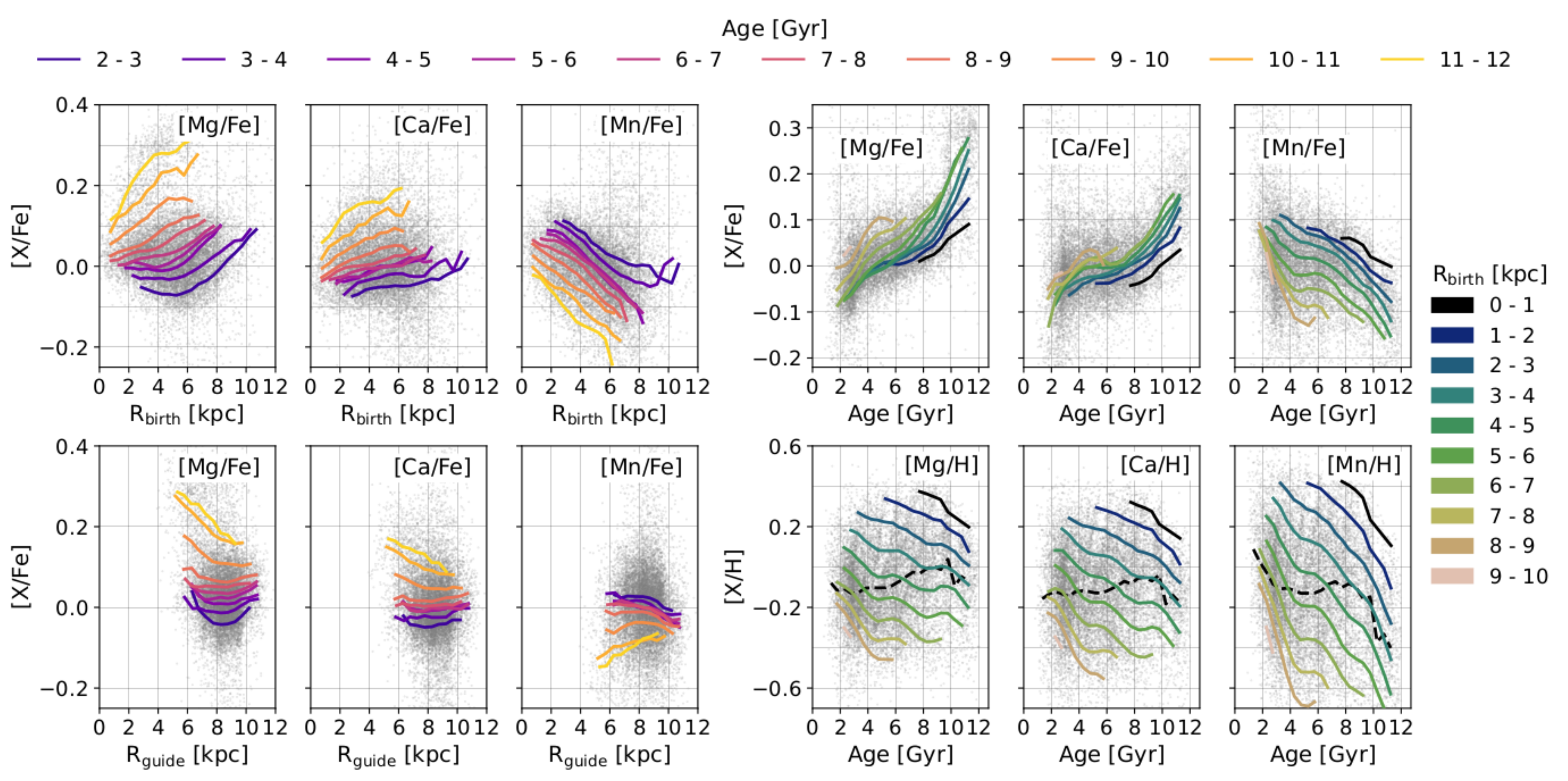}
\caption{Comparison with isochrone ages derived with \sh. \edits{{\bf Left 3 panels:} Mono-age populations of 12,949 \apogee\ DR17 subgiant branch stars in the \textbf{Top:} \rb--\xfe\ and \textbf{Bottom:} \rguide--\xfe\ planes. {\bf Right 3 panels:} Mono-\rb\ populations in the {\bf Top:} age--\xfe\ plane and {\bf Bottom:} age--\xh\ plane of the \sh\ sample. Consistencies seen between these planes and the ones in Figures \ref{fig:xfe_Rb} and \ref{fig:xfe_age_r} suggest that the conclusions drawn in Section \ref{sec:results} are not an artifact of stellar age derivation, and are representative of Milky Way evolution.}}
\label{fig:xfe_Rb_sh}
\end{figure*}

\begin{figure*}
     \centering
     \includegraphics[width=.25\textwidth]{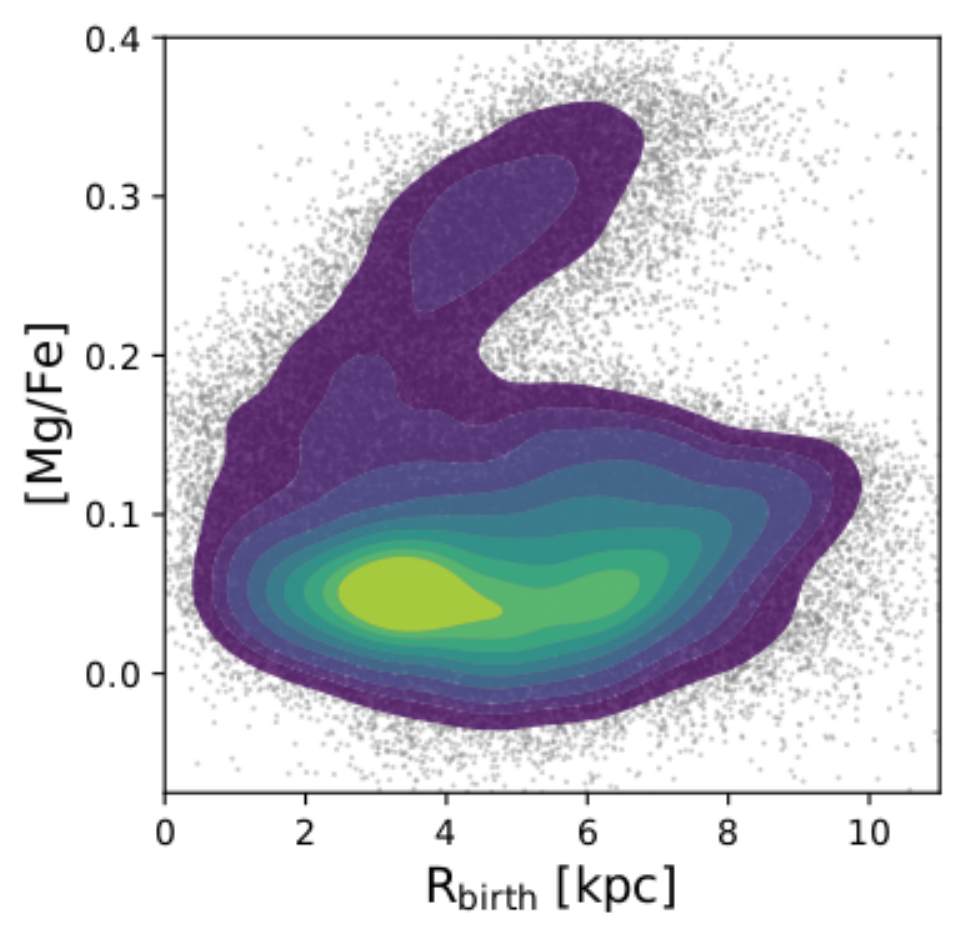}
     \includegraphics[width=.26\textwidth]{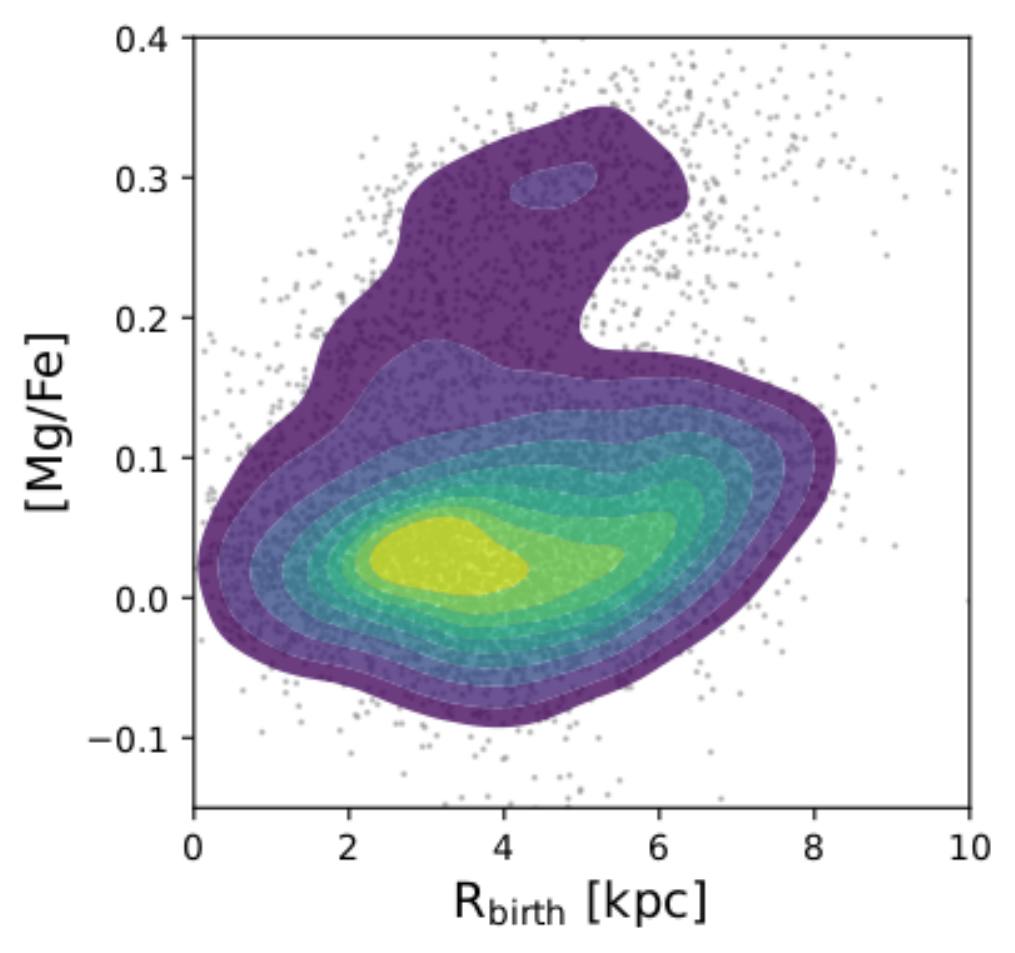}\\
\caption{The \rb--\mgfe\ plane of \textbf{Left:} \numstar\ \apogee\ DR17 red giant stars with ages derived from abundances, and \textbf{Right:} 12,949 \apogee\ DR17 subgiant branch stars with ages derived from isochrone fitting. Since \sh\ has few stars with ages $<2$ Gyr, the left figure has been cut to exclude the younger stars for a consistent comparison. Despite different evolutionary states and age derivations, the structure of the high- and low-$\alpha$ sequences is apparent in both figures, suggesting that the structure is real. }
\label{fig:mgfe_rb_compare}
\end{figure*}

After removing stars with flagged abundances and $\xfe_\text{err}>0.05$ dex, we perform this comparison analysis for three elements --- \mgfe, \cafe, and \mnfe --- of nearly 13,000 subgiant branch stars. Figure \ref{fig:xfe_Rb_sh} shows the \xfe\ radial trends as a function of both \rb\ and \rguide\ for mono-age populations. Similar to Figure \ref{fig:xfe_Rb}, the inherent radial gradients tend to be steeper than looking at the present-day Galactic disk. The bimodal structure found in the \rb--\alphafe\ plane is also still preserved (Figure \ref{fig:mgfe_rb_compare}). While the distribution of the low-$\alpha$ sequence in the \rb--\mgfe\ plane contain minimal differences ---possibly due to a difference in sample size or evolutionary state --- the overall shape of this plane is quite similar. The similarities \edits{we find in all results --- specifically the multiple fluctuations in \gradFeh, the radial \xfe\ and \xh\ gradients being steeper with \rb\ than \rguide, the structure in the \rb--\xfe\ plane, mono-\rb\ populations lying on steeper tracks in the age--\xh\ and age--\xfe\ planes, and the non-monotonic increase in [X/H] at a given \rb\ ---} show that the conclusions drawn in this work are not an artifact of stellar age catalog or evolutionary state, and are indicative of the Milky Way's chemical evolution.

\section{cerium results using BAWLAS abundances}\label{sec:bawlas}

In Section \ref{sec:results}, we used \cefe\ abundances given by \aspcap\ in the \apogee\ DR17 catalog. Here, we compare our results using \cefe\ provided in the BACCHUS Analysis of Weak Lines in APOGEE Spectra \citep[BAWLAS;][]{Hayes2022_bawlas} catalog. For this analysis, we keep disk ($|z|\leq 1$ kpc, eccentricity $<0.5$) stars with $-1\leq \cefe \leq 0.5$, and a Ce empirical error cut $<0.07$ dex. The cut of \cefe$<0.5$ is chosen to remove s-process enhanced stars \citep{Hayes2022_bawlas} (though our results are robust to the inclusion of them). This leaves us with 36,239 stars for this analysis.

\begin{figure*}
     \centering
     \includegraphics[width=.8\textwidth]{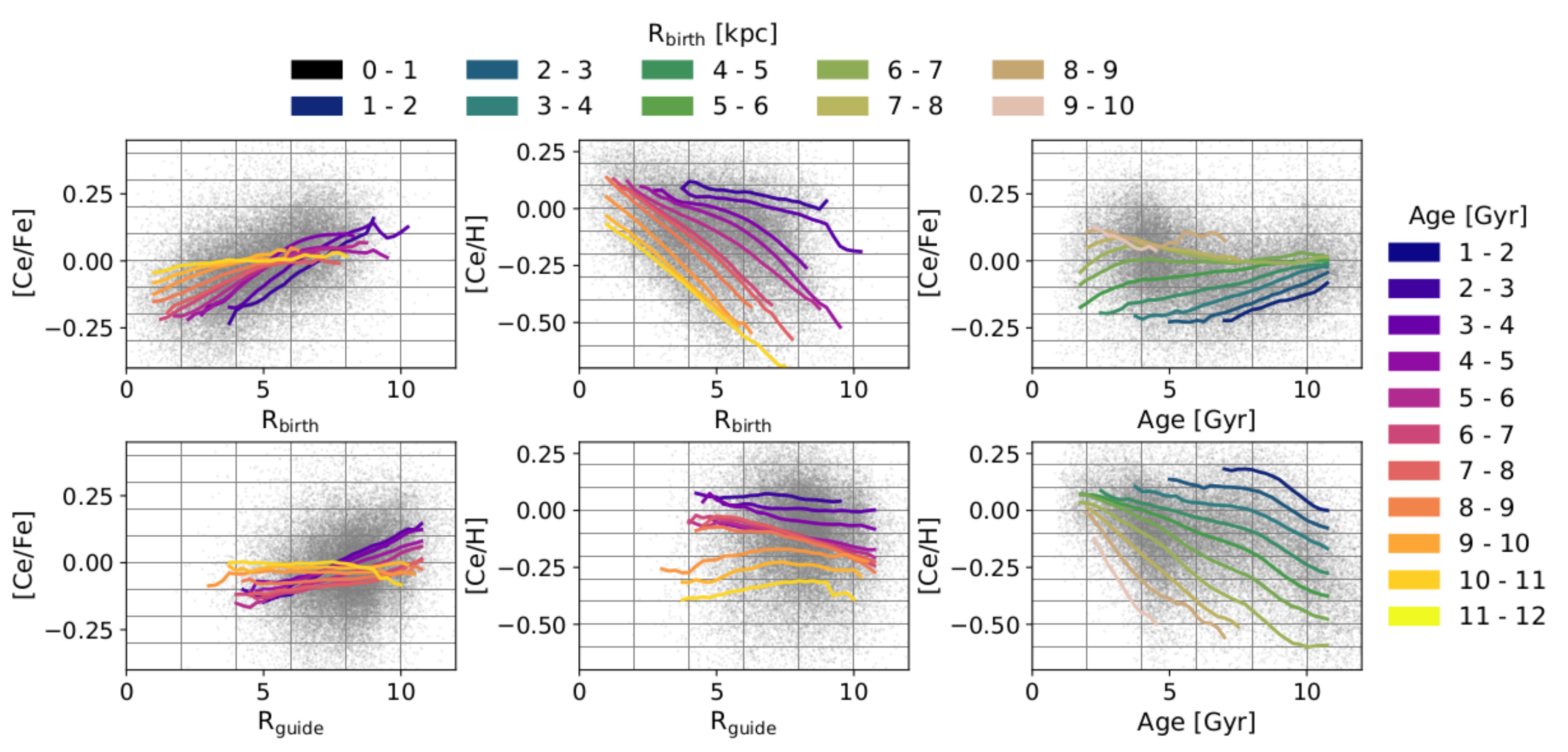}
\caption{Comparison of Ce results using \cefe\ abundances given in the BAWLAS catalog. The running means of mono-age populations are in the \textbf{top left:} \rb--\cefe, \textbf{bottom left:} \rguide--\cefe, \textbf{top middle:} \rb--[Ce/H], and  \textbf{bottom middle:} \rguide--[Ce/H] planes for 36,239 stars. The running means of mono-\rb\ populations in the age--\cefe\ and age--[Ce/H] planes are shown in the top right and bottom right panels respectively. The overall trends are similar to Figures \ref{fig:xh_r}, \ref{fig:xfe_Rb}, and \ref{fig:xfe_age_r}, with mono-\rb\ populations showing more separation in the age--\cefe\ plane. }
\label{fig:bawlas}
\end{figure*}

Figure \ref{fig:bawlas} provides the main figures reproduced with the new abundances. Overall, the results are consistent with the results shown in Section \ref{sec:results}. In particular, \cefe\ has a weak radial birth gradient for older stars which becomes stronger with decreasing age (top left panel of Figure \ref{fig:bawlas}). However, the existence of a break point is no longer noticeable for the youngest and oldest populations, but seems to still exist for the $3-10$ Gyr populations. The [Ce/H] gradient also sees significant flattening with decreasing age (top middle panel of Figure \ref{fig:bawlas}), in agreement with Figure \ref{fig:xh_r}. In Figure \ref{fig:xfe_age_r}, we saw that the oldest end of each mono-\rb\ population in the age--\cefe\ plane decreased in \cefe. Now, with \cefe\ derived using BAWLAS, we find even clearer separation of mono-\rb\ populations in the age--\cefe\ plane, which do not show this downward trend. However, the correlation between \cefe\ and age is still dependent on \rb. The similarities between \cefe\ abundances in the \apogee\ DR17 and BAWLAS catalogs suggest that the trends we find are not an artifact of pipeline, and are an inherent property of Ce evolution.

\section{Consistency when varying the strength of the metallicity gradient} \label{sec:varyGrad}

\begin{figure*}
     \centering
     \includegraphics[width=.725\textwidth]{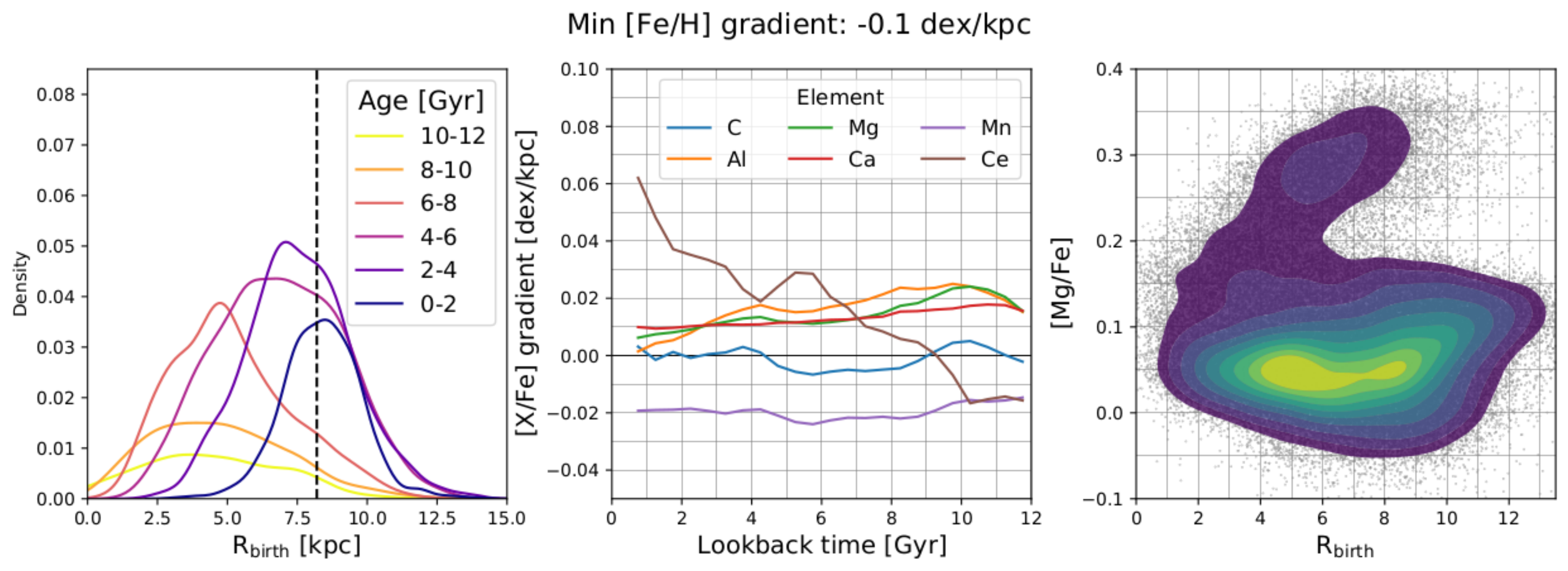}\\
     \includegraphics[width=.725\textwidth]{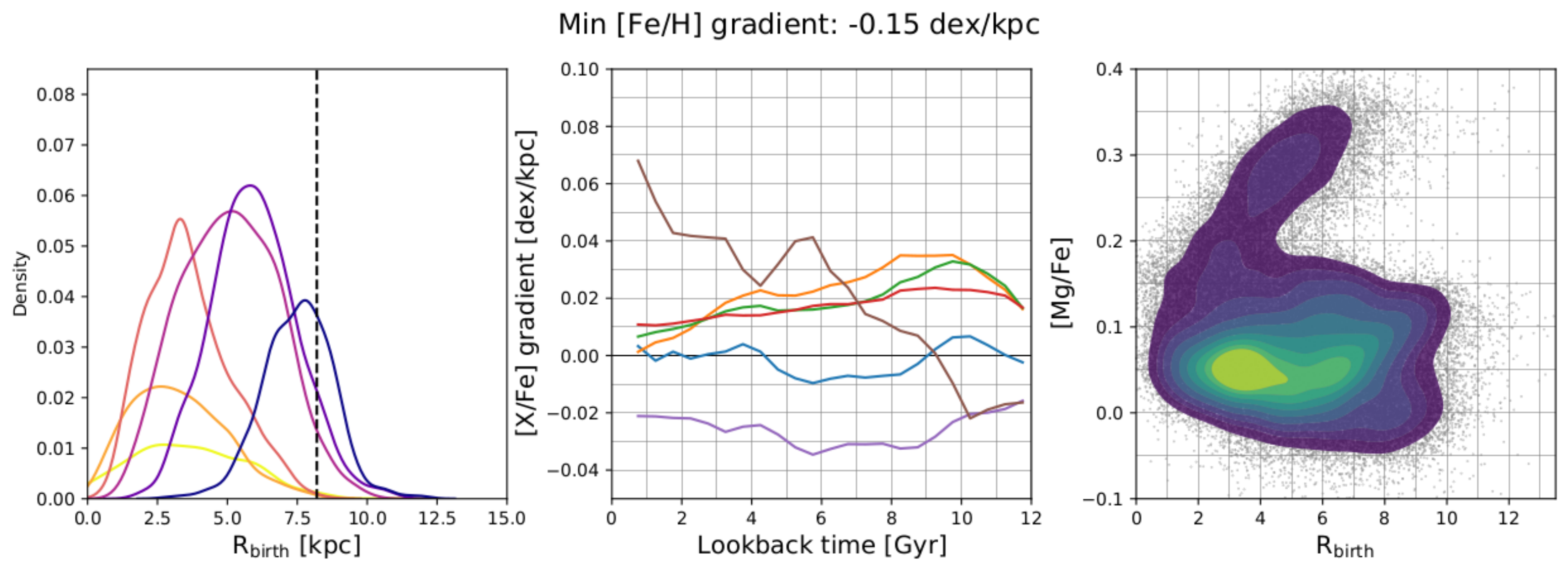}\\
     \includegraphics[width=.725\textwidth]{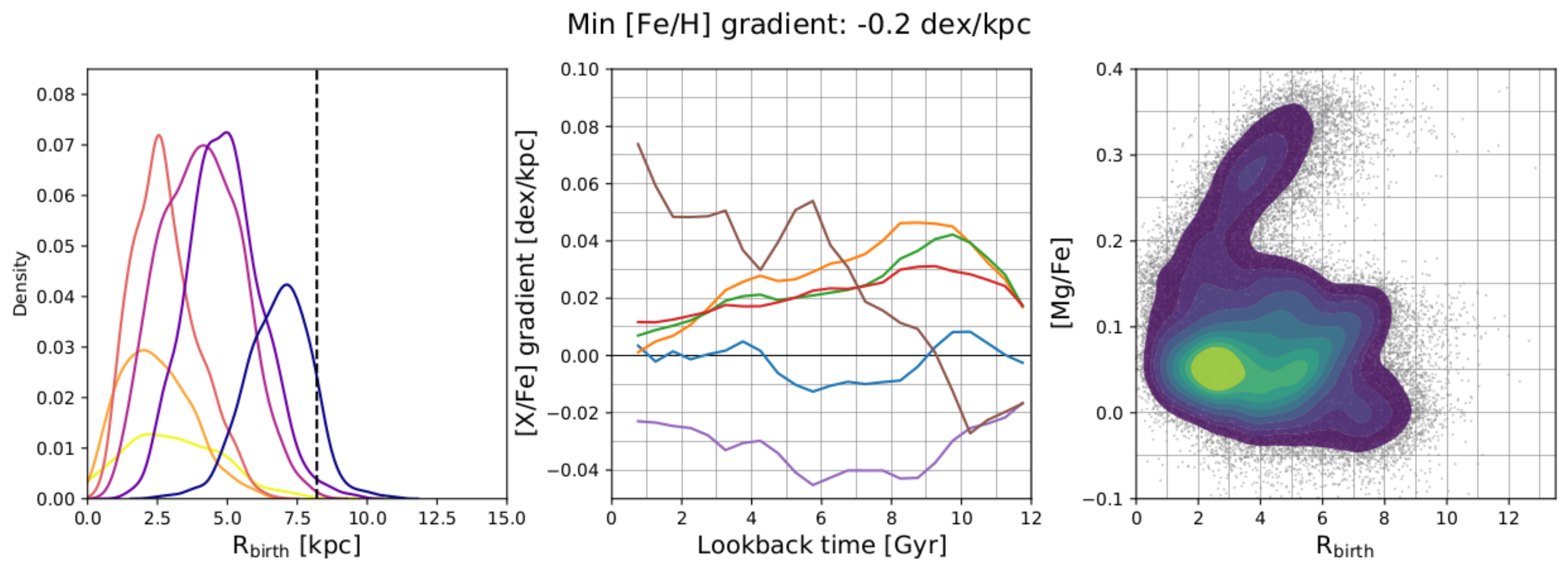}
\caption{The \textbf{Left:} distribution of \rb\ for stars in the solar neighborhood, \textbf{Middle:} \xfe\ radial gradient over lookback time ($\Delta$lookback time = 1 Gyr), and \textbf{Right:} \rb--\mgfe\ plane when \rb\ is derived using a minimum metallicity gradient of \textbf{Top row:} $-0.1$ dex/kpc, \textbf{Middle row:} $-0.15$ dex/kpc (used in the main body of this work), and \textbf{Bottom row:} $-0.2$ dex/kpc. The results presented in Section \ref{sec:results} are consistent for varying minimum \feh\ radial gradients, with the overall trends in each plane being similar and the numerical values shifting.}
\label{fig:maxGrads}
\end{figure*}

As discussed in Section \ref{sec:Rb_method}, we chose min(\gradFeh) = $-0.15$ dex/kpc based on previous works \citep{2018Minchev_rbirth, Lu2022_Rb}. The left panels of Figure \ref{fig:maxGrads} show the distribution of \rb\ for mono-age populations (similar to Figure \ref{fig:SN_Rbdist}) when \rb\ are derived with min(\gradFeh) = $-0.1, -0.15$, and $-0.2$ dex/kpc. When the minimum \feh\ gradient is $-0.1$ dex/kpc, the \rb\ distributions in the solar neighborhood become too wide, and implies most of the $0-2$ Gyr stars have migrated inwards, while a minimum \feh\ gradient of $-0.2$ dex/kpc indicates the younger aged stars have migrated outwards by over 3 kpc on average. Thus, our choice of min(\gradFeh) = $-0.15$ dex/kpc aligns most with expectations. 

The middle and right panels of Figure \ref{fig:maxGrads} illustrate that even if our assumption of min(\gradFeh) is not optimal, the overall trends of our results are consistent when this assumption is varied. A more shallow gradient (min(\gradFeh) = $-0.1$ dex/kpc) leads to flatter \xfe\ radial gradients over time, whereas a steeper gradient (min(\gradFeh) = $-0.2$ dex/kpc) creates steeper \xfe\ radial gradients for a given lookback time. Despite the numerical differences between the \xfe\ radial gradients derived using different gradient strengths, the relative trends among the gradients is preserved. Similarly, the bimodal structure of the high- and low-$\alpha$ sequences in the \rb--\mgfe\ plane found in Section \ref{sec:xfe} is still visible under different metallicity gradient strengths. The third clump of younger aged stars is more prominent under the steeper \gradFeh,  but can be seen using the different strengths.

\section{Radial abundance gradients with lookback time for many abundances}

Table \ref{tab:xfe_grad} presents the numerical values of the \xfe\ radial gradients with lookback time. For each abundance, the number of stars used to determine the gradients is given in the second column. Only stars with unflagged \xfe\ with $|\xfe| < 1$ and $\xfe_\text{err}<0.1$ dex for \cefe, or $<0.05$ dex for all other \xfe\ are used.  

\begin{table*}
\begin{center}
\begin{tabular}{ c |c |c c c c c c c c c c c c}
 & & \multicolumn{12}{c}{\xfe\ radial birth gradient [dex/kpc] for given lookback time [Gyr]}\\
 \xfe\ &  $n$ & $0-1$  & $1-2$ & $2-3$  & $3-4$  & $4-5$  & $5-6$  & $6-7$  & $7-8$  & $8-9$  & $9-10$ & $10-11$ & $11-12$  \\ \hline
$\text{[C/Fe]}$ & 144070 & 0.0 & 0.0 & 0.0 & 0.01 & -0.0 & -0.01 & -0.01 & -0.01 & -0.0 & 0.01 & 0.01 & -0.0 \\
$\text{[N/Fe]}$ & 143986 & -0.02 & -0.02 & -0.01 & -0.02 & -0.02 & -0.03 & -0.03 & -0.02 & -0.01 & -0.01 & -0.01 & -0.01 \\
$\text{[Al/Fe]}$ & 144072 & 0.0 & 0.01 & 0.01 & 0.02 & 0.02 & 0.02 & 0.02 & 0.03 & 0.04 & 0.04 & 0.03 & 0.02 \\
$\text{[Mg/Fe]}$ & 144087 & 0.01 & 0.01 & 0.01 & 0.02 & 0.02 & 0.02 & 0.02 & 0.02 & 0.03 & 0.03 & 0.03 & 0.02 \\
$\text{[O/Fe]}$ & 142406 & 0.01 & 0.01 & 0.02 & 0.03 & 0.02 & 0.02 & 0.03 & 0.03 & 0.04 & 0.05 & 0.04 & 0.04 \\
$\text{[Si/Fe]}$ & 144090 & 0.0 & 0.0 & 0.01 & 0.01 & 0.01 & 0.01 & 0.01 & 0.02 & 0.02 & 0.03 & 0.03 & 0.02 \\
$\text{[Ca/Fe]}$ & 144089 & 0.01 & 0.01 & 0.01 & 0.01 & 0.01 & 0.02 & 0.02 & 0.02 & 0.02 & 0.02 & 0.02 & 0.02 \\
$\text{[Mn/Fe]}$ & 144090 & -0.02 & -0.02 & -0.02 & -0.02 & -0.03 & -0.03 & -0.03 & -0.03 & -0.03 & -0.03 & -0.02 & -0.02 \\
$\text{[Cr/Fe]}$ & 89091 & -0.0 & -0.0 & -0.01 & -0.01 & -0.01 & -0.01 & -0.01 & -0.01 & -0.01 & -0.02 & -0.02 & -0.01 \\
$\text{[Co/Fe]}$ & 33877 & -0.04 & -0.04 & -0.03 & -0.04 & -0.04 & -0.04 & -0.03 & -0.03 & -0.02 & -0.01 & 0.01 & 0.01 \\
$\text{[Ce/Fe]}$ & 93123 & 0.06 & 0.04 & 0.04 & 0.04 & 0.03 & 0.04 & 0.03 & 0.01 & 0.01 & -0.0 & -0.02 & -0.02 \\
\end{tabular}
\caption{\label{tab:xfe_grad} Numerical \xfe\ radial gradients over lookback time. Gradients are found by fitting a linear model to a mono-age population.}
\end{center}
\end{table*}

\section{The relationship between age, birth radii, and abundances}\label{sec:chemicalTag}

Figure \ref{fig:xfe_Rb} suggests that there is a relationship between \xfe, age, and the \rb\ we derived in this work. In this section we examine this relationship to test if abundances, age, and \rb\ have a unique link.

\begin{figure*}
     \centering
     \includegraphics[width=.89\textwidth]{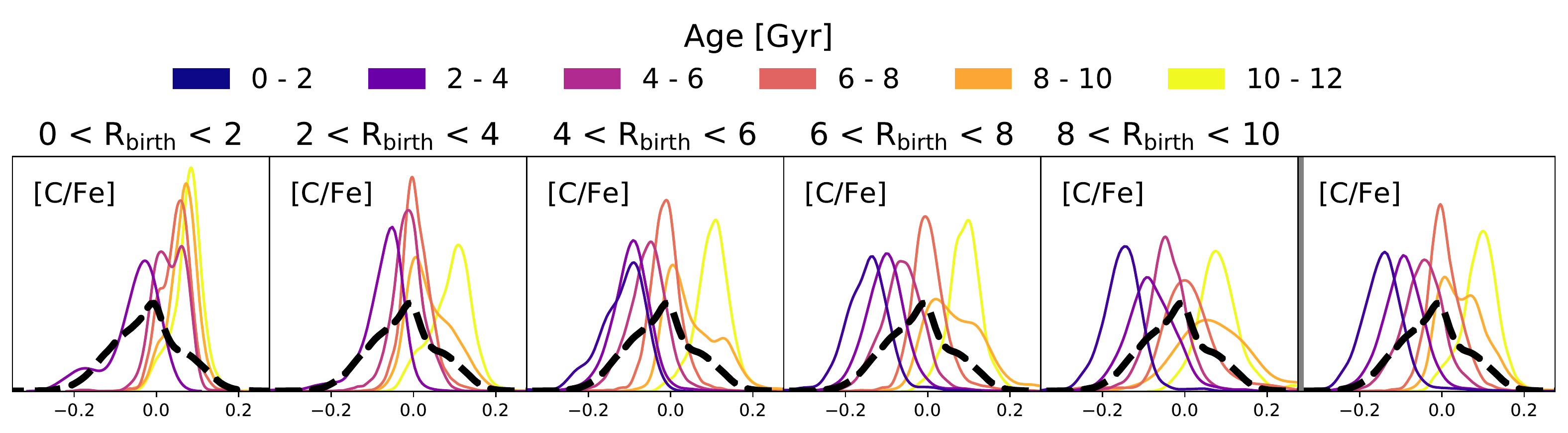}\\
     \includegraphics[width=.89\textwidth]{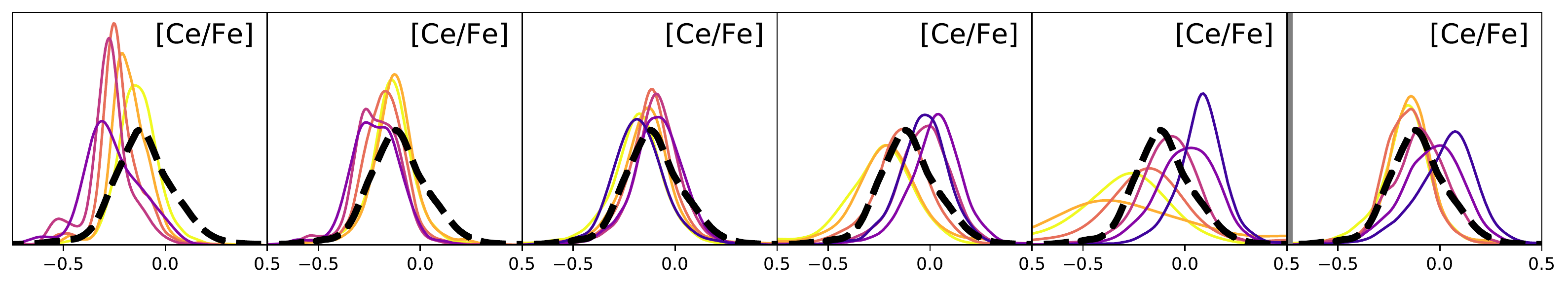}\\
     \includegraphics[width=.89\textwidth]{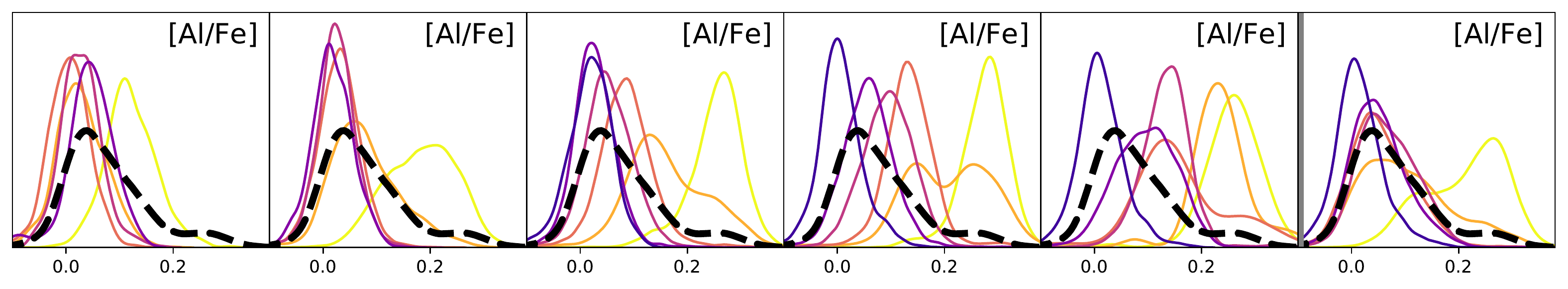}\\
     \includegraphics[width=.89\textwidth]{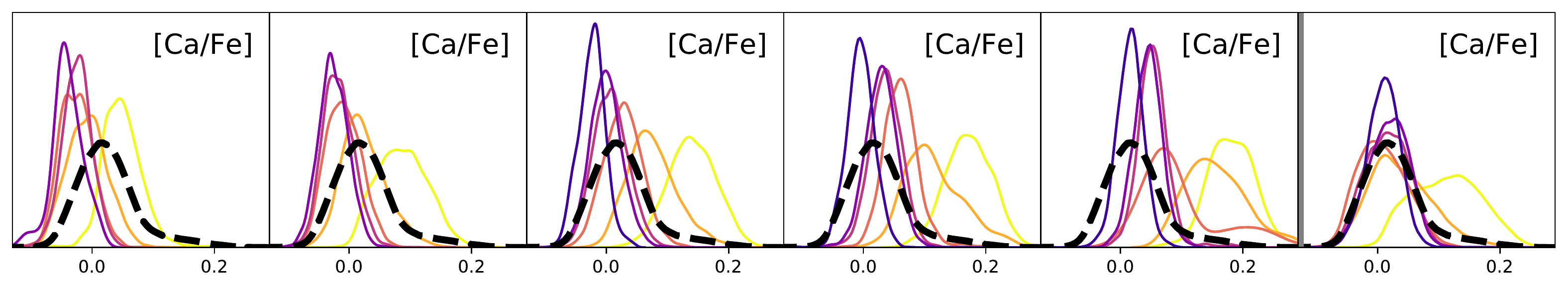}\\
     \includegraphics[width=.89\textwidth]{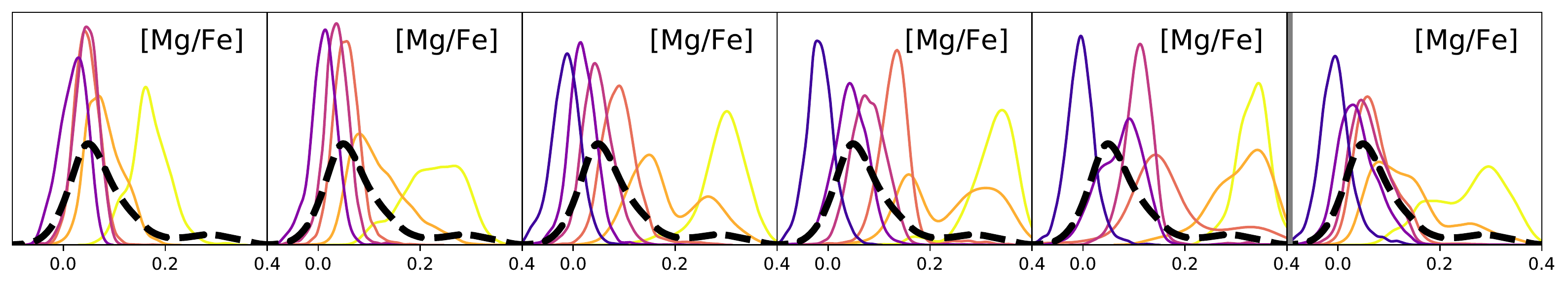}\\
     \includegraphics[width=.89\textwidth]{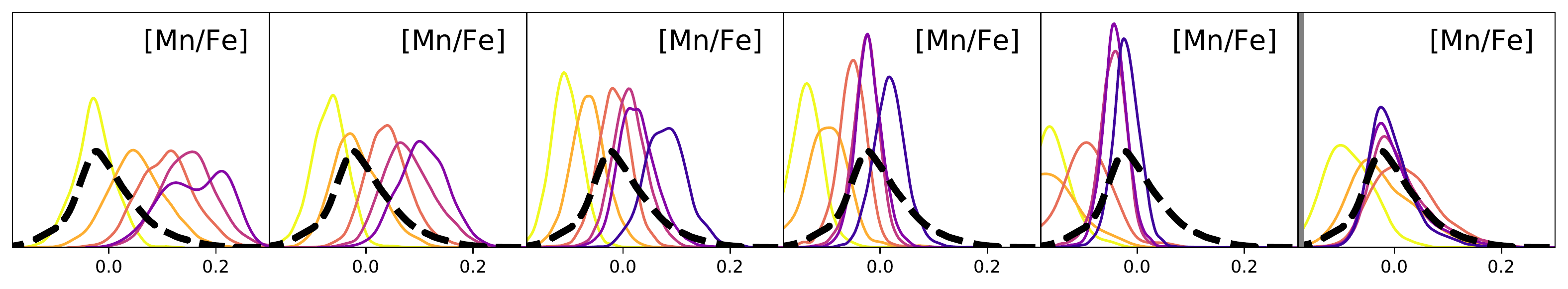}\\
\caption{\textbf{Left 5 columns:} \xfe\ distributions for mono-age, mono-\rb\ populations alongside the overall \xfe\ distribution for the entire sample (black dashed line). \textbf{Right:} \xfe\ distributions for mono-age populations alongside the distribution for the entire sample (black dashed line). Most abundances show minimal separation when only conditioning on age (e.g.\mnfe, \cafe, \alfe), but show distinct differences when additionally conditioning on place of birth.
}
\label{fig:xfe_dists}
\end{figure*}

The left five columns of Figure \ref{fig:xfe_dists} show the \xfe\ distribution conditioned on age and birth radius alongside the \xfe\ distribution for the entire sample (black dashed line). Even when the stellar sample shows skew  (\alfe, \mnfe), long tails (\mgfe, \cafe), or other non-normal traits (\cfe), most \xfe\ for a given age and \rb\ are normally distributed with little variance. This implies that a given birth time and place is reasonably chemically homogeneous in these abundances. The few non-normally distributed \xfe\ (e.g. long tails in \mgfe\ for $2 < \rb < 4$ kpc, bimodality in \alfe\ for $4 < \rb < 6$ kpc) tend to be for older aged stars. These traits may be a consequence of poor age estimates of older stars \citep[as it is difficult to measure the difference between an e.g. 10 and 12 Gyr old star;][]{Mackereth2019}, or that period in time corresponds to a quickly enriching environment in the inner disk. We note that the $8-10$ Gyr age bin encompasses both high- and low-$\alpha$ stars. 

The old ($10-12$ Gyr) stars show a clear separation from the other age distributions for all mono-\rb\ populations for \alfe, \mgfe, \mnfe\ and \cafe, and across larger \rb\ for \cfe. This distinction roughly corresponds to the high- and low-$\alpha$ sequences, and demonstrates that the sequences are most similar in \xfe\ at smaller \rb. This idea is clear in the \feh--\mgfe\ plane, where the sequences converge for small \rb\ and have no obvious distinction (Figure \ref{fig:mgfe_feh}). For larger \rb, some elements have an average separation between the older and younger age groups of 0.2 dex, with minimal stars between the oldest and younger age groups (e.g. \mgfe\ and \cafe\ at $6<\rb<8$). Figure \ref{fig:xfe_dists} clearly illustrates that the older age stellar population has a different chemical composition than the younger populations, which suggests that there was a major enrichment event that quickly diluted the interstellar medium and caused a difference in these abundances.

To demonstrate the additional information gained with knowledge of \rb, the rightmost column of Figure \ref{fig:xfe_dists} presents each \xfe\ distribution only conditioned on stellar age. For half of the abundances considered in this work (\alfe, \cafe, and \mnfe), there is very little separation between age bins beyond a bulk difference between the oldest ($10-12$ Gyr) and younger aged populations. For these elements, the $<10$ Gyr age bins peak at the same location as the overall population (black dashed line), with only the $10-12$ Gyr age bin having a wider, sometimes bimodal, distribution with a different peak. \mgfe\ also has a similar trend to these three abundances, though the younger aged populations have slight variations in the location of their peak. Once utilizing birth radii information, the mono-age populations separate as a function of \rb. The \mgfe\ bimodality of the $10-12$ Gyr population becomes, for most \rb, more normally distributed with less variation. The younger aged populations also begin to show less variability in \mgfe\ when additionally conditioning on \rb, with the peak of each mono-age, mono-\rb\ population being slightly shifted.  

Opposite to the four abundances listed so far, the light element abundance \cfe\ behaves differently than the iron-peak and $\alpha$-elements. With only conditioning on age, \cfe\ shows clear separation between different age bins, and additionally conditioning on \rb\ does not reveal any additional distinction between the groups. This again highlights the strong relationship between \cfe\ and age. The s-process abundance \cefe\ also behaves differently than the iron-peak and $\alpha$-elements. \cefe\ has very little difference in the older mono-age populations, with the younger ($< 4$ Gyr) stars showing the most deviation. Once binned in \rb, the older populations are separated for smaller \rb, and show minimal separation for $\rb > 2$ kpc. All the mono-age, mono-\rb\ populations show no differences in \cefe\ for $4 < \rb < 6$ kpc. This location in the disk is where we find the pivot point of the \rb--\cefe\ radial gradients (Figure \ref{fig:xfe_Rb}), suggesting that the inner and outer disk had different \cefe\ enrichment histories. The second row of Figure \ref{fig:xfe_dists} also depicts that the lower \cefe\ values are in older stars produced in the inner disk, while the higher \cefe\ values are in younger stars born in the outer disk.

Within each mono-\rb\ population, the relative distribution of the mono-age populations reveals the \xfe\ time gradient at a given \rb. As discussed above, the mono-age populations show distinction in the light element \cfe, with minimal additional separation when also conditioning on place of birth. Within a mono-\rb\ population, the \xfe\ time gradient is fairly consistent, particularly for $\rb > 6$ kpc. For $\rb < 6$ kpc, the younger populations have higher \cfe, showing a weaker gradient with time. 

\alfe, \cafe, and \mgfe\ also show weaker time gradients closer to the Galactic center, with the gradients steepening at larger \rb. Specifically, for $\rb <2$ kpc, the mono-age populations are clustered about the peak of the non-conditional sample (black dashed line), showing that the \xfe\ distribution does not vary much near the Galactic center. For larger \rb, the difference between \xfe\ for different aged stars becomes clearer, especially between the $10-12$ Gyr age group and the others. Even within the $<8$ Gyr populations, the distributions peak at slightly different locations, indicating a smoothly continuous evolution with time. We also find for these three abundances that for any given \rb, a wide range of \xfe\ values can be created.

While the $\alpha$- and light elements show weaker time gradients for smaller \rb, \mnfe\ does the opposite. For $\rb < 4$ kpc, the different age populations are structured with time. However for $\rb > 6$ kpc, the $< 8$ Gyr stellar groups show little variation with lookback time. This indicates that nearly all values of \mnfe\ are formed at lower radii, but only stars with a small range of \mnfe\ are born further out in the disk.

\section{extra plots}

\begin{figure}
    \centering
     \includegraphics[width=.275\textwidth]{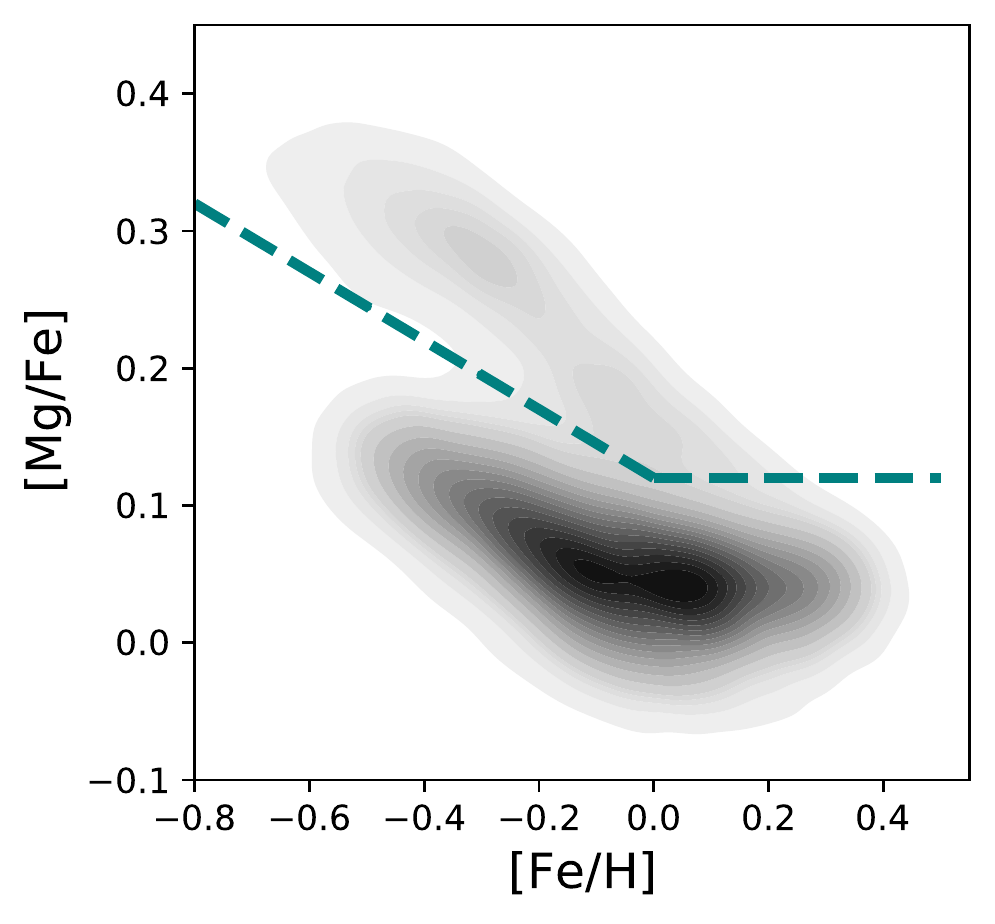}
\caption{2D density distribution of \numstar\ \apogee\ DR17 red giant stars in the [Fe/H]–-[Mg/Fe] abundance plane. The dashed teal line represents our simple by eye division into the high- and low-$\alpha$ sequence.}
\label{fig:alphaSeq_def}
\end{figure}

\begin{figure}
    \centering
     \includegraphics[width=.4\textwidth]{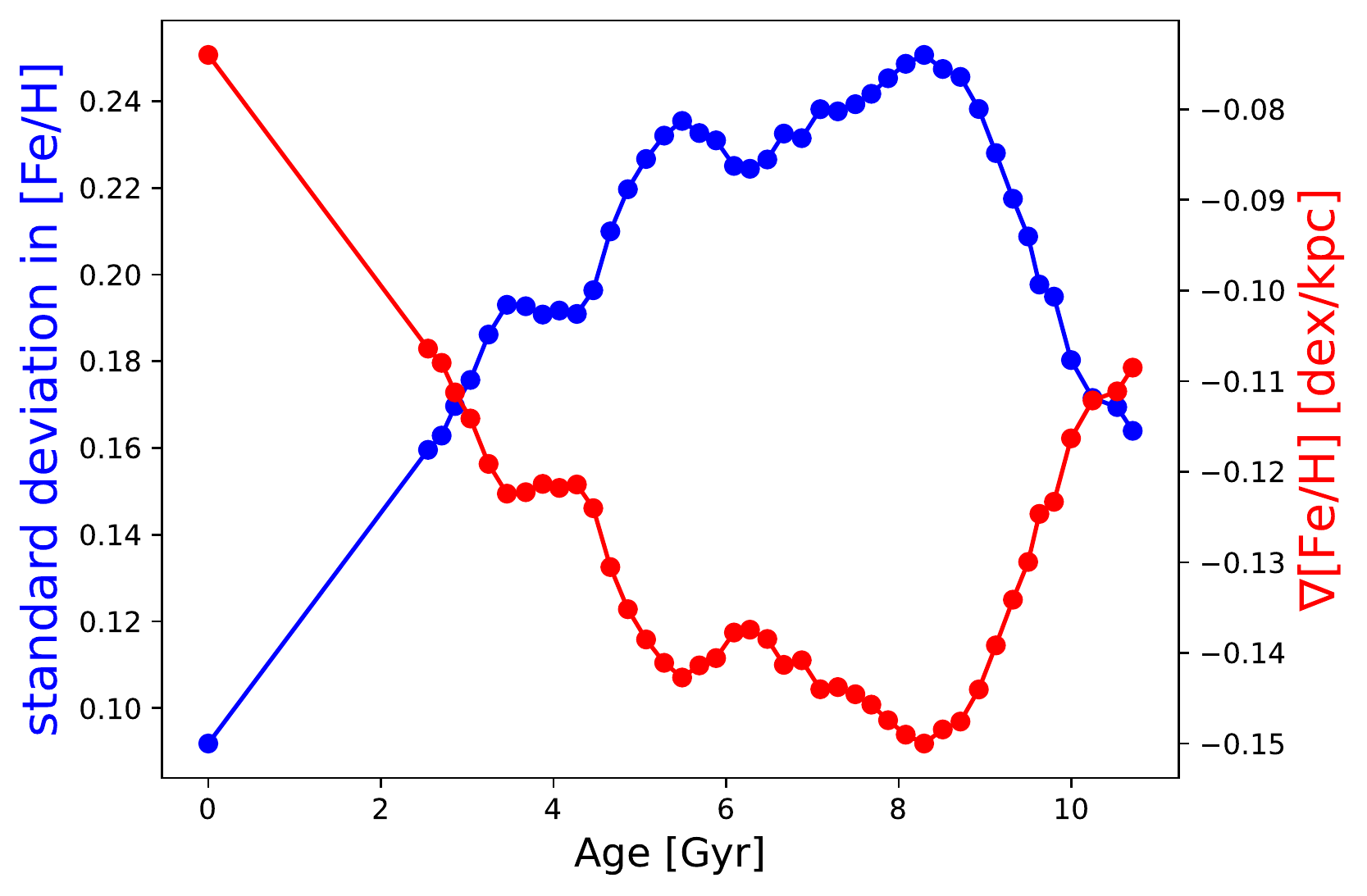}
\caption{\edits{Inferred metallicity gradient (red) and scatter in \feh\ (blue) using GALAH DR3 subgiant branch disk ($|z| < 1$, eccentricity $<0.5$, $|\feh| < 1$) stars with ages derived by isochrone fitting from the \sh\ catalog. We additionally remove flagged \feh\ and age measurements and keep stars with \feh$_\text{err} < 0.1$ and age$_{84}$ - age$_{16} < 3$. We calculate the gradient using a bin size of 1 Gyr every 0.2 Gyr (similar as done in Figure \ref{fig:samplingGrad}). The 3 fluctuations found in Section \ref{sec:Rb_method} are again recovered, showing their robustness across survey, age derivation, and evolutionary state.}}
\label{fig:galah_gradFeh}
\end{figure}

\begin{figure*}
     \centering
     \includegraphics[width=.7\textwidth]{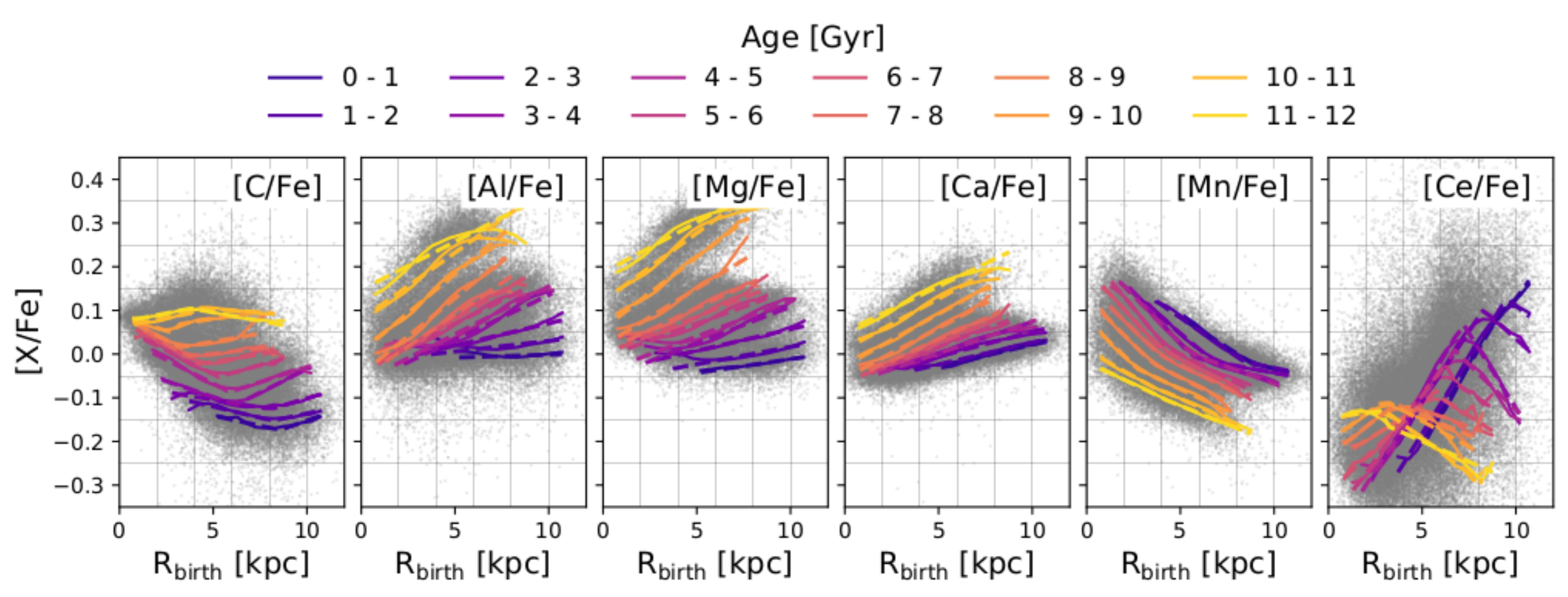}\\
\caption{Running mean of mono-age populations in the \rb--\xfe\ plane (solid lines) with the estimated fits (dashed lines). Mono-aged populations are fit as a linear function in \mgfe, \cafe, and \alfe, while the mono-age groups show non-linear trends between \rb\ and \cfe, \cefe, and \mnfe. Therefore we fit these three abundances as a function of two linear functions, one before a break point and one after. The break points are different for each mono-age group, as well as each \xfe.}
\label{fig:linearFit}
\end{figure*}

In this section we provide additional plots. Figure \ref{fig:alphaSeq_def} indicates the cut we use to define the high- and low-$\alpha$ sequences by eye. \edits{Figure \ref{fig:galah_gradFeh} illustrates the robustness of the fluctuations in \gradFeh\ by showing the gradient recovered using GALAH DR3 \citep{Buder2021} subgiant branch disk stars with ages derived by isochrone fitting from the \sh\ catalog \citep{Queiroz2023_SH}.} Figure \ref{fig:linearFit} shows the linear fits for each mono-age population in the \rb--\xfe\ plane, with \cfe, \cefe, and \mnfe\ being fit with 2 linear models (one before and one after the defined break points).

\bibliographystyle{mnras}
\bibliography{Ratcliffe}{}

\begin{thebibliography}{}
\makeatletter
\relax
\def\mn@urlcharsother{\let\do\@makeother \do\$\do\&\do\#\do\^\do\_\do\%\do\~}
\def\mn@doi{\begingroup\mn@urlcharsother \@ifnextchar [ {\mn@doi@}
  {\mn@doi@[]}}
\def\mn@doi@[#1]#2{\def\@tempa{#1}\ifx\@tempa\@empty \href
  {http://dx.doi.org/#2} {doi:#2}\else \href {http://dx.doi.org/#2} {#1}\fi
  \endgroup}
\def\mn@eprint#1#2{\mn@eprint@#1:#2::\@nil}
\def\mn@eprint@arXiv#1{\href {http://arxiv.org/abs/#1} {{\tt arXiv:#1}}}
\def\mn@eprint@dblp#1{\href {http://dblp.uni-trier.de/rec/bibtex/#1.xml}
  {dblp:#1}}
\def\mn@eprint@#1:#2:#3:#4\@nil{\def\@tempa {#1}\def\@tempb {#2}\def\@tempc
  {#3}\ifx \@tempc \@empty \let \@tempc \@tempb \let \@tempb \@tempa \fi \ifx
  \@tempb \@empty \def\@tempb {arXiv}\fi \@ifundefined
  {mn@eprint@\@tempb}{\@tempb:\@tempc}{\expandafter \expandafter \csname
  mn@eprint@\@tempb\endcsname \expandafter{\@tempc}}}

\bibitem[\protect\citeauthoryear{{Abdurro'uf} et~al.,}{{Abdurro'uf}
  et~al.}{2022}]{apogeeDR17}
{Abdurro'uf} et~al., 2022, \mn@doi [\apjs] {10.3847/1538-4365/ac4414}, \href
  {https://ui.adsabs.harvard.edu/abs/2022ApJS..259...35A} {259, 35}

\bibitem[\protect\citeauthoryear{{Adibekyan}, {Sousa}, {Santos}, {Delgado
  Mena}, {Gonz{\'a}lez Hern{\'a}ndez}, {Israelian}, {Mayor}  \&
  {Khachatryan}}{{Adibekyan} et~al.}{2012}]{Adibekyan2012}
{Adibekyan} V.~Z.,  {Sousa} S.~G.,  {Santos} N.~C.,  {Delgado Mena} E.,
  {Gonz{\'a}lez Hern{\'a}ndez} J.~I.,  {Israelian} G.,  {Mayor} M.,
  {Khachatryan} G.,  2012, \mn@doi [\aap] {10.1051/0004-6361/201219401}, \href
  {https://ui.adsabs.harvard.edu/abs/2012A&A...545A..32A} {545, A32}

\bibitem[\protect\citeauthoryear{{Agertz} et~al.,}{{Agertz}
  et~al.}{2021}]{Agertz2021_vintergatanI}
{Agertz} O.,  et~al., 2021, \mn@doi [\mnras] {10.1093/mnras/stab322}, \href
  {https://ui.adsabs.harvard.edu/abs/2021MNRAS.503.5826A} {503, 5826}

\bibitem[\protect\citeauthoryear{{Anders} et~al.,}{{Anders}
  et~al.}{2014}]{Anders2014}
{Anders} F.,  et~al., 2014, \mn@doi [\aap] {10.1051/0004-6361/201323038}, \href
  {https://ui.adsabs.harvard.edu/abs/2014A&A...564A.115A} {564, A115}

\bibitem[\protect\citeauthoryear{{Anders} et~al.,}{{Anders}
  et~al.}{2017}]{Anders2017}
{Anders} F.,  et~al., 2017, \mn@doi [\aap] {10.1051/0004-6361/201629363}, \href
  {https://ui.adsabs.harvard.edu/abs/2017A&A...600A..70A} {600, A70}

\bibitem[\protect\citeauthoryear{{Anders}, {Chiappini}, {Santiago},
  {Matijevi{\v{c}}}, {Queiroz}, {Steinmetz}  \& {Guiglion}}{{Anders}
  et~al.}{2018}]{Anders2018}
{Anders} F.,  {Chiappini} C.,  {Santiago} B.~X.,  {Matijevi{\v{c}}} G.,
  {Queiroz} A.~B.,  {Steinmetz} M.,   {Guiglion} G.,  2018, \mn@doi [\aap]
  {10.1051/0004-6361/201833099}, \href
  {https://ui.adsabs.harvard.edu/abs/2018A&A...619A.125A} {619, A125}

\bibitem[\protect\citeauthoryear{{Anders} et~al.,}{{Anders}
  et~al.}{2023}]{Anders2023_ages}
{Anders} F.,  et~al., 2023, \mn@doi [arXiv e-prints]
  {10.48550/arXiv.2304.08276}, \href
  {https://ui.adsabs.harvard.edu/abs/2023arXiv230408276A} {p. arXiv:2304.08276}

\bibitem[\protect\citeauthoryear{{Annem} \& {Khoperskov}}{{Annem} \&
  {Khoperskov}}{2022}]{Annem2022}
{Annem} B.,  {Khoperskov} S.,  2022, \mn@doi [arXiv e-prints]
  {10.48550/arXiv.2210.17054}, \href
  {https://ui.adsabs.harvard.edu/abs/2022arXiv221017054A} {p. arXiv:2210.17054}

\bibitem[\protect\citeauthoryear{{Arellano-C{\'o}rdova}, {Esteban},
  {Garc{\'\i}a-Rojas}  \& {M{\'e}ndez-Delgado}}{{Arellano-C{\'o}rdova}
  et~al.}{2021}]{ArellanoCordova2021}
{Arellano-C{\'o}rdova} K.~Z.,  {Esteban} C.,  {Garc{\'\i}a-Rojas} J.,
  {M{\'e}ndez-Delgado} J.~E.,  2021, \mn@doi [\mnras] {10.1093/mnras/staa3903},
  \href {https://ui.adsabs.harvard.edu/abs/2021MNRAS.502..225A} {502, 225}

\bibitem[\protect\citeauthoryear{{Asplund}, {Amarsi}  \& {Grevesse}}{{Asplund}
  et~al.}{2021}]{Asplund2021}
{Asplund} M.,  {Amarsi} A.~M.,   {Grevesse} N.,  2021, \mn@doi [\aap]
  {10.1051/0004-6361/202140445}, \href
  {https://ui.adsabs.harvard.edu/abs/2021A&A...653A.141A} {653, A141}

\bibitem[\protect\citeauthoryear{{Battistini} \& {Bensby}}{{Battistini} \&
  {Bensby}}{2016}]{2016Battistini}
{Battistini} C.,  {Bensby} T.,  2016, \mn@doi [\aap]
  {10.1051/0004-6361/201527385}, \href
  {https://ui.adsabs.harvard.edu/abs/2016A&A...586A..49B} {586, A49}

\bibitem[\protect\citeauthoryear{{Bellardini}, {Wetzel}, {Loebman}  \&
  {Bailin}}{{Bellardini} et~al.}{2022}]{Bellardini2022}
{Bellardini} M.~A.,  {Wetzel} A.,  {Loebman} S.~R.,   {Bailin} J.,  2022,
  \mn@doi [\mnras] {10.1093/mnras/stac1637}, \href
  {https://ui.adsabs.harvard.edu/abs/2022MNRAS.514.4270B} {514, 4270}

\bibitem[\protect\citeauthoryear{{Belokurov} \& {Kravtsov}}{{Belokurov} \&
  {Kravtsov}}{2022}]{Belokurov2022}
{Belokurov} V.,  {Kravtsov} A.,  2022, \mn@doi [\mnras]
  {10.1093/mnras/stac1267}, \href
  {https://ui.adsabs.harvard.edu/abs/2022MNRAS.514..689B} {514, 689}

\bibitem[\protect\citeauthoryear{{Belokurov}, {Erkal}, {Evans}, {Koposov}  \&
  {Deason}}{{Belokurov} et~al.}{2018}]{Belokurov2018}
{Belokurov} V.,  {Erkal} D.,  {Evans} N.~W.,  {Koposov} S.~E.,   {Deason}
  A.~J.,  2018, \mn@doi [\mnras] {10.1093/mnras/sty982}, \href
  {https://ui.adsabs.harvard.edu/abs/2018MNRAS.478..611B} {478, 611}

\bibitem[\protect\citeauthoryear{{Bennett} \& {Bovy}}{{Bennett} \&
  {Bovy}}{2019}]{Bennett2019}
{Bennett} M.,  {Bovy} J.,  2019, \mn@doi [\mnras] {10.1093/mnras/sty2813},
  \href {https://ui.adsabs.harvard.edu/abs/2019MNRAS.482.1417B} {482, 1417}

\bibitem[\protect\citeauthoryear{{Bensby}, {Alves-Brito}, {Oey}, {Yong}  \&
  {Mel{\'e}ndez}}{{Bensby} et~al.}{2011}]{Bensby2011}
{Bensby} T.,  {Alves-Brito} A.,  {Oey} M.~S.,  {Yong} D.,   {Mel{\'e}ndez} J.,
  2011, \mn@doi [\apjl] {10.1088/2041-8205/735/2/L46}, \href
  {https://ui.adsabs.harvard.edu/abs/2011ApJ...735L..46B} {735, L46}

\bibitem[\protect\citeauthoryear{{Bensby}, {Alves-Brito}, {Oey}, {Yong}  \&
  {Mel{\'e}ndez}}{{Bensby} et~al.}{2012}]{2012bensby}
{Bensby} T.,  {Alves-Brito} A.,  {Oey} M.~S.,  {Yong} D.,   {Mel{\'e}ndez} J.,
  2012, in {Aoki} W.,  {Ishigaki} M.,  {Suda} T.,  {Tsujimoto} T.,   {Arimoto}
  N.,  eds,  Astronomical Society of the Pacific Conference Series Vol. 458,
  Galactic Archaeology: Near-Field Cosmology and the Formation of the Milky
  Way. p.~171 (\mn@eprint {arXiv} {1201.2009})

\bibitem[\protect\citeauthoryear{{Blancato}, {Ness}, {Johnston}, {Rybizki}  \&
  {Bedell}}{{Blancato} et~al.}{2019}]{blancato2019variations}
{Blancato} K.,  {Ness} M.,  {Johnston} K.~V.,  {Rybizki} J.,   {Bedell} M.,
  2019, \mn@doi [\apj] {10.3847/1538-4357/ab39e5}, \href
  {https://ui.adsabs.harvard.edu/abs/2019ApJ...883...34B} {883, 34}

\bibitem[\protect\citeauthoryear{{Bland-Hawthorn}, {Krumholz}  \&
  {Freeman}}{{Bland-Hawthorn} et~al.}{2010}]{BH2010}
{Bland-Hawthorn} J.,  {Krumholz} M.~R.,   {Freeman} K.,  2010, \mn@doi [\apj]
  {10.1088/0004-637X/713/1/166}, \href
  {http://adsabs.harvard.edu/abs/2010ApJ...713..166B} {713, 166}

\bibitem[\protect\citeauthoryear{Blanton et~al.,}{Blanton
  et~al.}{2017}]{blanton2017sloan}
Blanton M.~R.,  et~al., 2017, The Astronomical Journal, 154, 28

\bibitem[\protect\citeauthoryear{Bovy, Rix  \& Hogg}{Bovy
  et~al.}{2012a}]{bovy2012milky}
Bovy J.,  Rix H.-W.,   Hogg D.~W.,  2012a, The Astrophysical Journal, 751, 131

\bibitem[\protect\citeauthoryear{{Bovy} et~al.,}{{Bovy}
  et~al.}{2012b}]{Bovy2012_velocityCurve}
{Bovy} J.,  et~al., 2012b, \mn@doi [\apj] {10.1088/0004-637X/759/2/131}, \href
  {https://ui.adsabs.harvard.edu/abs/2012ApJ...759..131B} {759, 131}

\bibitem[\protect\citeauthoryear{{Bowen} \& {Vaughan}}{{Bowen} \&
  {Vaughan}}{1973}]{Bowen1973_apogeeTelescope}
{Bowen} I.~S.,  {Vaughan} A.~H. J.,  1973, \mn@doi [\ao]
  {10.1364/AO.12.001430}, \href
  {https://ui.adsabs.harvard.edu/abs/1973ApOpt..12.1430B} {12, 1430}

\bibitem[\protect\citeauthoryear{{Bragan{\c{c}}a} et~al.,}{{Bragan{\c{c}}a}
  et~al.}{2019}]{Braganca2019}
{Bragan{\c{c}}a} G.~A.,  et~al., 2019, \mn@doi [\aap]
  {10.1051/0004-6361/201834554}, \href
  {https://ui.adsabs.harvard.edu/abs/2019A&A...625A.120B} {625, A120}

\bibitem[\protect\citeauthoryear{{Buck}}{{Buck}}{2020}]{2020_buckchemical}
{Buck} T.,  2020, \mn@doi [\mnras] {10.1093/mnras/stz3289}, \href
  {https://ui.adsabs.harvard.edu/abs/2020MNRAS.491.5435B} {491, 5435}

\bibitem[\protect\citeauthoryear{{Buck}, {Macci{\`o}}, {Dutton}, {Obreja}  \&
  {Frings}}{{Buck} et~al.}{2019}]{Buck2019}
{Buck} T.,  {Macci{\`o}} A.~V.,  {Dutton} A.~A.,  {Obreja} A.,   {Frings} J.,
  2019, \mn@doi [\mnras] {10.1093/mnras/sty2913}, \href
  {https://ui.adsabs.harvard.edu/abs/2019MNRAS.483.1314B} {483, 1314}

\bibitem[\protect\citeauthoryear{{Buck}, {Obreja}, {Macci{\`o}}, {Minchev},
  {Dutton}  \& {Ostriker}}{{Buck} et~al.}{2020}]{2020buck_NIHAO-UHD}
{Buck} T.,  {Obreja} A.,  {Macci{\`o}} A.~V.,  {Minchev} I.,  {Dutton} A.~A.,
  {Ostriker} J.~P.,  2020, \mn@doi [\mnras] {10.1093/mnras/stz3241}, \href
  {https://ui.adsabs.harvard.edu/abs/2020MNRAS.491.3461B} {491, 3461}

\bibitem[\protect\citeauthoryear{{Buck}, {Rybizki}, {Buder}, {Obreja},
  {Macci{\`o}}, {Pfrommer}, {Steinmetz}  \& {Ness}}{{Buck}
  et~al.}{2021}]{2021BuckHD_chemEnrich}
{Buck} T.,  {Rybizki} J.,  {Buder} S.,  {Obreja} A.,  {Macci{\`o}} A.~V.,
  {Pfrommer} C.,  {Steinmetz} M.,   {Ness} M.,  2021, \mn@doi [\mnras]
  {10.1093/mnras/stab2736}, \href
  {https://ui.adsabs.harvard.edu/abs/2021MNRAS.508.3365B} {508, 3365}

\bibitem[\protect\citeauthoryear{{Buck}, {Obreja}, {Ratcliffe}, {Yuxi}, {Lu},
  {Minchev}  \& {Macci{\`o}}}{{Buck} et~al.}{2023}]{buck2023}
{Buck} T.,  {Obreja} A.,  {Ratcliffe} B.,  {Yuxi} {Lu} {Minchev} I.,
  {Macci{\`o}} A.~V.,  2023, \mn@doi [arXiv e-prints]
  {10.48550/arXiv.2305.13759}, \href
  {https://ui.adsabs.harvard.edu/abs/2023arXiv230513759B} {p. arXiv:2305.13759}

\bibitem[\protect\citeauthoryear{{Buder} et~al.,}{{Buder}
  et~al.}{2021}]{Buder2021}
{Buder} S.,  et~al., 2021, \mn@doi [\mnras] {10.1093/mnras/stab1242}, \href
  {https://ui.adsabs.harvard.edu/abs/2021MNRAS.506..150B} {506, 150}

\bibitem[\protect\citeauthoryear{{Carr}, {Johnston}, {Laporte}  \&
  {Ness}}{{Carr} et~al.}{2022}]{Carr2022}
{Carr} C.,  {Johnston} K.~V.,  {Laporte} C. F.~P.,   {Ness} M.~K.,  2022,
  \mn@doi [\mnras] {10.1093/mnras/stac2403}, \href
  {https://ui.adsabs.harvard.edu/abs/2022MNRAS.516.5067C} {516, 5067}

\bibitem[\protect\citeauthoryear{{Carrera} \& {Pancino}}{{Carrera} \&
  {Pancino}}{2011}]{Carrera2011}
{Carrera} R.,  {Pancino} E.,  2011, \mn@doi [\aap]
  {10.1051/0004-6361/201117473}, \href
  {https://ui.adsabs.harvard.edu/abs/2011A&A...535A..30C} {535, A30}

\bibitem[\protect\citeauthoryear{{Carrillo}, {Ness}, {Hawkins}, {Sanderson},
  {Wang}, {Wetzel}  \& {Bellardini}}{{Carrillo} et~al.}{2023}]{Carrillo2022}
{Carrillo} A.,  {Ness} M.~K.,  {Hawkins} K.,  {Sanderson} R.~E.,  {Wang} K.,
  {Wetzel} A.,   {Bellardini} M.~A.,  2023, \mn@doi [\apj]
  {10.3847/1538-4357/aca1c7}, \href
  {https://ui.adsabs.harvard.edu/abs/2023ApJ...942...35C} {942, 35}

\bibitem[\protect\citeauthoryear{{Casagrande}, {Sch{\"o}nrich}, {Asplund},
  {Cassisi}, {Ram{\'\i}rez}, {Mel{\'e}ndez}, {Bensby}  \&
  {Feltzing}}{{Casagrande} et~al.}{2011}]{Casagrande2011}
{Casagrande} L.,  {Sch{\"o}nrich} R.,  {Asplund} M.,  {Cassisi} S.,
  {Ram{\'\i}rez} I.,  {Mel{\'e}ndez} J.,  {Bensby} T.,   {Feltzing} S.,  2011,
  \mn@doi [\aap] {10.1051/0004-6361/201016276}, \href
  {https://ui.adsabs.harvard.edu/abs/2011A&A...530A.138C} {530, A138}

\bibitem[\protect\citeauthoryear{{Casamiquela}, {Castro-Ginard}, {Anders}  \&
  {Soubiran}}{{Casamiquela} et~al.}{2021}]{Casamiquela2021}
{Casamiquela} L.,  {Castro-Ginard} A.,  {Anders} F.,   {Soubiran} C.,  2021,
  \mn@doi [\aap] {10.1051/0004-6361/202141779}, \href
  {https://ui.adsabs.harvard.edu/abs/2021A&A...654A.151C} {654, A151}

\bibitem[\protect\citeauthoryear{{Chen}, {Hou}  \& {Wang}}{{Chen}
  et~al.}{2003}]{Chen2003}
{Chen} L.,  {Hou} J.~L.,   {Wang} J.~J.,  2003, \mn@doi [\aj] {10.1086/367911},
  \href {https://ui.adsabs.harvard.edu/abs/2003AJ....125.1397C} {125, 1397}

\bibitem[\protect\citeauthoryear{{Chiappini}, {Matteucci}  \&
  {Gratton}}{{Chiappini} et~al.}{1997}]{Chiappini1997}
{Chiappini} C.,  {Matteucci} F.,   {Gratton} R.,  1997, \mn@doi [\apj]
  {10.1086/303726}, \href
  {https://ui.adsabs.harvard.edu/abs/1997ApJ...477..765C} {477, 765}

\bibitem[\protect\citeauthoryear{{Chiappini}, {Matteucci}  \&
  {Romano}}{{Chiappini} et~al.}{2001}]{Chiappini2001}
{Chiappini} C.,  {Matteucci} F.,   {Romano} D.,  2001, \mn@doi [\apj]
  {10.1086/321427}, \href
  {https://ui.adsabs.harvard.edu/abs/2001ApJ...554.1044C} {554, 1044}

\bibitem[\protect\citeauthoryear{{Chiappini} et~al.,}{{Chiappini}
  et~al.}{2015}]{Chiappini2015}
{Chiappini} C.,  et~al., 2015, \mn@doi [\aap] {10.1051/0004-6361/201525865},
  \href {https://ui.adsabs.harvard.edu/abs/2015A&A...576L..12C} {576, L12}

\bibitem[\protect\citeauthoryear{{Ciuc{\u{a}}} et~al.,}{{Ciuc{\u{a}}}
  et~al.}{2022}]{Ciuca2022}
{Ciuc{\u{a}}} I.,  et~al., 2022, arXiv e-prints, \href
  {https://ui.adsabs.harvard.edu/abs/2022arXiv221101006C} {p. arXiv:2211.01006}

\bibitem[\protect\citeauthoryear{{Clarke} et~al.,}{{Clarke}
  et~al.}{2019}]{Clarke2019}
{Clarke} A.~J.,  et~al., 2019, \mn@doi [\mnras] {10.1093/mnras/stz104}, \href
  {http://adsabs.harvard.edu/abs/2019MNRAS.484.3476C} {484, 3476}

\bibitem[\protect\citeauthoryear{{Conroy} et~al.,}{{Conroy}
  et~al.}{2022}]{Conroy2022}
{Conroy} C.,  et~al., 2022, arXiv e-prints, \href
  {https://ui.adsabs.harvard.edu/abs/2022arXiv220402989C} {p. arXiv:2204.02989}

\bibitem[\protect\citeauthoryear{{Cunha} et~al.,}{{Cunha}
  et~al.}{2017}]{Cunha2017}
{Cunha} K.,  et~al., 2017, \mn@doi [\apj] {10.3847/1538-4357/aa7beb}, \href
  {https://ui.adsabs.harvard.edu/abs/2017ApJ...844..145C} {844, 145}

\bibitem[\protect\citeauthoryear{{Daflon} \& {Cunha}}{{Daflon} \&
  {Cunha}}{2004}]{Daflon2004}
{Daflon} S.,  {Cunha} K.,  2004, \mn@doi [\apj] {10.1086/425607}, \href
  {https://ui.adsabs.harvard.edu/abs/2004ApJ...617.1115D} {617, 1115}

\bibitem[\protect\citeauthoryear{{De Pascale}, {Worley}, {de Laverny},
  {Recio-Blanco}, {Hill}  \& {Bijaoui}}{{De Pascale}
  et~al.}{2014}]{DePascale2014_ambre}
{De Pascale} M.,  {Worley} C.~C.,  {de Laverny} P.,  {Recio-Blanco} A.,  {Hill}
  V.,   {Bijaoui} A.,  2014, \mn@doi [\aap] {10.1051/0004-6361/201423767},
  \href {https://ui.adsabs.harvard.edu/abs/2014A&A...570A..68D} {570, A68}

\bibitem[\protect\citeauthoryear{{Deharveng}, {Pe{\~n}a}, {Caplan}  \&
  {Costero}}{{Deharveng} et~al.}{2000}]{Deharveng2000}
{Deharveng} L.,  {Pe{\~n}a} M.,  {Caplan} J.,   {Costero} R.,  2000, \mn@doi
  [\mnras] {10.1046/j.1365-8711.2000.03030.x}, \href
  {https://ui.adsabs.harvard.edu/abs/2000MNRAS.311..329D} {311, 329}

\bibitem[\protect\citeauthoryear{{Delgado Mena}, {Tsantaki}, {Adibekyan},
  {Sousa}, {Santos}, {Gonz{\'a}lez Hern{\'a}ndez}  \& {Israelian}}{{Delgado
  Mena} et~al.}{2017}]{DelgadoMena2017}
{Delgado Mena} E.,  {Tsantaki} M.,  {Adibekyan} V.~Z.,  {Sousa} S.~G.,
  {Santos} N.~C.,  {Gonz{\'a}lez Hern{\'a}ndez} J.~I.,   {Israelian} G.,  2017,
  \mn@doi [\aap] {10.1051/0004-6361/201730535}, \href
  {https://ui.adsabs.harvard.edu/abs/2017A&A...606A..94D} {606, A94}

\bibitem[\protect\citeauthoryear{{Di Cintio}, {Mostoghiu}, {Knebe}  \&
  {Navarro}}{{Di Cintio} et~al.}{2021}]{DiCintio2021}
{Di Cintio} A.,  {Mostoghiu} R.,  {Knebe} A.,   {Navarro} J.~F.,  2021, \mn@doi
  [\mnras] {10.1093/mnras/stab1682}, \href
  {https://ui.adsabs.harvard.edu/abs/2021MNRAS.506..531D} {506, 531}

\bibitem[\protect\citeauthoryear{{Donor} et~al.,}{{Donor}
  et~al.}{2018}]{Donor2018}
{Donor} J.,  et~al., 2018, \mn@doi [\aj] {10.3847/1538-3881/aad635}, \href
  {https://ui.adsabs.harvard.edu/abs/2018AJ....156..142D} {156, 142}

\bibitem[\protect\citeauthoryear{{Dotter}, {Conroy}, {Cargile}  \&
  {Asplund}}{{Dotter} et~al.}{2017}]{2017ApJ...840...99D}
{Dotter} A.,  {Conroy} C.,  {Cargile} P.,   {Asplund} M.,  2017, \mn@doi [\apj]
  {10.3847/1538-4357/aa6d10}, \href
  {https://ui.adsabs.harvard.edu/abs/2017ApJ...840...99D} {840, 99}

\bibitem[\protect\citeauthoryear{{Eilers}, {Hogg}, {Rix}, {Ness},
  {Price-Whelan}, {M{\'e}sz{\'a}ros}  \& {Nitschelm}}{{Eilers}
  et~al.}{2022}]{Eilers2022}
{Eilers} A.-C.,  {Hogg} D.~W.,  {Rix} H.-W.,  {Ness} M.~K.,  {Price-Whelan}
  A.~M.,  {M{\'e}sz{\'a}ros} S.,   {Nitschelm} C.,  2022, \mn@doi [\apj]
  {10.3847/1538-4357/ac54ad}, \href
  {https://ui.adsabs.harvard.edu/abs/2022ApJ...928...23E} {928, 23}

\bibitem[\protect\citeauthoryear{{Esteban}, {Fang}, {Garc{\'\i}a-Rojas}  \&
  {Toribio San Cipriano}}{{Esteban} et~al.}{2017}]{Esteban2017}
{Esteban} C.,  {Fang} X.,  {Garc{\'\i}a-Rojas} J.,   {Toribio San Cipriano} L.,
   2017, \mn@doi [\mnras] {10.1093/mnras/stx1624}, \href
  {https://ui.adsabs.harvard.edu/abs/2017MNRAS.471..987E} {471, 987}

\bibitem[\protect\citeauthoryear{{Esteban}, {M{\'e}ndez-Delgado},
  {Garc{\'\i}a-Rojas}  \& {Arellano-C{\'o}rdova}}{{Esteban}
  et~al.}{2022}]{Esteban2022}
{Esteban} C.,  {M{\'e}ndez-Delgado} J.~E.,  {Garc{\'\i}a-Rojas} J.,
  {Arellano-C{\'o}rdova} K.~Z.,  2022, \mn@doi [\apj]
  {10.3847/1538-4357/ac6b38}, \href
  {https://ui.adsabs.harvard.edu/abs/2022ApJ...931...92E} {931, 92}

\bibitem[\protect\citeauthoryear{{Feltzing}, {Bowers}  \& {Agertz}}{{Feltzing}
  et~al.}{2020}]{2020Feltzing}
{Feltzing} S.,  {Bowers} J.~B.,   {Agertz} O.,  2020, \mn@doi [\mnras]
  {10.1093/mnras/staa340}, \href
  {https://ui.adsabs.harvard.edu/abs/2020MNRAS.493.1419F} {493, 1419}

\bibitem[\protect\citeauthoryear{{Frankel}, {Rix}, {Ting}, {Ness}  \&
  {Hogg}}{{Frankel} et~al.}{2018}]{Frankel2018}
{Frankel} N.,  {Rix} H.-W.,  {Ting} Y.-S.,  {Ness} M.,   {Hogg} D.~W.,  2018,
  \mn@doi [\apj] {10.3847/1538-4357/aadba5}, \href
  {https://ui.adsabs.harvard.edu/abs/2018ApJ...865...96F} {865, 96}

\bibitem[\protect\citeauthoryear{{Freeman} \& {Bland-Hawthorn}}{{Freeman} \&
  {Bland-Hawthorn}}{2002}]{2002freeman-BH}
{Freeman} K.,  {Bland-Hawthorn} J.,  2002, \mn@doi [\araa]
  {10.1146/annurev.astro.40.060401.093840}, \href
  {https://ui.adsabs.harvard.edu/abs/2002ARA&A..40..487F} {40, 487}

\bibitem[\protect\citeauthoryear{{Friel}}{{Friel}}{1995}]{Friel1995}
{Friel} E.~D.,  1995, \mn@doi [\araa] {10.1146/annurev.aa.33.090195.002121},
  \href {https://ui.adsabs.harvard.edu/abs/1995ARA&A..33..381F} {33, 381}

\bibitem[\protect\citeauthoryear{{Fuhrmann} \& {Chini}}{{Fuhrmann} \&
  {Chini}}{2017}]{Fuhrmann2017}
{Fuhrmann} K.,  {Chini} R.,  2017, \mn@doi [\apj]
  {10.3847/1538-4357/834/2/114}, \href
  {https://ui.adsabs.harvard.edu/abs/2017ApJ...834..114F} {834, 114}

\bibitem[\protect\citeauthoryear{{GRAVITY Collaboration} et~al.,}{{GRAVITY
  Collaboration} et~al.}{2018}]{Gravity2018}
{GRAVITY Collaboration} et~al., 2018, \mn@doi [\aap]
  {10.1051/0004-6361/201833718}, \href
  {https://ui.adsabs.harvard.edu/abs/2018A&A...615L..15G} {615, L15}

\bibitem[\protect\citeauthoryear{{Gaia Collaboration} et~al.,}{{Gaia
  Collaboration} et~al.}{2022}]{GaiaCollaboration2022RB}
{Gaia Collaboration} et~al., 2022, \mn@doi [arXiv e-prints]
  {10.48550/arXiv.2206.05534}, \href
  {https://ui.adsabs.harvard.edu/abs/2022arXiv220605534G} {p. arXiv:2206.05534}

\bibitem[\protect\citeauthoryear{{Garc{\'{\i}}a P{\'e}rez}
  et~al.,}{{Garc{\'{\i}}a P{\'e}rez} et~al.}{2016}]{GP2016}
{Garc{\'{\i}}a P{\'e}rez} A.~E.,  et~al., 2016, \mn@doi [\aj]
  {10.3847/0004-6256/151/6/144}, \href
  {http://adsabs.harvard.edu/abs/2016AJ....151..144G} {151, 144}

\bibitem[\protect\citeauthoryear{{Genovali} et~al.,}{{Genovali}
  et~al.}{2014}]{Genovali2014}
{Genovali} K.,  et~al., 2014, \mn@doi [\aap] {10.1051/0004-6361/201323198},
  \href {https://ui.adsabs.harvard.edu/abs/2014A&A...566A..37G} {566, A37}

\bibitem[\protect\citeauthoryear{{Grand}, {Kawata}  \& {Cropper}}{{Grand}
  et~al.}{2012}]{Grand2012}
{Grand} R. J.~J.,  {Kawata} D.,   {Cropper} M.,  2012, \mn@doi [\mnras]
  {10.1111/j.1365-2966.2012.20411.x}, \href
  {https://ui.adsabs.harvard.edu/abs/2012MNRAS.421.1529G} {421, 1529}

\bibitem[\protect\citeauthoryear{{Grenon}}{{Grenon}}{1987}]{Grenon1987}
{Grenon} M.,  1987, \mn@doi [Journal of Astrophysics and Astronomy]
  {10.1007/BF02714310}, \href
  {https://ui.adsabs.harvard.edu/abs/1987JApA....8..123G} {8, 123}

\bibitem[\protect\citeauthoryear{{Griffith} et~al.,}{{Griffith}
  et~al.}{2021}]{2021Griffith}
{Griffith} E.,  et~al., 2021, \mn@doi [\apj] {10.3847/1538-4357/abd6be}, \href
  {https://ui.adsabs.harvard.edu/abs/2021ApJ...909...77G} {909, 77}

\bibitem[\protect\citeauthoryear{{Grisoni}, {Spitoni}, {Matteucci},
  {Recio-Blanco}, {de Laverny}, {Hayden}, {Mikolaitis}  \& {Worley}}{{Grisoni}
  et~al.}{2017}]{Grisoni2017}
{Grisoni} V.,  {Spitoni} E.,  {Matteucci} F.,  {Recio-Blanco} A.,  {de Laverny}
  P.,  {Hayden} M.,  {Mikolaitis} {\^{S}}.,   {Worley} C.~C.,  2017, \mn@doi
  [\mnras] {10.1093/mnras/stx2201}, \href
  {https://ui.adsabs.harvard.edu/abs/2017MNRAS.472.3637G} {472, 3637}

\bibitem[\protect\citeauthoryear{{Gruyters}, {Nordlander}  \&
  {Korn}}{{Gruyters} et~al.}{2014}]{Gruyters2014}
{Gruyters} P.,  {Nordlander} T.,   {Korn} A.~J.,  2014, \mn@doi [\aap]
  {10.1051/0004-6361/201423590}, \href
  {https://ui.adsabs.harvard.edu/abs/2014A&A...567A..72G} {567, A72}

\bibitem[\protect\citeauthoryear{{Gunn} et~al.,}{{Gunn}
  et~al.}{2006}]{Gunn2006_apogeeTelescope}
{Gunn} J.~E.,  et~al., 2006, \mn@doi [\aj] {10.1086/500975}, \href
  {https://ui.adsabs.harvard.edu/abs/2006AJ....131.2332G} {131, 2332}

\bibitem[\protect\citeauthoryear{{Guszejnov}, {Hopkins}  \&
  {Graus}}{{Guszejnov} et~al.}{2019}]{Guszejnov2019}
{Guszejnov} D.,  {Hopkins} P.~F.,   {Graus} A.~S.,  2019, \mn@doi [\mnras]
  {10.1093/mnras/stz736}, \href
  {https://ui.adsabs.harvard.edu/abs/2019MNRAS.485.4852G} {485, 4852}

\bibitem[\protect\citeauthoryear{Hayden et~al.,}{Hayden
  et~al.}{2015}]{hayden2015chemical}
Hayden M.~R.,  et~al., 2015, The Astrophysical Journal, 808, 132

\bibitem[\protect\citeauthoryear{{Hayden}, {Recio-Blanco}, {de Laverny},
  {Mikolaitis}  \& {Worley}}{{Hayden} et~al.}{2017}]{Hayden2017_ambre}
{Hayden} M.~R.,  {Recio-Blanco} A.,  {de Laverny} P.,  {Mikolaitis} S.,
  {Worley} C.~C.,  2017, \mn@doi [\aap] {10.1051/0004-6361/201731494}, \href
  {https://ui.adsabs.harvard.edu/abs/2017A&A...608L...1H} {608, L1}

\bibitem[\protect\citeauthoryear{{Hayden} et~al.,}{{Hayden}
  et~al.}{2022}]{Hayden2022}
{Hayden} M.~R.,  et~al., 2022, \mn@doi [\mnras] {10.1093/mnras/stac2787}, \href
  {https://ui.adsabs.harvard.edu/abs/2022MNRAS.517.5325H} {517, 5325}

\bibitem[\protect\citeauthoryear{{Hayes} et~al.,}{{Hayes}
  et~al.}{2022}]{Hayes2022_bawlas}
{Hayes} C.~R.,  et~al., 2022, \mn@doi [\apjs] {10.3847/1538-4365/ac839f}, \href
  {https://ui.adsabs.harvard.edu/abs/2022ApJS..262...34H} {262, 34}

\bibitem[\protect\citeauthoryear{{Haywood}, {Di Matteo}, {Lehnert}, {Katz}  \&
  {G{\'o}mez}}{{Haywood} et~al.}{2013}]{Haywood2013}
{Haywood} M.,  {Di Matteo} P.,  {Lehnert} M.~D.,  {Katz} D.,   {G{\'o}mez} A.,
  2013, \mn@doi [\aap] {10.1051/0004-6361/201321397}, \href
  {http://adsabs.harvard.edu/abs/2013A%26A...560A.109H} {560, A109}

\bibitem[\protect\citeauthoryear{{Haywood}, {Di Matteo}, {Snaith}  \&
  {Lehnert}}{{Haywood} et~al.}{2015}]{2015Haywood}
{Haywood} M.,  {Di Matteo} P.,  {Snaith} O.,   {Lehnert} M.~D.,  2015, \mn@doi
  [\aap] {10.1051/0004-6361/201425459}, \href
  {https://ui.adsabs.harvard.edu/abs/2015A&A...579A...5H} {579, A5}

\bibitem[\protect\citeauthoryear{{Haywood}, {Snaith}, {Lehnert}, {Di Matteo}
  \& {Khoperskov}}{{Haywood} et~al.}{2019}]{Haywood2019}
{Haywood} M.,  {Snaith} O.,  {Lehnert} M.~D.,  {Di Matteo} P.,   {Khoperskov}
  S.,  2019, \mn@doi [\aap] {10.1051/0004-6361/201834155}, \href
  {https://ui.adsabs.harvard.edu/abs/2019A&A...625A.105H} {625, A105}

\bibitem[\protect\citeauthoryear{{He}, {Luo}  \& {Chen}}{{He}
  et~al.}{2022}]{He2022}
{He} X.-J.,  {Luo} A.~L.,   {Chen} Y.-Q.,  2022, \mn@doi [\mnras]
  {10.1093/mnras/stac484}, \href
  {https://ui.adsabs.harvard.edu/abs/2022MNRAS.512.1710H} {512, 1710}

\bibitem[\protect\citeauthoryear{{Helmi}, {Babusiaux}, {Koppelman}, {Massari},
  {Veljanoski}  \& {Brown}}{{Helmi} et~al.}{2018}]{Helmi2018_gse}
{Helmi} A.,  {Babusiaux} C.,  {Koppelman} H.~H.,  {Massari} D.,  {Veljanoski}
  J.,   {Brown} A. G.~A.,  2018, \mn@doi [\nat] {10.1038/s41586-018-0625-x},
  \href {https://ui.adsabs.harvard.edu/abs/2018Natur.563...85H} {563, 85}

\bibitem[\protect\citeauthoryear{{Hemler} et~al.,}{{Hemler}
  et~al.}{2021}]{Hemler2021}
{Hemler} Z.~S.,  et~al., 2021, \mn@doi [\mnras] {10.1093/mnras/stab1803}, \href
  {https://ui.adsabs.harvard.edu/abs/2021MNRAS.506.3024H} {506, 3024}

\bibitem[\protect\citeauthoryear{{Holtzman} et~al.,}{{Holtzman}
  et~al.}{2015}]{Holtzman2015}
{Holtzman} J.~A.,  et~al., 2015, \mn@doi [\aj] {10.1088/0004-6256/150/5/148},
  \href {http://adsabs.harvard.edu/abs/2015AJ....150..148H} {150, 148}

\bibitem[\protect\citeauthoryear{{Horta} et~al.,}{{Horta}
  et~al.}{2022a}]{horta2022_haloStreams}
{Horta} D.,  et~al., 2022a, \mn@doi [\mnras] {10.1093/mnras/stac3179}, \href
  {https://ui.adsabs.harvard.edu/abs/2022MNRAS.tmp.3011H} {}

\bibitem[\protect\citeauthoryear{{Horta}, {Ness}, {Rybizki}, {Schiavon}  \&
  {Buder}}{{Horta} et~al.}{2022b}]{Horta2021}
{Horta} D.,  {Ness} M.~K.,  {Rybizki} J.,  {Schiavon} R.~P.,   {Buder} S.,
  2022b, \mn@doi [\mnras] {10.1093/mnras/stac953}, \href
  {https://ui.adsabs.harvard.edu/abs/2022MNRAS.513.5477H} {513, 5477}

\bibitem[\protect\citeauthoryear{{Ibata}, {Gilmore}  \& {Irwin}}{{Ibata}
  et~al.}{1994}]{Ibata1994}
{Ibata} R.~A.,  {Gilmore} G.,   {Irwin} M.~J.,  1994, \mn@doi [\nat]
  {10.1038/370194a0}, \href
  {https://ui.adsabs.harvard.edu/abs/1994Natur.370..194I} {370, 194}

\bibitem[\protect\citeauthoryear{{Izzard}, {Preece}, {Jofre}, {Halabi},
  {Masseron}  \& {Tout}}{{Izzard} et~al.}{2018}]{Izzard2018}
{Izzard} R.~G.,  {Preece} H.,  {Jofre} P.,  {Halabi} G.~M.,  {Masseron} T.,
  {Tout} C.~A.,  2018, \mn@doi [\mnras] {10.1093/mnras/stx2355}, \href
  {https://ui.adsabs.harvard.edu/abs/2018MNRAS.473.2984I} {473, 2984}

\bibitem[\protect\citeauthoryear{{Janes}}{{Janes}}{1979}]{Janes1979}
{Janes} K.~A.,  1979, \mn@doi [\apjs] {10.1086/190568}, \href
  {https://ui.adsabs.harvard.edu/abs/1979ApJS...39..135J} {39, 135}

\bibitem[\protect\citeauthoryear{{Jofr{\'e}} et~al.,}{{Jofr{\'e}}
  et~al.}{2016}]{Jofre2016_BSS}
{Jofr{\'e}} P.,  et~al., 2016, \mn@doi [\aap] {10.1051/0004-6361/201629356},
  \href {https://ui.adsabs.harvard.edu/abs/2016A&A...595A..60J} {595, A60}

\bibitem[\protect\citeauthoryear{{Jofr{\'e}}, {Heiter}  \&
  {Soubiran}}{{Jofr{\'e}} et~al.}{2019}]{2019Jofre}
{Jofr{\'e}} P.,  {Heiter} U.,   {Soubiran} C.,  2019, \mn@doi [\araa]
  {10.1146/annurev-astro-091918-104509}, \href
  {https://ui.adsabs.harvard.edu/abs/2019ARA&A..57..571J} {57, 571}

\bibitem[\protect\citeauthoryear{{Jofr{\'e}} et~al.,}{{Jofr{\'e}}
  et~al.}{2023}]{Jofre2023}
{Jofr{\'e}} P.,  et~al., 2023, \mn@doi [\aap] {10.1051/0004-6361/202244524},
  \href {https://ui.adsabs.harvard.edu/abs/2023A&A...671A..21J} {671, A21}

\bibitem[\protect\citeauthoryear{{Johnson} et~al.,}{{Johnson}
  et~al.}{2021}]{2021Johnson}
{Johnson} J.~W.,  et~al., 2021, \mn@doi [\mnras] {10.1093/mnras/stab2718},
  \href {https://ui.adsabs.harvard.edu/abs/2021MNRAS.508.4484J} {508, 4484}

\bibitem[\protect\citeauthoryear{{J{\"o}nsson} et~al.,}{{J{\"o}nsson}
  et~al.}{2020}]{Jonsson2020}
{J{\"o}nsson} H.,  et~al., 2020, \mn@doi [\aj] {10.3847/1538-3881/aba592},
  \href {https://ui.adsabs.harvard.edu/abs/2020AJ....160..120J} {160, 120}

\bibitem[\protect\citeauthoryear{{Katz}, {G{\'o}mez}, {Haywood}, {Snaith}  \&
  {Di Matteo}}{{Katz} et~al.}{2021}]{Katz2021}
{Katz} D.,  {G{\'o}mez} A.,  {Haywood} M.,  {Snaith} O.,   {Di Matteo} P.,
  2021, \mn@doi [\aap] {10.1051/0004-6361/202140453}, \href
  {https://ui.adsabs.harvard.edu/abs/2021A&A...655A.111K} {655, A111}

\bibitem[\protect\citeauthoryear{{Khoperskov}, {Haywood}, {Snaith}, {Di
  Matteo}, {Lehnert}, {Vasiliev}, {Naroenkov}  \& {Berczik}}{{Khoperskov}
  et~al.}{2021}]{Khoperskov2021}
{Khoperskov} S.,  {Haywood} M.,  {Snaith} O.,  {Di Matteo} P.,  {Lehnert} M.,
  {Vasiliev} E.,  {Naroenkov} S.,   {Berczik} P.,  2021, \mn@doi [\mnras]
  {10.1093/mnras/staa3996}, \href
  {https://ui.adsabs.harvard.edu/abs/2021MNRAS.501.5176K} {501, 5176}

\bibitem[\protect\citeauthoryear{{Khoperskov} et~al.,}{{Khoperskov}
  et~al.}{2022a}]{Khoperskov2022a}
{Khoperskov} S.,  et~al., 2022a, \mn@doi [arXiv e-prints]
  {10.48550/arXiv.2206.04521}, \href
  {https://ui.adsabs.harvard.edu/abs/2022arXiv220604521K} {p. arXiv:2206.04521}

\bibitem[\protect\citeauthoryear{{Khoperskov} et~al.,}{{Khoperskov}
  et~al.}{2022b}]{Khoperskov2022c}
{Khoperskov} S.,  et~al., 2022b, arXiv e-prints, \href
  {https://ui.adsabs.harvard.edu/abs/2022arXiv220605491K} {p. arXiv:2206.05491}

\bibitem[\protect\citeauthoryear{{Kobayashi}, {Karakas}  \&
  {Lugaro}}{{Kobayashi} et~al.}{2020}]{Kobayashi2020}
{Kobayashi} C.,  {Karakas} A.~I.,   {Lugaro} M.,  2020, \mn@doi [\apj]
  {10.3847/1538-4357/abae65}, \href
  {https://ui.adsabs.harvard.edu/abs/2020ApJ...900..179K} {900, 179}

\bibitem[\protect\citeauthoryear{{Kordopatis} et~al.,}{{Kordopatis}
  et~al.}{2015}]{Kordopatis2015}
{Kordopatis} G.,  et~al., 2015, \mn@doi [\mnras] {10.1093/mnras/stu2726}, \href
  {https://ui.adsabs.harvard.edu/abs/2015MNRAS.447.3526K} {447, 3526}

\bibitem[\protect\citeauthoryear{{Kubryk}, {Prantzos}  \&
  {Athanassoula}}{{Kubryk} et~al.}{2013}]{Kubryk2013}
{Kubryk} M.,  {Prantzos} N.,   {Athanassoula} E.,  2013, \mn@doi [\mnras]
  {10.1093/mnras/stt1667}, \href
  {https://ui.adsabs.harvard.edu/abs/2013MNRAS.436.1479K} {436, 1479}

\bibitem[\protect\citeauthoryear{{Kubryk}, {Prantzos}  \&
  {Athanassoula}}{{Kubryk} et~al.}{2015}]{Kubryk2015}
{Kubryk} M.,  {Prantzos} N.,   {Athanassoula} E.,  2015, \mn@doi [\aap]
  {10.1051/0004-6361/201424599}, \href
  {https://ui.adsabs.harvard.edu/abs/2015A&A...580A.127K} {580, A127}

\bibitem[\protect\citeauthoryear{{Laporte}, {Johnston}, {G{\'o}mez},
  {Garavito-Camargo}  \& {Besla}}{{Laporte} et~al.}{2018}]{Laporte2018}
{Laporte} C. F.~P.,  {Johnston} K.~V.,  {G{\'o}mez} F.~A.,  {Garavito-Camargo}
  N.,   {Besla} G.,  2018, \mn@doi [\mnras] {10.1093/mnras/sty1574}, \href
  {https://ui.adsabs.harvard.edu/abs/2018MNRAS.481..286L} {481, 286}

\bibitem[\protect\citeauthoryear{{Law} \& {Majewski}}{{Law} \&
  {Majewski}}{2010}]{Law2010}
{Law} D.~R.,  {Majewski} S.~R.,  2010, \mn@doi [\apj]
  {10.1088/0004-637X/714/1/229}, \href
  {https://ui.adsabs.harvard.edu/abs/2010ApJ...714..229L} {714, 229}

\bibitem[\protect\citeauthoryear{{Leung} \& {Bovy}}{{Leung} \&
  {Bovy}}{2019}]{LeungBovy2019a}
{Leung} H.~W.,  {Bovy} J.,  2019, \mn@doi [\mnras] {10.1093/mnras/sty3217},
  \href {https://ui.adsabs.harvard.edu/abs/2019MNRAS.483.3255L} {483, 3255}

\bibitem[\protect\citeauthoryear{{Leung}, {Bovy}, {Mackereth}  \&
  {Miglio}}{{Leung} et~al.}{2023}]{Leung2023}
{Leung} H.~W.,  {Bovy} J.,  {Mackereth} J.~T.,   {Miglio} A.,  2023, \mn@doi
  [arXiv e-prints] {10.48550/arXiv.2302.05479}, \href
  {https://ui.adsabs.harvard.edu/abs/2023arXiv230205479L} {p. arXiv:2302.05479}

\bibitem[\protect\citeauthoryear{{Lian} et~al.,}{{Lian}
  et~al.}{2020}]{2020Lian_alphaDichotomy}
{Lian} J.,  et~al., 2020, \mn@doi [\mnras] {10.1093/mnras/staa2078}, \href
  {https://ui.adsabs.harvard.edu/abs/2020MNRAS.497.2371L} {497, 2371}

\bibitem[\protect\citeauthoryear{{Lian} et~al.,}{{Lian}
  et~al.}{2022}]{Lian2022}
{Lian} J.,  et~al., 2022, \mn@doi [\mnras] {10.1093/mnras/stac479}, \href
  {https://ui.adsabs.harvard.edu/abs/2022MNRAS.511.5639L} {511, 5639}

\bibitem[\protect\citeauthoryear{{Libeskind} et~al.,}{{Libeskind}
  et~al.}{2020}]{Libeskind2020}
{Libeskind} N.~I.,  et~al., 2020, \mn@doi [\mnras] {10.1093/mnras/staa2541},
  \href {https://ui.adsabs.harvard.edu/abs/2020MNRAS.498.2968L} {498, 2968}

\bibitem[\protect\citeauthoryear{{Lin} et~al.,}{{Lin} et~al.}{2020}]{Lin2020}
{Lin} J.,  et~al., 2020, \mn@doi [\mnras] {10.1093/mnras/stz3048}, \href
  {https://ui.adsabs.harvard.edu/abs/2020MNRAS.491.2043L} {491, 2043}

\bibitem[\protect\citeauthoryear{{Liu}, {Asplund}, {Yong}, {Feltzing},
  {Dotter}, {Mel{\'e}ndez}  \& {Ram{\'\i}rez}}{{Liu} et~al.}{2019}]{Liu2019}
{Liu} F.,  {Asplund} M.,  {Yong} D.,  {Feltzing} S.,  {Dotter} A.,
  {Mel{\'e}ndez} J.,   {Ram{\'\i}rez} I.,  2019, \mn@doi [\aap]
  {10.1051/0004-6361/201935306}, \href
  {https://ui.adsabs.harvard.edu/abs/2019A&A...627A.117L} {627, A117}

\bibitem[\protect\citeauthoryear{{Lu}, {Minchev}, {Buck}, {Khoperskov},
  {Steinmetz}, {Libeskind}, {Cescutti}  \& {Freeman}}{{Lu}
  et~al.}{2022a}]{Lu2022_Rb}
{Lu} Y.,  {Minchev} I.,  {Buck} T.,  {Khoperskov} S.,  {Steinmetz} M.,
  {Libeskind} N.,  {Cescutti} G.,   {Freeman} K.~C.,  2022a, arXiv e-prints,
  \href {https://ui.adsabs.harvard.edu/abs/2022arXiv221204515Y} {p.
  arXiv:2212.04515}

\bibitem[\protect\citeauthoryear{{Lu}, {Buck}, {Minchev}  \& {Ness}}{{Lu}
  et~al.}{2022b}]{Lu2022_sims}
{Lu} Y.,  {Buck} T.,  {Minchev} I.,   {Ness} M.~K.,  2022b, \mn@doi [\mnras]
  {10.1093/mnrasl/slac065}, \href
  {https://ui.adsabs.harvard.edu/abs/2022MNRAS.515L..34L} {515, L34}

\bibitem[\protect\citeauthoryear{{Mackereth} \& {Bovy}}{{Mackereth} \&
  {Bovy}}{2018}]{Mackereth2018}
{Mackereth} J.~T.,  {Bovy} J.,  2018, \mn@doi [\pasp]
  {10.1088/1538-3873/aadcdd}, \href
  {https://ui.adsabs.harvard.edu/abs/2018PASP..130k4501M} {130, 114501}

\bibitem[\protect\citeauthoryear{Mackereth, Crain, Schiavon, Schaye, Theuns  \&
  Schaller}{Mackereth et~al.}{2018}]{mackereth2018origin}
Mackereth J.~T.,  Crain R.~A.,  Schiavon R.~P.,  Schaye J.,  Theuns T.,
  Schaller M.,  2018, Monthly Notices of the Royal Astronomical Society, 477,
  5072

\bibitem[\protect\citeauthoryear{{Mackereth} et~al.,}{{Mackereth}
  et~al.}{2019}]{Mackereth2019}
{Mackereth} J.~T.,  et~al., 2019, \mn@doi [\mnras] {10.1093/mnras/stz1521},
  \href {https://ui.adsabs.harvard.edu/abs/2019MNRAS.489..176M} {489, 176}

\bibitem[\protect\citeauthoryear{{Magrini}, {Sestito}, {Randich}  \&
  {Galli}}{{Magrini} et~al.}{2009}]{Magrini2009}
{Magrini} L.,  {Sestito} P.,  {Randich} S.,   {Galli} D.,  2009, \mn@doi [\aap]
  {10.1051/0004-6361:200810634}, \href
  {https://ui.adsabs.harvard.edu/abs/2009A&A...494...95M} {494, 95}

\bibitem[\protect\citeauthoryear{{Magrini} et~al.,}{{Magrini}
  et~al.}{2023}]{Magrini2023}
{Magrini} L.,  et~al., 2023, \mn@doi [\aap] {10.1051/0004-6361/202244957},
  \href {https://ui.adsabs.harvard.edu/abs/2023A&A...669A.119M} {669, A119}

\bibitem[\protect\citeauthoryear{{Majewski} et~al.,}{{Majewski}
  et~al.}{2017}]{Majewski2017}
{Majewski} S.~R.,  et~al., 2017, \mn@doi [\aj] {10.3847/1538-3881/aa784d},
  \href {http://adsabs.harvard.edu/abs/2017AJ....154...94M} {154, 94}

\bibitem[\protect\citeauthoryear{{Martig} et~al.,}{{Martig}
  et~al.}{2015}]{Martig2015}
{Martig} M.,  et~al., 2015, \mn@doi [\mnras] {10.1093/mnras/stv1071}, \href
  {https://ui.adsabs.harvard.edu/abs/2015MNRAS.451.2230M} {451, 2230}

\bibitem[\protect\citeauthoryear{{Matteucci}}{{Matteucci}}{2012}]{Matteucci2012}
{Matteucci} F.,  2012, {Chemical Evolution of Galaxies}.
Springer Science \& Business Media, \mn@doi{10.1007/978-3-642-22491-1}

\bibitem[\protect\citeauthoryear{{Matteucci}}{{Matteucci}}{2021}]{Matteucci2021}
{Matteucci} F.,  2021, \mn@doi [\aapr] {10.1007/s00159-021-00133-8}, \href
  {https://ui.adsabs.harvard.edu/abs/2021A&ARv..29....5M} {29, 5}

\bibitem[\protect\citeauthoryear{{Matteucci} \& {Francois}}{{Matteucci} \&
  {Francois}}{1989}]{Matteucci1989}
{Matteucci} F.,  {Francois} P.,  1989, \mn@doi [\mnras]
  {10.1093/mnras/239.3.885}, \href
  {https://ui.adsabs.harvard.edu/abs/1989MNRAS.239..885M} {239, 885}

\bibitem[\protect\citeauthoryear{{Mayor}}{{Mayor}}{1976}]{Mayor1976}
{Mayor} M.,  1976, \aap, \href
  {https://ui.adsabs.harvard.edu/abs/1976A&A....48..301M} {48, 301}

\bibitem[\protect\citeauthoryear{{Miglio} et~al.,}{{Miglio}
  et~al.}{2021}]{Miglio2021}
{Miglio} A.,  et~al., 2021, \mn@doi [\aap] {10.1051/0004-6361/202038307}, \href
  {https://ui.adsabs.harvard.edu/abs/2021A&A...645A..85M} {645, A85}

\bibitem[\protect\citeauthoryear{{Minchev} \& {Famaey}}{{Minchev} \&
  {Famaey}}{2010}]{Minchev2010}
{Minchev} I.,  {Famaey} B.,  2010, \mn@doi [\apj]
  {10.1088/0004-637X/722/1/112}, \href
  {https://ui.adsabs.harvard.edu/abs/2010ApJ...722..112M} {722, 112}

\bibitem[\protect\citeauthoryear{{Minchev}, {Famaey}, {Quillen}  \&
  {Dehnen}}{{Minchev} et~al.}{2012}]{Minchev2012a}
{Minchev} I.,  {Famaey} B.,  {Quillen} A.~C.,   {Dehnen} W.,  2012, in European
  Physical Journal Web of Conferences. p. 07002 (\mn@eprint {arXiv}
  {1111.0195}), \mn@doi{10.1051/epjconf/20121907002}

\bibitem[\protect\citeauthoryear{{Minchev}, {Chiappini}  \& {Martig}}{{Minchev}
  et~al.}{2013}]{Minchev2013}
{Minchev} I.,  {Chiappini} C.,   {Martig} M.,  2013, \mn@doi [\aap]
  {10.1051/0004-6361/201220189}, \href
  {https://ui.adsabs.harvard.edu/abs/2013A&A...558A...9M} {558, A9}

\bibitem[\protect\citeauthoryear{{Minchev}, {Steinmetz}, {Chiappini}, {Martig},
  {Anders}, {Matijevic}  \& {de Jong}}{{Minchev} et~al.}{2017}]{Minchev2017}
{Minchev} I.,  {Steinmetz} M.,  {Chiappini} C.,  {Martig} M.,  {Anders} F.,
  {Matijevic} G.,   {de Jong} R.~S.,  2017, \mn@doi [\apj]
  {10.3847/1538-4357/834/1/27}, \href
  {https://ui.adsabs.harvard.edu/abs/2017ApJ...834...27M} {834, 27}

\bibitem[\protect\citeauthoryear{{Minchev} et~al.,}{{Minchev}
  et~al.}{2018}]{2018Minchev_rbirth}
{Minchev} I.,  et~al., 2018, \mn@doi [\mnras] {10.1093/mnras/sty2033}, \href
  {https://ui.adsabs.harvard.edu/abs/2018MNRAS.481.1645M} {481, 1645}

\bibitem[\protect\citeauthoryear{{Minchev} et~al.,}{{Minchev}
  et~al.}{2019}]{2019minchev}
{Minchev} I.,  et~al., 2019, \mn@doi [\mnras] {10.1093/mnras/stz1239}, \href
  {https://ui.adsabs.harvard.edu/abs/2019MNRAS.487.3946M} {487, 3946}

\bibitem[\protect\citeauthoryear{{Mor}, {Robin}, {Figueras}, {Roca-F{\`a}brega}
   \& {Luri}}{{Mor} et~al.}{2019}]{Mor2019}
{Mor} R.,  {Robin} A.~C.,  {Figueras} F.,  {Roca-F{\`a}brega} S.,   {Luri} X.,
  2019, \mn@doi [\aap] {10.1051/0004-6361/201935105}, \href
  {https://ui.adsabs.harvard.edu/abs/2019A&A...624L...1M} {624, L1}

\bibitem[\protect\citeauthoryear{{Myers} et~al.,}{{Myers}
  et~al.}{2022}]{Myers2022}
{Myers} N.,  et~al., 2022, \mn@doi [\aj] {10.3847/1538-3881/ac7ce5}, \href
  {https://ui.adsabs.harvard.edu/abs/2022AJ....164...85M} {164, 85}

\bibitem[\protect\citeauthoryear{{Ness}, {Johnston}, {Blancato}, {Rix},
  {Beane}, {Bird}  \& {Hawkins}}{{Ness} et~al.}{2019}]{Ness2019}
{Ness} M.~K.,  {Johnston} K.~V.,  {Blancato} K.,  {Rix} H.~W.,  {Beane} A.,
  {Bird} J.~C.,   {Hawkins} K.,  2019, \mn@doi [\apj]
  {10.3847/1538-4357/ab3e3c}, \href
  {https://ui.adsabs.harvard.edu/abs/2019ApJ...883..177N} {883, 177}

\bibitem[\protect\citeauthoryear{{Ness}, {Wheeler}, {McKinnon}, {Horta},
  {Casey}, {Cunningham}  \& {Price-Whelan}}{{Ness} et~al.}{2022}]{2022Ness}
{Ness} M.~K.,  {Wheeler} A.~J.,  {McKinnon} K.,  {Horta} D.,  {Casey} A.~R.,
  {Cunningham} E.~C.,   {Price-Whelan} A.~M.,  2022, \mn@doi [\apj]
  {10.3847/1538-4357/ac4754}, \href
  {https://ui.adsabs.harvard.edu/abs/2022ApJ...926..144N} {926, 144}

\bibitem[\protect\citeauthoryear{{Netopil}, {Paunzen}, {Heiter}  \&
  {Soubiran}}{{Netopil} et~al.}{2016}]{Netopil2016}
{Netopil} M.,  {Paunzen} E.,  {Heiter} U.,   {Soubiran} C.,  2016, \mn@doi
  [\aap] {10.1051/0004-6361/201526370}, \href
  {https://ui.adsabs.harvard.edu/abs/2016A&A...585A.150N} {585, A150}

\bibitem[\protect\citeauthoryear{{Netopil}, {Oralhan}, {{\c{C}}akmak}, {Michel}
   \& {Karata{\c{s}}}}{{Netopil} et~al.}{2022}]{Netopil2022}
{Netopil} M.,  {Oralhan} {\.I}.~A.,  {{\c{C}}akmak} H.,  {Michel} R.,
  {Karata{\c{s}}} Y.,  2022, \mn@doi [\mnras] {10.1093/mnras/stab2961}, \href
  {https://ui.adsabs.harvard.edu/abs/2022MNRAS.509..421N} {509, 421}

\bibitem[\protect\citeauthoryear{Nidever et~al.,}{Nidever
  et~al.}{2014}]{nidever2014tracing}
Nidever D.~L.,  et~al., 2014, The Astrophysical Journal, 796, 38

\bibitem[\protect\citeauthoryear{{Nidever} et~al.,}{{Nidever}
  et~al.}{2015}]{Nidever2015_apogeePipeline}
{Nidever} D.~L.,  et~al., 2015, \mn@doi [\aj] {10.1088/0004-6256/150/6/173},
  \href {https://ui.adsabs.harvard.edu/abs/2015AJ....150..173N} {150, 173}

\bibitem[\protect\citeauthoryear{{Nieva} \& {Przybilla}}{{Nieva} \&
  {Przybilla}}{2012}]{Nieva2012}
{Nieva} M.~F.,  {Przybilla} N.,  2012, \mn@doi [\aap]
  {10.1051/0004-6361/201118158}, \href
  {https://ui.adsabs.harvard.edu/abs/2012A&A...539A.143N} {539, A143}

\bibitem[\protect\citeauthoryear{{Pagel}}{{Pagel}}{2009}]{Pagel2009}
{Pagel} B. E.~J.,  2009, {Nucleosynthesis and Chemical Evolution of Galaxies}.
Cambridge University Press

\bibitem[\protect\citeauthoryear{{Palla}, {Santos-Peral}, {Recio-Blanco}  \&
  {Matteucci}}{{Palla} et~al.}{2022}]{Palla2022}
{Palla} M.,  {Santos-Peral} P.,  {Recio-Blanco} A.,   {Matteucci} F.,  2022,
  \mn@doi [\aap] {10.1051/0004-6361/202142645}, \href
  {https://ui.adsabs.harvard.edu/abs/2022A&A...663A.125P} {663, A125}

\bibitem[\protect\citeauthoryear{{Pilkington} et~al.,}{{Pilkington}
  et~al.}{2012}]{Pilkington2012}
{Pilkington} K.,  et~al., 2012, \mn@doi [\aap] {10.1051/0004-6361/201117466},
  \href {https://ui.adsabs.harvard.edu/abs/2012A&A...540A..56P} {540, A56}

\bibitem[\protect\citeauthoryear{{Prantzos} \& {Boissier}}{{Prantzos} \&
  {Boissier}}{2000}]{PrantzosBoissier2000}
{Prantzos} N.,  {Boissier} S.,  2000, \mn@doi [\mnras]
  {10.1046/j.1365-8711.2000.03228.x}, \href
  {https://ui.adsabs.harvard.edu/abs/2000MNRAS.313..338P} {313, 338}

\bibitem[\protect\citeauthoryear{{Price-Jones} \& {Bovy}}{{Price-Jones} \&
  {Bovy}}{2018}]{PJ2018}
{Price-Jones} N.,  {Bovy} J.,  2018, \mn@doi [\mnras] {10.1093/mnras/stx3198},
  \href {http://adsabs.harvard.edu/abs/2018MNRAS.475.1410P} {475, 1410}

\bibitem[\protect\citeauthoryear{{Price-Jones} et~al.,}{{Price-Jones}
  et~al.}{2020}]{PriceJones2020}
{Price-Jones} N.,  et~al., 2020, \mn@doi [\mnras] {10.1093/mnras/staa1905},
  \href {https://ui.adsabs.harvard.edu/abs/2020MNRAS.496.5101P} {496, 5101}

\bibitem[\protect\citeauthoryear{{Queiroz} et~al.,}{{Queiroz}
  et~al.}{2018}]{Queiroz2018_starhorse}
{Queiroz} A.~B.~A.,  et~al., 2018, \mn@doi [\mnras] {10.1093/mnras/sty330},
  \href {https://ui.adsabs.harvard.edu/abs/2018MNRAS.476.2556Q} {476, 2556}

\bibitem[\protect\citeauthoryear{{Queiroz} et~al.,}{{Queiroz}
  et~al.}{2020}]{Queiroz2020}
{Queiroz} A.~B.~A.,  et~al., 2020, \mn@doi [\aap]
  {10.1051/0004-6361/201937364}, \href
  {https://ui.adsabs.harvard.edu/abs/2020A&A...638A..76Q} {638, A76}

\bibitem[\protect\citeauthoryear{{Queiroz} et~al.,}{{Queiroz}
  et~al.}{2023}]{Queiroz2023_SH}
{Queiroz} A. B.~A.,  et~al., 2023, \mn@doi [arXiv e-prints]
  {10.48550/arXiv.2303.09926}, \href
  {https://ui.adsabs.harvard.edu/abs/2023arXiv230309926Q} {p. arXiv:2303.09926}

\bibitem[\protect\citeauthoryear{{Quillen}, {Minchev}, {Bland-Hawthorn}  \&
  {Haywood}}{{Quillen} et~al.}{2009}]{Quillen2009}
{Quillen} A.~C.,  {Minchev} I.,  {Bland-Hawthorn} J.,   {Haywood} M.,  2009,
  \mn@doi [\mnras] {10.1111/j.1365-2966.2009.15054.x}, \href
  {https://ui.adsabs.harvard.edu/abs/2009MNRAS.397.1599Q} {397, 1599}

\bibitem[\protect\citeauthoryear{{Ratcliffe} \& {Ness}}{{Ratcliffe} \&
  {Ness}}{2023}]{Ratcliffe2022_conditional}
{Ratcliffe} B.~L.,  {Ness} M.~K.,  2023, \mn@doi [\apj]
  {10.3847/1538-4357/aca8a1}, \href
  {https://ui.adsabs.harvard.edu/abs/2023ApJ...943...92R} {943, 92}

\bibitem[\protect\citeauthoryear{{Ratcliffe}, {Ness}, {Johnston}  \&
  {Sen}}{{Ratcliffe} et~al.}{2020}]{2020Ratcliffe}
{Ratcliffe} B.~L.,  {Ness} M.~K.,  {Johnston} K.~V.,   {Sen} B.,  2020, \mn@doi
  [\apj] {10.3847/1538-4357/abac61}, \href
  {https://ui.adsabs.harvard.edu/abs/2020ApJ...900..165R} {900, 165}

\bibitem[\protect\citeauthoryear{{Ratcliffe}, {Ness}, {Buck}, {Johnston},
  {Sen}, {Beraldo e Silva}  \& {Debattista}}{{Ratcliffe}
  et~al.}{2022}]{2022Ratcliffe}
{Ratcliffe} B.~L.,  {Ness} M.~K.,  {Buck} T.,  {Johnston} K.~V.,  {Sen} B.,
  {Beraldo e Silva} L.,   {Debattista} V.~P.,  2022, \mn@doi [\apj]
  {10.3847/1538-4357/ac3481}, \href
  {https://ui.adsabs.harvard.edu/abs/2022ApJ...924...60R} {924, 60}

\bibitem[\protect\citeauthoryear{{Renaud}, {Agertz}, {Read}, {Ryde},
  {Andersson}, {Bensby}, {Rey}  \& {Feuillet}}{{Renaud}
  et~al.}{2021a}]{Renaud2021_vintergatanII}
{Renaud} F.,  {Agertz} O.,  {Read} J.~I.,  {Ryde} N.,  {Andersson} E.~P.,
  {Bensby} T.,  {Rey} M.~P.,   {Feuillet} D.~K.,  2021a, \mn@doi [\mnras]
  {10.1093/mnras/stab250}, \href
  {https://ui.adsabs.harvard.edu/abs/2021MNRAS.503.5846R} {503, 5846}

\bibitem[\protect\citeauthoryear{{Renaud}, {Agertz}, {Andersson}, {Read},
  {Ryde}, {Bensby}, {Rey}  \& {Feuillet}}{{Renaud}
  et~al.}{2021b}]{Renaud2021_vintergatanIII}
{Renaud} F.,  {Agertz} O.,  {Andersson} E.~P.,  {Read} J.~I.,  {Ryde} N.,
  {Bensby} T.,  {Rey} M.~P.,   {Feuillet} D.~K.,  2021b, \mn@doi [\mnras]
  {10.1093/mnras/stab543}, \href
  {https://ui.adsabs.harvard.edu/abs/2021MNRAS.503.5868R} {503, 5868}

\bibitem[\protect\citeauthoryear{{Ro{\v s}kar}, {Debattista}, {Quinn},
  {Stinson}  \& {Wadsley}}{{Ro{\v s}kar} et~al.}{2008}]{Roskar2008}
{Ro{\v s}kar} R.,  {Debattista} V.~P.,  {Quinn} T.~R.,  {Stinson} G.~S.,
  {Wadsley} J.,  2008, \mn@doi [\apjl] {10.1086/592231}, \href
  {http://adsabs.harvard.edu/abs/2008ApJ...684L..79R} {684, L79}

\bibitem[\protect\citeauthoryear{{Ruiz-Lara}, {Gallart}, {Bernard}  \&
  {Cassisi}}{{Ruiz-Lara} et~al.}{2020}]{RuizLara2020}
{Ruiz-Lara} T.,  {Gallart} C.,  {Bernard} E.~J.,   {Cassisi} S.,  2020, \mn@doi
  [Nature Astronomy] {10.1038/s41550-020-1097-0}, \href
  {https://ui.adsabs.harvard.edu/abs/2020NatAs...4..965R} {4, 965}

\bibitem[\protect\citeauthoryear{{Rybizki}, {Just}  \& {Rix}}{{Rybizki}
  et~al.}{2017}]{Rybizki2017_chempy}
{Rybizki} J.,  {Just} A.,   {Rix} H.-W.,  2017, \mn@doi [\aap]
  {10.1051/0004-6361/201730522}, \href
  {https://ui.adsabs.harvard.edu/abs/2017A&A...605A..59R} {605, A59}

\bibitem[\protect\citeauthoryear{{Sahlholdt}, {Feltzing}  \&
  {Feuillet}}{{Sahlholdt} et~al.}{2022}]{Sahlholdt2022}
{Sahlholdt} C.~L.,  {Feltzing} S.,   {Feuillet} D.~K.,  2022, \mn@doi [\mnras]
  {10.1093/mnras/stab3681}, \href
  {https://ui.adsabs.harvard.edu/abs/2022MNRAS.510.4669S} {510, 4669}

\bibitem[\protect\citeauthoryear{{Salaris}, {Pietrinferni}, {Piersimoni}  \&
  {Cassisi}}{{Salaris} et~al.}{2015}]{Salaris2015}
{Salaris} M.,  {Pietrinferni} A.,  {Piersimoni} A.~M.,   {Cassisi} S.,  2015,
  \mn@doi [\aap] {10.1051/0004-6361/201526951}, \href
  {https://ui.adsabs.harvard.edu/abs/2015A&A...583A..87S} {583, A87}

\bibitem[\protect\citeauthoryear{{Sales-Silva} et~al.,}{{Sales-Silva}
  et~al.}{2022}]{SalesSilva2022}
{Sales-Silva} J.~V.,  et~al., 2022, \mn@doi [\apj] {10.3847/1538-4357/ac4254},
  \href {https://ui.adsabs.harvard.edu/abs/2022ApJ...926..154S} {926, 154}

\bibitem[\protect\citeauthoryear{{Sanders} \& {Binney}}{{Sanders} \&
  {Binney}}{2015}]{sanders15}
{Sanders} J.~L.,  {Binney} J.,  2015, \mn@doi [\mnras] {10.1093/mnras/stv578},
  \href {https://ui.adsabs.harvard.edu/abs/2015MNRAS.449.3479S} {449, 3479}

\bibitem[\protect\citeauthoryear{{Sanderson} et~al.,}{{Sanderson}
  et~al.}{2020}]{2020FIRE_sanderson}
{Sanderson} R.~E.,  et~al., 2020, \mn@doi [\apjs] {10.3847/1538-4365/ab5b9d},
  \href {https://ui.adsabs.harvard.edu/abs/2020ApJS..246....6S} {246, 6}

\bibitem[\protect\citeauthoryear{{Sch{\"o}nrich} \& {Binney}}{{Sch{\"o}nrich}
  \& {Binney}}{2009a}]{Schonrich2009}
{Sch{\"o}nrich} R.,  {Binney} J.,  2009a, \mn@doi [\mnras]
  {10.1111/j.1365-2966.2009.14750.x}, \href
  {https://ui.adsabs.harvard.edu/abs/2009MNRAS.396..203S} {396, 203}

\bibitem[\protect\citeauthoryear{{Sch{\"o}nrich} \& {Binney}}{{Sch{\"o}nrich}
  \& {Binney}}{2009b}]{2009schonrichBinney}
{Sch{\"o}nrich} R.,  {Binney} J.,  2009b, \mn@doi [\mnras]
  {10.1111/j.1365-2966.2009.15365.x}, \href
  {https://ui.adsabs.harvard.edu/abs/2009MNRAS.399.1145S} {399, 1145}

\bibitem[\protect\citeauthoryear{{Sch{\"o}nrich}, {Binney}  \&
  {Dehnen}}{{Sch{\"o}nrich} et~al.}{2010}]{Schonrich2010_lsr}
{Sch{\"o}nrich} R.,  {Binney} J.,   {Dehnen} W.,  2010, \mn@doi [\mnras]
  {10.1111/j.1365-2966.2010.16253.x}, \href
  {https://ui.adsabs.harvard.edu/abs/2010MNRAS.403.1829S} {403, 1829}

\bibitem[\protect\citeauthoryear{{Sellwood} \& {Binney}}{{Sellwood} \&
  {Binney}}{2002}]{Selwood2002}
{Sellwood} J.~A.,  {Binney} J.~J.,  2002, \mn@doi [\mnras]
  {10.1046/j.1365-8711.2002.05806.x}, \href
  {https://ui.adsabs.harvard.edu/abs/2002MNRAS.336..785S} {336, 785}

\bibitem[\protect\citeauthoryear{{Sharma} et~al.,}{{Sharma}
  et~al.}{2018}]{Sharma2018}
{Sharma} S.,  et~al., 2018, \mn@doi [\mnras] {10.1093/mnras/stx2582}, \href
  {https://ui.adsabs.harvard.edu/abs/2018MNRAS.473.2004S} {473, 2004}

\bibitem[\protect\citeauthoryear{{Sharma}, {Hayden}  \&
  {Bland-Hawthorn}}{{Sharma} et~al.}{2021}]{2021Sharma}
{Sharma} S.,  {Hayden} M.~R.,   {Bland-Hawthorn} J.,  2021, \mn@doi [\mnras]
  {10.1093/mnras/stab2015}, \href
  {https://ui.adsabs.harvard.edu/abs/2021MNRAS.507.5882S} {507, 5882}

\bibitem[\protect\citeauthoryear{{Sharma} et~al.,}{{Sharma}
  et~al.}{2022}]{Sharma2022}
{Sharma} S.,  et~al., 2022, \mn@doi [\mnras] {10.1093/mnras/stab3341}, \href
  {https://ui.adsabs.harvard.edu/abs/2022MNRAS.510..734S} {510, 734}

\bibitem[\protect\citeauthoryear{{Shetrone} et~al.,}{{Shetrone}
  et~al.}{2019}]{Shetrone2019}
{Shetrone} M.,  et~al., 2019, \mn@doi [\apj] {10.3847/1538-4357/aaff66}, \href
  {https://ui.adsabs.harvard.edu/abs/2019ApJ...872..137S} {872, 137}

\bibitem[\protect\citeauthoryear{Simpson}{Simpson}{1951}]{simpson1951}
Simpson E.~H.,  1951, Journal of the Royal Statistical Society: Series B
  (Methodological), 13, 238

\bibitem[\protect\citeauthoryear{{Smith} et~al.,}{{Smith}
  et~al.}{2021}]{Smith2021}
{Smith} V.~V.,  et~al., 2021, \mn@doi [\aj] {10.3847/1538-3881/abefdc}, \href
  {https://ui.adsabs.harvard.edu/abs/2021AJ....161..254S} {161, 254}

\bibitem[\protect\citeauthoryear{{Souto} et~al.,}{{Souto}
  et~al.}{2018}]{Souto2018}
{Souto} D.,  et~al., 2018, \mn@doi [\apj] {10.3847/1538-4357/aab612}, \href
  {https://ui.adsabs.harvard.edu/abs/2018ApJ...857...14S} {857, 14}

\bibitem[\protect\citeauthoryear{{Spina} et~al.,}{{Spina}
  et~al.}{2021}]{Spina2021}
{Spina} L.,  et~al., 2021, \mn@doi [\mnras] {10.1093/mnras/stab471}, \href
  {https://ui.adsabs.harvard.edu/abs/2021MNRAS.503.3279S} {503, 3279}

\bibitem[\protect\citeauthoryear{{Spina}, {Magrini}  \& {Cunha}}{{Spina}
  et~al.}{2022}]{2022spina}
{Spina} L.,  {Magrini} L.,   {Cunha} K.,  2022, \mn@doi [Universe]
  {10.3390/universe8020087}, \href
  {https://ui.adsabs.harvard.edu/abs/2022Univ....8...87S} {8, 87}

\bibitem[\protect\citeauthoryear{{Spitoni} et~al.,}{{Spitoni}
  et~al.}{2023}]{Spitoni2023}
{Spitoni} E.,  et~al., 2023, \mn@doi [\aap] {10.1051/0004-6361/202244349},
  \href {https://ui.adsabs.harvard.edu/abs/2023A&A...670A.109S} {670, A109}

\bibitem[\protect\citeauthoryear{{Steinmetz} \& {Mueller}}{{Steinmetz} \&
  {Mueller}}{1994}]{Steinmetz1994}
{Steinmetz} M.,  {Mueller} E.,  1994, \mn@doi [\aap]
  {10.48550/arXiv.astro-ph/9312010}, \href
  {https://ui.adsabs.harvard.edu/abs/1994A&A...281L..97S} {281, L97}

\bibitem[\protect\citeauthoryear{{Tepper-Garc{\'\i}a} \&
  {Bland-Hawthorn}}{{Tepper-Garc{\'\i}a} \&
  {Bland-Hawthorn}}{2018}]{TepperGarcia2018}
{Tepper-Garc{\'\i}a} T.,  {Bland-Hawthorn} J.,  2018, \mn@doi [\mnras]
  {10.1093/mnras/sty1359}, \href
  {https://ui.adsabs.harvard.edu/abs/2018MNRAS.478.5263T} {478, 5263}

\bibitem[\protect\citeauthoryear{Ting, Freeman, Kobayashi, De~Silva  \&
  Bland-Hawthorn}{Ting et~al.}{2012}]{ting2012principal}
Ting Y.-S.,  Freeman K.~C.,  Kobayashi C.,  De~Silva G.~M.,   Bland-Hawthorn
  J.,  2012, Monthly Notices of the Royal Astronomical Society, 421, 1231

\bibitem[\protect\citeauthoryear{{Tinsley}}{{Tinsley}}{1979}]{Tinsley1979}
{Tinsley} B.~M.,  1979, \mn@doi [\apj] {10.1086/157039}, \href
  {https://ui.adsabs.harvard.edu/abs/1979ApJ...229.1046T} {229, 1046}

\bibitem[\protect\citeauthoryear{{Vincenzo} \& {Kobayashi}}{{Vincenzo} \&
  {Kobayashi}}{2018}]{Vincenzo2018}
{Vincenzo} F.,  {Kobayashi} C.,  2018, \mn@doi [\mnras]
  {10.1093/mnras/sty1047}, \href
  {https://ui.adsabs.harvard.edu/abs/2018MNRAS.478..155V} {478, 155}

\bibitem[\protect\citeauthoryear{{Vincenzo} \& {Kobayashi}}{{Vincenzo} \&
  {Kobayashi}}{2020}]{Vincenzo2020}
{Vincenzo} F.,  {Kobayashi} C.,  2020, \mn@doi [\mnras]
  {10.1093/mnras/staa1451}, \href
  {https://ui.adsabs.harvard.edu/abs/2020MNRAS.496...80V} {496, 80}

\bibitem[\protect\citeauthoryear{{Vincenzo}, {Spitoni}, {Calura}, {Matteucci},
  {Silva Aguirre}, {Miglio}  \& {Cescutti}}{{Vincenzo}
  et~al.}{2019}]{Vincenzo2019}
{Vincenzo} F.,  {Spitoni} E.,  {Calura} F.,  {Matteucci} F.,  {Silva Aguirre}
  V.,  {Miglio} A.,   {Cescutti} G.,  2019, \mn@doi [\mnras]
  {10.1093/mnrasl/slz070}, \href
  {https://ui.adsabs.harvard.edu/abs/2019MNRAS.487L..47V} {487, L47}

\bibitem[\protect\citeauthoryear{{Weinberg} et~al.,}{{Weinberg}
  et~al.}{2019}]{Weinberg2019}
{Weinberg} D.~H.,  et~al., 2019, \mn@doi [\apj] {10.3847/1538-4357/ab07c7},
  \href {https://ui.adsabs.harvard.edu/abs/2019ApJ...874..102W} {874, 102}

\bibitem[\protect\citeauthoryear{{Wilson} et~al.,}{{Wilson}
  et~al.}{2019}]{Wilson2019_apogeeRes}
{Wilson} J.~C.,  et~al., 2019, \mn@doi [\pasp] {10.1088/1538-3873/ab0075},
  \href {https://ui.adsabs.harvard.edu/abs/2019PASP..131e5001W} {131, 055001}

\bibitem[\protect\citeauthoryear{{Yong}, {Carney}  \& {Friel}}{{Yong}
  et~al.}{2012}]{Yong2012}
{Yong} D.,  {Carney} B.~W.,   {Friel} E.~D.,  2012, \mn@doi [\aj]
  {10.1088/0004-6256/144/4/95}, \href
  {https://ui.adsabs.harvard.edu/abs/2012AJ....144...95Y} {144, 95}

\bibitem[\protect\citeauthoryear{{Zasowski} et~al.,}{{Zasowski}
  et~al.}{2017}]{Zasowski2017_apogee2}
{Zasowski} G.,  et~al., 2017, \mn@doi [\aj] {10.3847/1538-3881/aa8df9}, \href
  {https://ui.adsabs.harvard.edu/abs/2017AJ....154..198Z} {154, 198}

\bibitem[\protect\citeauthoryear{{Zhang}, {Chen}  \& {Zhao}}{{Zhang}
  et~al.}{2021}]{Zhang2021}
{Zhang} H.,  {Chen} Y.,   {Zhao} G.,  2021, \mn@doi [\apj]
  {10.3847/1538-4357/ac0e92}, \href
  {https://ui.adsabs.harvard.edu/abs/2021ApJ...919...52Z} {919, 52}

\makeatother
\end{thebibliography}

\bsp	
\label{lastpage}
\end{document}